\documentclass[fleqn,usenatbib]{mnras}

\usepackage{newtxtext,newtxmath}

\usepackage[T1]{fontenc}
\usepackage{ae,aecompl}

\usepackage{graphicx}	
\usepackage{amsmath}	

\def\2dfdr{{\sc 2dfDR}}

\newcommand{\vmf}{$V_{\rm{m4}}$}
\newcommand{\smf}{$\sigma_{\rm{m4}}$}
\newcommand{\vmt}{$V_{\rm{m2}}$}
\newcommand{\smt}{$\sigma_{\rm{m2}}$}
\newcommand{\sobs}{\ifmmode{\sigma_{\rm{obs}}}\else{$\sigma_{\rm{obs}}$}\fi}

\newcommand{\lr}{\ifmmode{\lambda_R}\else{$\lambda_{R}$}\fi}
\newcommand{\lre}{\ifmmode{\lambda_{R_{\rm{e}}}}\else{$\lambda_{R_{\rm{e}}}$}\fi}
\newcommand{\re}{\ifmmode{R_{\rm{e}}}\else{$R_{\rm{e}}$}\fi}
\newcommand{\vs}{\ifmmode{V / \sigma}\else{$V / \sigma$}\fi}
\newcommand{\vse}{\ifmmode{(V / \sigma)_{\rm{e}}}\else{$(V / \sigma)_{\rm{e}}$}\fi}

\newcommand{\kms}{\ifmmode{\,\rm{km}\,\rm{s}^{-1}}\else{$\,$km$\,$s$^{-1}$}\fi}
\newcommand{\msun}{\ifmmode{\,\rm{M}_{\odot}}\else{M$_{\odot}$}\fi}
\newcommand{\mstar}{\ifmmode{M_{\star}}\else{$M_{\star}$}\fi}
\newcommand{\logm}{\ifmmode{\log(M_{\star}/M_{\odot})}\else{$\log(M_{\star}/M_{\odot})$}\fi}
\newcommand{\mm}{\ifmmode{M_{\star}/M_{\odot}}\else{$M_{\star}/M_{\odot}$}\fi}

\newcommand{\kinemetry}{\textsc{kinemetry}}
\newcommand{\mk}{\ifmmode{\overline{k_5 / k_1}}\else{$\overline{k_5 / k_1}$}\fi}
\newcommand{\ha}{H\,$\alpha$}
\newcommand{\hb}{H\,$\beta$}
\newcommand{\hc}{H\,$\gamma$}
\newcommand{\hd}{H\,$\delta$}
\newcommand{\nii}{[N\,{\small II}]}
\newcommand{\sii}{[S\,{\small II}]}
\newcommand{\oii}{[O\,{\small II}]}
\newcommand{\oiii}{[O\,{\small III}]}
\def \update {}

\title[The SAMI Galaxy Survey: DR3]{The SAMI Galaxy Survey: the third and final data release}

\author[S. M. Croom et al.]{Scott M. Croom$^{1,2}$\thanks{E-mail: scott.croom@sydney.edu.au (SMC)},
Matt S. Owers$^{3, 4}$,
Nicholas Scott$^{1,2}$,
Henry Poetrodjojo$^{1,2}$, \newauthor
Brent Groves$^{5,6,2}$,
Jesse van de Sande$^{1,2}$,
Tania M. Barone$^{6,1,2}$, 
Luca Cortese$^{5,2}$,\newauthor
Francesco D'Eugenio$^{7}$,
Joss Bland-Hawthorn$^{1,2}$,
Julia Bryant$^{1,2}$,
Sree Oh$^{6,2}$, \newauthor
Sarah Brough$^{8,2}$,
James Agostino$^{9}$,
Sarah Casura$^{10}$,
Barbara Catinella$^{5,2}$,\newauthor
Matthew Colless$^{6,2}$, 
Gerald Cecil$^{11}$,
Roger L. Davies$^{12}$,
Michael J. Drinkwater$^{13}$,\newauthor
Simon P. Driver$^{5}$,
Ignacio Ferreras$^{14,15,16}$,
Caroline Foster$^{1,2}$,
Amelia Fraser-McKelvie$^{5,2}$,\newauthor
Jon Lawrence$^{17}$,
Sarah K. Leslie$^{18,2}$,
Jochen Liske$^{10}$,
\'Angel R. L\'opez-S\'anchez$^{3,4,2}$,\newauthor
Nuria P. F. Lorente$^{17}$,
Rebecca McElroy$^{1,2}$,
Anne M. Medling$^{9,2}$,
Danail Obreschkow$^{5,2}$,\newauthor
Samuel N. Richards$^{19}$,
Rob Sharp$^{6}$,
Sarah M. Sweet$^{12,2}$,
Dan S. Taranu$^{20}$,\newauthor
Edward N. Taylor$^{21}$,
Edoardo Tescari$^{22}$,
Adam D. Thomas$^{6,2}$,
James Tocknell$^{17}$,\newauthor
Sam P. Vaughan$^{1,2}$
\\
$^{1}$Sydney Institute for Astronomy (SIfA), School of Physics, The University of Sydney, NSW, 2006, Australia\\
$^{2}$ARC Centre of Excellence for All Sky Astrophysics in 3 Dimensions (ASTRO 3D)\\
$^{3}$Department of Physics and Astronomy, Macquarie University, NSW, 2109, Australia\\
$^{4}$Astronomy, Astrophysics and Astrophotonics Research Centre, Macquarie University, Sydney, NSW 2109, Australia\\
$^{5}$ICRAR, The University of Western Australia, Crawley WA 6009, Australia\\
$^{6}$Research School of Astronomy and Astrophysics, Australian
National University, Canberra, ACT 2611, Australia\\
$^{7}$Sterrenkundig Observatorium, Universiteit Gent, Krijgslaan 281 S9, B-9000 Gent, Belgium\\
$^{8}$School of Physics, University of New South Wales, NSW 2052,
Australia\\
$^{9}$Ritter Astrophysical Research Center, University of Toledo,
Mail Stop 111, Toledo, OH, 43606, United States\\
$^{10}$ Hamburger Sternwarte, Universit{\"a}t Hamburg, Gojenbergsweg 112, 21029 Hamburg, Germany\\
$^{11}$Dept. Physics and Astronomy University of North Carolina Chapel
Hill, NC 27599 USA\\
$^{12}$Astrophysics, Department of Physics, University of Oxford, Denys Wilkinson Building, Keble Rd., Oxford, OX1 3RH, UK.\\
$^{13}$School of Mathematics and Physics, University of Queensland,
Brisbane, QLD 4072, Australia\\
$^{14}$ Department of Physics and Astronomy, University College London, Gower Street, London WC1E 6BT, UK\\ 
$^{15}$ Instituto de Astrofisica de Canarias, Calle Via Lactea s/n, E38205 La Laguna, Tenerife, Spain\\
$^{16}$ Departamento de Astrofisica, Universidad de La Laguna (ULL), La Laguna, E-38206 Tenerife, Spain\\ 
$^{17}$Australian Astronomical Optics - Macquarie, Macquarie
University, NSW 2109, Australia\\
$^{18}$ Leiden Observatory, Leiden University, PO Box 9513, NL-2300 RA Leiden, the Netherlands\\
$^{19}$SOFIA Science Center, USRA, NASA Ames Research Center, Building
N232, M/S 232-12, P.O. Box 1, Moffett Field, CA 94035-0001, USA\\
$^{20}$ Department of Astrophysical Sciences, Princeton University, 4 Ivy Lane, Princeton, NJ 08544, USA\\
$^{21}$Centre for Astrophysics and Supercomputing, Swinburne University of Technology, John Street, Hawthorn, VIC 3122, Australia\\
$^{22}$ Melbourne Data Analytics Platform (MDAP), The University of Melbourne, Parkville, VIC 3010, Australia
}

\date{Accepted XXX. Received YYY; in original form ZZZ}

\pubyear{2021}

\begin{document}
\label{firstpage}
\pagerange{\pageref{firstpage}--\pageref{lastpage}}
\maketitle

\begin{abstract}
We have entered a new era where integral-field spectroscopic surveys of galaxies are sufficiently large to adequately sample large-scale structure over a cosmologically significant volume. This was the primary design goal of the SAMI Galaxy Survey. Here, in Data Release~3 (DR3), we release data for the full sample of 3068 unique galaxies observed.  This includes the SAMI cluster sample of 888 unique galaxies for the first time. For each galaxy, there are two primary spectral cubes covering the blue (370--570\,nm) and red (630--740\,nm) optical wavelength ranges at spectral resolving power of $R=1808$ and 4304 respectively. For each primary cube, we also provide three spatially binned spectral cubes and a set of standardized aperture spectra. For each galaxy, we include complete 2D maps from parameterized fitting to the emission-line and absorption-line spectral data. These maps provide information on the gas ionization and kinematics, stellar kinematics and populations, and more. All data are available online through Australian Astronomical Optics (AAO) Data Central.
\end{abstract}




\section{Introduction}

Our understanding of how galaxies form and evolve has taken great strides in the last few decades, but we are far from a complete picture. No two galaxies are the same, as illustrated by many attempts to find close analogues of the Milky Way or M31 \citep[e.g.][]{2016ARA&A..54..529B,2020MNRAS.498.4943B}.
This complexity is apparent in the range of distinct components within galaxies, e.g.\ truncated or extended dark matter haloes, one or more disks, long bars and rings, short bars/bulges, smooth or structured stellar bulges and haloes, central star clusters and nuclear disks, or a massive black hole, as well as different gas/dust (molecular, atomic, ionized) phases. 
To further complicate matters, the different components or phases interact in a variety of ways: gas cooling to form stars, stellar  or supernovae feedback, feedback from a central super-massive black hole, dynamical mixing from bars, etc.  It is a daunting task to form a coherent narrative for all of these processes over cosmic time, and to demonstrate the robustness of this narrative with a consistent set of cosmological simulations. Yet only when this is achieved can we begin to claim a solid understanding of the primary processes involved in a galaxy's life cycle.

In recent years, questions have been raised about the limitations of finite resolution in cosmological simulations, as well as chaotic-like behaviour
--- the `butterfly effect' --- particularly in relation to the inherent complexity of so many competing processes and whether we will ever be able to track these interactions meaningfully \citep{2019ApJ...871...21G,2019MNRAS.482.2244K,2020arXiv200403608K}.  To make robust tests of these simulations requires us to acquire observations with sufficient information on each galaxy.  Coupled with this, we need to sample galaxies over a cosmologically representative volume.

We know that mass is a fundamental parameter in controlling galaxy properties.  For example, star formation, mean stellar age and morphology are strongly linked to stellar mass \citep[e.g.][]{2006MNRAS.373..469B}. We also now understand that a galaxy's life is shaped by its surroundings. Environment is known to modify galaxy morphology \citep[e.g.][]{1980ApJ...236..351D}, current star formation \citep[e.g.][]{2002MNRAS.334..673L} and star formation history \citep[e.g.][]{2009ARA&A..47..159B}.  In some cases it appears that the trends in mass and environment may be separable \citep[e.g.][]{2010ApJ...721..193P}.  

Other properties of galaxies also have important evolutionary roles, such as angular momentum, binding energy, gas and dark matter fractions.  A challenge for us is to separate the trends in the driving parameters from stochastic effects.  In the fullness of time, it may even be possible to reveal the drivers behind apparently stochastic evolution, for example, uncovering the detailed merger history of a galaxy.  However, for now we often have to average over the stochasticity, and this is one of the requirements that drives us to large samples of galaxies. To complicate matters further, we observe galaxies over a randomized distribution of viewing angles. This also pushes us to larger samples in order to determine internal properties from projected properties. 

The requirement for larger samples has driven the large-scale multi-fibre spectroscopic surveys of the last two decades \citep{2000AJ....120.1579Y,2001MNRAS.328.1039C,2011MNRAS.413..971D}, that have characterized the local galaxy distribution very effectively. However, the need to understand the internal structure of galaxies has driven the development of integral-field spectroscopic surveys. This was pioneered by the SAURON project \citep{2001MNRAS.326...23B} and followed by ATLAS3D \citep{2011MNRAS.413..813C}, both focussed on early-type galaxies. The CALIFA survey was the first to cover a large number ($\sim600$) of galaxies of all types \citep{2012A&A...538A...8S}. The complex multi-parameter nature of galaxy formation drives us to much larger samples, which motivated the step to multi-object integral field spectrographs. The SAMI Galaxy Survey \citep{2012MNRAS.421..872C,2015MNRAS.447.2857B} was the first of these large-scale multiplexed projects, followed by MaNGA \citep{2015ApJ...798....7B}. Future projects such as Hector \citep{2016SPIE.9908E..1FB} will further extend the reach of integral-field spectroscopic surveys.

The SAMI Galaxy Survey \citep{2012MNRAS.421..872C,2015MNRAS.447.2857B} aimed to span the plane of mass and environment with a large sample of galaxies, each with spatially-resolved structural and kinematic measurements.  The specific science goals for the survey were to answer the questions i) what is the physical role of environment in galaxy evolution? ii) What is the interplay between gas flows and galaxy evolution? iii) How are mass and angular momentum built up in galaxies?  As part of this we aimed to compare and contrast our 3D integral field data cubes for each galaxy with synthetic galaxies emerging from cosmological simulations sampled in the same fashion. Our first detailed comparisons reveal that all simulators are able to match a subset of the galaxy parameters, but often at the expense of other parameters \citep{2019MNRAS.484..869V}.  The inconsistencies are due largely to our limited understanding of the many complex baryonic processes that work together over billions of years.

The SAMI Galaxy Survey observations took place between 2013 and 2018, obtaining data for over 3000 galaxies. These data have been used in a wide variety of scientific analyses, including studies of galactic winds \citep{2012ApJ...761..169F,2014MNRAS.444.3894H,2016MNRAS.457.1257H}, the relationship of angular momentum and spin to galaxy properties and environment \citep{2014MNRAS.443..485F,2016MNRAS.463..170C,vandesande2017a,2017ApJ...844...59B,2017MNRAS.472..966F,2020MNRAS.491.2864W}, stellar populations \citep{2017MNRAS.472.2833S,2018ApJ...856...64B,2019MNRAS.489..608F,2020ApJ...896...75S}, star formation and quenching \citep{2018MNRAS.475.5194M,2019MNRAS.483.2851S,2019ApJ...873...52O,2019MNRAS.485.2656C,2020MNRAS.495.2265V}, gas-phase metallicity, \citep{2018MNRAS.479.5235P,2019MNRAS.484.3042S} kinematic and structural scaling relations \citep{2014ApJ...795L..37C,2017MNRAS.472.1809B,2019MNRAS.487.2924B,2020MNRAS.495.4638O}, detailed comparison to simulations \citep{2019MNRAS.484..869V,2020ApJ...894..106K}, and much more. The SAMI Galaxy Survey team have provided regular data releases \citep{2015MNRAS.446.1567A,2018MNRAS.475.716G,2018MNRAS.481.2299S}. In this current paper we present the third and final data release (DR3) of all SAMI observations, together with value-added products such as stellar kinematics, stellar populations and emission line fits.  This is the first SAMI data release to contain data from the eight massive galaxy clusters observed as part of the SAMI Galaxy Survey.  We also include for the first time environmental metrics for the entire SAMI sample.

\begin{table*}
\centering
\caption{The coordinates and number of objects in the GAMA regions. For each region we list the right ascension (RA) and declination (Dec.). For both the primary and secondary samples we list the number of observed galaxies $N_{\rm obs}$, the number of good targets $N_{\rm good}$ (i.e. those not flagged as bad for photometric reasons such as a bright star in the field-of-view) and the number of all targets $N_{\rm all}$ (including objects with bad flags). The listed completeness for the primary sample is $N_{\rm obs}$/$N_{\rm good}$.\label{table:gama}}
    \begin{tabular}{ccccccc}
    \hline
Region & RA (J2000) & Dec. (J2000) & Primary & Secondary & Primary \\
& deg & deg & $N_{\rm obs}$/$N_{\rm good}$/$N_{\rm all}$ & $N_{\rm obs}$/$N_{\rm good}$/$N_{\rm all}$ &  Completeness\\
\hline
GAMA 09h & 129.0 to 141.0 &	$-1$ to $+3$ & 575/683/806 & 82/699/820 & 84.2\% \\
GAMA 12h & 174.0 to 186.0 &	$-2$ to $+2$ & 637/728/805 & 65/900/982 & 87.5\% \\
GAMA 15h & 211.5 to 223.5 &	$-2$ to $+2$ & 728/995/1127 & 13/915/996 & 73.2\% \\
\hline
Total & & & 1940/2406/2738 & 160/2514/2798 & 80.6\%\\
\hline
\end{tabular}
\end{table*}

In Section~\ref{sec:survey} we outline the survey input catalogues, the instrument and the observations. In Section~\ref{sec:qual} we discuss improvements in data reduction and assessments of data quality. Section~\ref{sec:primedata} describes the primary data products from the survey, including cubes (with various binning schemes) and aperture spectra.  Catalogues based on photometric data of SAMI targets are discussed in Section \ref{sec:cat_phot}.  The stellar kinematics and stellar population products are discussed in Sections~\ref{sec:kin} and~\ref{sec:pop} respectively. Emission line  products are presented in Section~\ref{sec:lzifu}. In Section~\ref{sec:env} we discuss environmental metrics within the SAMI Galaxy Survey. Discussion of data access is provided in Section~\ref{sec:access} and we summarize the paper in Section~\ref{sec:summary}. Throughout the paper we assume a cosmology with $\Omega_{\rm m}=0.3$, $\Omega_\Lambda=0.7$ and  $H_0=70$\,\kms\,Mpc$^{-1}$. All magnitudes are in the AB system (Oke \& Gunn 1983) and stellar masses and star-formation rates assume a Chabrier IMF \citep{2003PASP..115..763C}.

\section{The SAMI Galaxy Survey}
\label{sec:survey}

\begin{figure*}
    \centering
    \includegraphics[width=18cm]{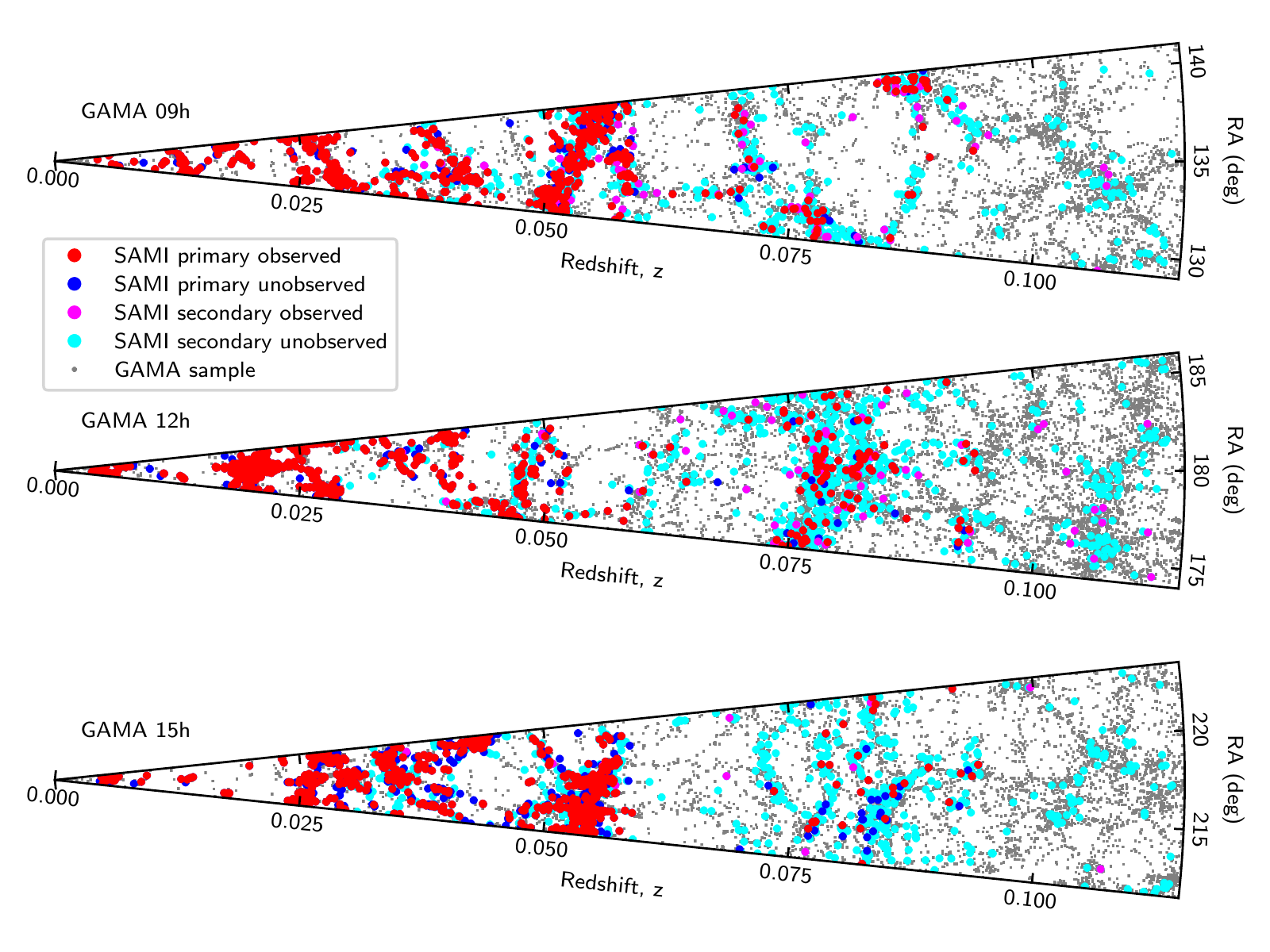}
    \caption{The distribution of SAMI galaxies within the three GAMA regions. We show the original GAMA redshift survey (small gray points), observed (red) and unobserved (blue) primary SAMI galaxies, observed (magenta) and unobserved (cyan) secondary SAMI galaxies.}
    \label{fig:gama_wedge}
\end{figure*}

\subsection{The Input Catalogues}\label{section:Input_cats}
The input catalogues used for the SAMI Galaxy Survey are drawn from the three equatorial regions of the Galaxy And Mass Assembly (GAMA) Survey \citep{2011MNRAS.413..971D}, as described in \citet{2015MNRAS.447.2857B}, and eight cluster regions described in \citet{2017MNRAS.468.1824O}. In addition, a small number of observations were made of filler targets when not all the integral field units could be allocated to main survey targets. We describe the different input catalogues below.

\subsubsection{GAMA regions}

The majority of galaxies in the SAMI Galaxy Survey are within regions observed as part of the GAMA survey. This provides the substantial advantage of deep and complete spectroscopic coverage to select targets and define environment. The GAMA regions also contain excellent photometric data across 21 bands from ultraviolet to far-infrared \citep{2016MNRAS.455.3911D}. \citet{2015MNRAS.447.2857B} describe the selection of targets in the GAMA regions in detail, but for completeness we outline the selection here.

The SAMI Galaxy Survey targets in the GAMA regions were selected within three $4\times12$\,degree regions along the equator (declination $\simeq0$ degrees), centred at approximate right ascension (RA) of 9, 12 and 15 hours (see Table \ref{table:gama}). Galaxies were targeted based on cuts in the redshift-stellar mass plane, with a stellar mass proxy that used $i$-band magnitude and $g-i$ colour \citep{2011MNRAS.418.1587T,2015MNRAS.447.2857B}. Galaxies were selected within a series of four stepped volumes, with higher stellar mass limits at higher redshift \citep[see Fig. 4 of][]{2015MNRAS.447.2857B}. The primary sample is limited to redshift $z<0.095$. Observations aimed to have high and uniform completeness for this primary sample. A secondary sample was also defined that included high mass galaxies to higher redshift ($z<0.115$) and used fainter stellar mass limits. As the secondary targets were observed at lower priority, these are less complete (see Table \ref{table:gama}). Fig.\, \ref{fig:gama_wedge} shows the distribution of primary (observed: red, unobserved: blue) and secondary targets (observed: magenta, unobserved: cyan) in redshift and RA.  A set of filler galaxies, at lower priority that the secondary targets, was also defined.  These are discussed in Section \ref{sec:input_filler}.

Each potential target was visually checked to identify problems, such as bright nearby stars or sources being a sub-component of a larger galaxy. These were flagged within the input catalogue and their priority set so that they were not observed.  The column named BAD\_CLASS in the input catalogue (named InputCatGAMADR3) contains flags for different types of problem sources.  A full description of the flags is given by \citet{2015MNRAS.447.2857B}.  A small subset of the sources required adjustment of the coordinates for targeting (0.6 percent of the input catalogue). This was typically in order to place an asymmetric galaxy or a close pair fully into the integral field unit (IFU) and these objects have BAD\_CLASS=5. The input catalogue that is publicly available as part of DR3 contains both object coordinates and IFU pointing coordinates. 88.8 percent of the input catalogue remained as a target after flagging for problems.  These good objects to be observed had either BAD\_CLASS=0 or 5.  A further case of BAD\_CLASS=8 is also included in observations, but only the cluster regions contain galaxies with these values.  BAD\_CLASS=8 indicates a galaxy that is the smaller component of a close pair, where the more massive galaxy is outside of the field of view of the IFU.  The full GAMA region input sample is contained within the InputCatGAMADR3 catalogue.

A calibration star was observed in one IFU during every galaxy observation. This star allowed improved flux calibration and a good estimate of the spatial point spread function (PSF). These calibration stars were selected to be F-stars, based on their SDSS photometric properties.  \cite{2015MNRAS.447.2857B}  describes the selection of calibration stars in detail. The catalogue of calibration stars is released as part of DR3 and called FstarCatGAMA.

\subsubsection{Cluster regions}\label{subsec:clus_input}

The cluster targets were drawn from eight regions centred at the positions listed in Table~\ref{table:clusters}.   Photometry was based either on SDSS DR9 \citep{2012ApJS..203...21A} or the VLT Survey Telescope ATLAS Survey \citep{2015MNRAS.451.4238S}, with details of the selection described by \citet{2017MNRAS.468.1824O}.
Targets were selected to be within a redshift range defined by the relative velocity of the galaxy with respect to the cluster redshift, $z_{\rm clus}$, such that $|v_{\rm pec}|/\sigma_{200} < 3.5$.  $v_{\rm pec}$ is the peculiar velocity of the galaxy with respect to the cluster redshift and $\sigma_{200}$ is the velocity dispersion of the cluster of interest measured within the over-density radius $R_{200}$ [see Section~\ref{subsection:cluster_env} and \citet{2017MNRAS.468.1824O} for further details]. We note that this cut in redshift is less conservative than the criterion used to allocate cluster membership in \citet{2017MNRAS.468.1824O}, and so the input catalogue contains galaxies that are close to the cluster in redshift space, but may not be bona-fide members. 

\begin{figure*}
    \centering
    \includegraphics[width=17.5cm]{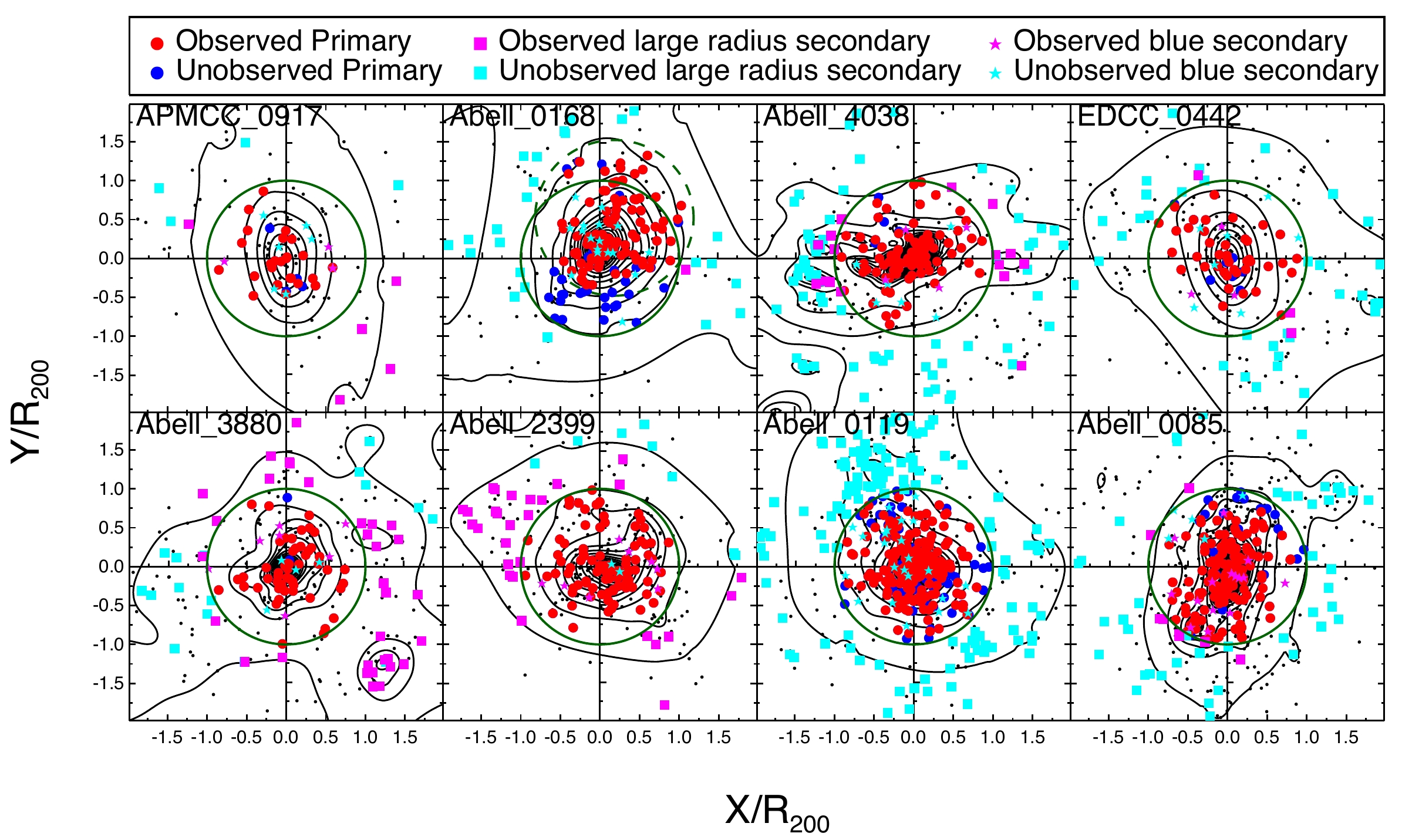}
    \caption{The distribution of SAMI targets in the cluster regions. The small gray points show cluster members selected from the SAMI Cluster Redshift Survey, which are used to define the galaxy surface density isopleths shown as black contours. Large circles show observed (red) and unobserved (blue) primary targets.  The magenta (cyan) points show observed (unobserved) secondary targets.  These secondary targets are indicated by either squares (for galaxies at large radius) or stars (for blue colour-selected galaxies). The green circles show R$_{200}$. For Abell~168, the dashed green circle shows the off-centre region used to include targets within R$_{200}$ of the northern BCG, which was used to define the cluster centre during the early parts of the survey.}
    \label{fig:clus_dist}
\end{figure*}

\begin{table*}
\centering
\caption{Properties of the eight clusters targeted during the SAMI Galaxy Survey. For each cluster, we list the name, RA and Dec.\ of the cluster centre, the cluster redshift, velocity dispersion, virial radius estimate, and virial mass as determined in \citet{2017MNRAS.468.1824O}. For the Primary and Secondary targets, we list the number of galaxies observed during the SAMI Galaxy Survey ($N_{\rm obs}$), the number of galaxies in the input catalogue ($N_{\rm all}$) and the number of good objects in the input catalogue ($N_{\rm good}$). The completeness of the Primary sample is determined as $N_{\rm obs}$/$N_{\rm good}$.
\label{table:clusters}}
    \begin{tabular}{cccccccccc}
    \hline
Region & RA (J2000) & Dec.\ (J2000) & z$_{\rm clus}$ & $\sigma_{200}$ & $R_{200}$& $M_{200}$  & Primary  &Secondary   & Primary \\
& deg & deg  & &(km/s) & (Mpc) &  ($10^{14}\,M_{\odot}$) & $N_{\rm obs}$/$N_{\rm good}$/$N_{\rm all}$ & $N_{\rm obs}$/$N_{\rm good}$/$N_{\rm all}$ &  Comp.\\
\hline
APMCC 917 & 355.397880 & -29.236351 & 0.0509 & 492  & 1.19 & 2.1 & 25/28/29 & 8/17/18 & 89\% \\
Abell 168 &  18.815777 &   0.213486 & 0.0449 & 546 & 1.32 & 3.0 & 95/126/130 & 1/40/43 & 75\%\\  
Abell 168 (N)          &       18.739974 &   0.430807       &       &              &     \\
Abell 4038 & 356.937810 & -28.140661 & 0.0293 & 597&1.46 & 2.9 & 100/104/111   & 19/85/87 & 96\%\\
EDCC 442 &   6.380680 & -33.046570 & 0.0498 & 583 & 1.41&3.6& 41/50/50   & 6/47/47 & 82\%\\
Abell 3880 & 336.977050 & -30.575371 & 0.0578 & 660 & 1.59& 4.6& 50/51/56   & 38/57/60 & 98\%\\ 
Abell 2399 & 329.372605 &  -7.795692 & 0.0580 & 690 & 1.66 &6.1& 91/92/94  & 37/48/49 & 99\%\\ 
Abell 119 &  14.067150 &  -1.255370 & 0.0442 & 840 & 2.04&9.7& 202/255/260  & 0/152/157 &79\%\\
Abell  85 &  10.460211 &  -9.303184 & 0.0549 & 1002 & 2.42 &17.0&  152/167/171 & 23/70/71 & 91\% \\
\hline
Total &  -- &  -- & -- & -- &  -- & -- &  756/873/901 & 132/516/532 & 87\%\\
\hline
    \end{tabular}
\end{table*}

In addition to meeting the aforementioned criteria, targets were further characterised into primary and secondary targets. Primary targets are defined as those that meet the following:
\begin{itemize}
    \item Cluster-centric distance $R < R_{200}$,
    \item for $0.045 < z_{\rm clus} < 0.06$, stellar mass $M^* > 10^{10}$\,M$_{\odot}$,
    \item for $z_{\rm clus} < 0.045$, stellar mass $M^* > 10^{9.5}$\,M$_{\odot}$.
\end{itemize}
Similar to the GAMA regions, lower priority secondary targets were also included, and were selected based on the following criteria:
\begin{itemize}
    \item Blue secondary targets selected as those galaxies with rest-frame $(g-i)_{\rm kcorr} < 0.9$, $R < R_{200}$, and stellar mass 0.5\,dex smaller than the stellar mass limit for the primary targets in the cluster of interest.
    \item large-radius secondary targets were selected to have $R_{200} < R < 2R_{200}$, and with stellar mass above the limit used for the selection of primary targets.
\end{itemize}
The total and observed number of primary and secondary targets for each cluster is listed in Table~\ref{table:clusters}. In Figure~\ref{fig:clus_dist}, the distribution of the primary and secondary targets is shown for each cluster, with positions measured relative to the centres defined in Table~\ref{table:clusters} and normalised by R$_{200}$ (green circle). The black contours show galaxy surface density isopleths generated using the cluster members defined in the SAMI Cluster Redshift Survey \citep[black points;][]{2017MNRAS.468.1824O}. The primary targets are shown as filled circles (observed in red, unobserved in blue), the colour-selected blue secondary galaxies as filled stars (observed in magenta, unobserved in cyan), and the large-radius secondaries as filled squares (observed in magenta, unobserved in cyan).

We note that there are two centres listed for Abell~168 in Table~\ref{table:clusters} because the ongoing merger in Abell~168 means that there are two sub-clusters.   Both the northern and southern sub-clusters contain bright cluster galaxies \citep{2004ApJ...610L..81H, 2014MNRAS.443..485F}, which leads to some ambiguity in defining the cluster centre. Initially, the cluster centre was defined at the position of the more massive bright cluster galaxy associated with the northern substructure (second listing in Table~\ref{table:clusters}; dashed green circle in Figure~\ref{fig:clus_dist}). Later, the centre was redefined to the southern bright cluster galaxy due to its proximity to the peak in the galaxy surface density \citep{2017MNRAS.468.1824O}. For consistency, we define the clustercentric distances of the targets in Abell~168 with respect to the southernmost coordinates listed in Table~\ref{table:clusters}. However, the galaxies initially allocated as primary targets using the northernmost coordinates maintain their status as primary targets, being within $R_{200}$ of the northernmost substructure.

Full information for the cluster region input sample is listed in the InputCatClustersDR3 catalogue.  Calibration stars in the cluster regions  are listed in the FstarCatClusters catalogue.

\begin{table}
\centering
\caption{Table of different filler targets defined by their FILLFLAG parameter that defines which filler sample they are from (see Section \ref{sec:input_filler} for details). The number of filler targets $N_{\rm all}$ and the number observed $N_{\rm obs}$ are listed.}\label{table:filler}
    \begin{tabular}{ccc}
    \hline
FILLFLAG  & $N_{\rm all}$ & $N_{\rm obs}$ \\
\hline
20 &  22 & 1\\
30 & 1800 & 36 \\
40 & 141 & 1\\
50 & 996 & 13\\
90 & 21 & 21\\
\hline
\end{tabular}
\end{table}

\subsubsection{Filler targets}\label{sec:input_filler}

In some fields there were not sufficient primary and secondary targets to fill all the IFUs. This was particularly the case for fields observed towards the end of the survey. In some cases spare IFUs were used to target primary and secondary targets that had already been observed. This was either because of low data quality in previous observations, or in order to obtain repeat observations for assessment of data quality. However, extra filler targets were also included for particular science cases in three categories.  The filler targets are listed in the catalogue InputCatFiller.  The FILLFLAG column in that catalogue identifies the particular class of objects used as fillers.  These are broken down into the following catagories:
\begin{itemize}
    \item Galaxies with 21cm H{\sc i} detections from the Arecibo Legacy Fast ALFA (ALFALFA) Survey \citep{2005AJ....130.2598G,2018ApJ...861...49H}, but that are lower mass than the SAMI selection limits (FILLFLAG=20). We include both high signal-to-noise ratio (S/N) and marginal H{\sc i} detections (i.e., ALFALFA detcode=1 and 2).
    \item Close pairs of galaxies identified from the GAMA Survey \citep{2014MNRAS.444.3986R} that fall outside of the SAMI selection limits. These are in two classes:  either both galaxies in the pair are outside of the selection boundaries (FILLFLAG=30), or one of the pair is within the main SAMI sample (FILLFLAG=40). 
    \item Typical star forming disk galaxies at $0.12 < z < 0.15$ (i.e., slightly beyond the SAMI redshift limit), to explore the potential for using velocity information as a means to precision weak lensing experiments \citep{deBurghDay_2015, Gurri_2020} (FILLFLAG=50).
\end{itemize}
The input catalogue for these filler objects (InputCatFiller) forms part of DR3, but is simplified compared to the main SAMI targets, containing only positions, redshifts and FILLFLAG. A small number of cluster galaxies (21) observed early in the survey, but that did not meet the final selection limits are also contained within the filler catalogue (with FILLFLAG=90).

\subsection{The SAMI Instrument}
The SAMI instrument is a multi-object integral field spectrograph comprising of 13 optical fibre integral field units feeding the AAOmega spectrograph \citep{2012MNRAS.421..872C,2015MNRAS.447.2857B}. It is installed on the 1 degree diameter prime focus of the Anglo-Australian Telescope (AAT) in NSW, Australia. The IFUs are {\it hexabundles} - optical fibre imaging bundles with >75\% of the light collected in fibre cores \citep{2011OpExpr.19.2649,2011MNRAS.415.2173}. These unique bundles are the product of a fusing technique that allows tight packing but without any extra loss of light through focal ratio degradation \citep{2014MNRAS.438.869}. The hexabundles each contain 61 fibres with core diameter of $105\mu$m or 1.6 arcsec, subtending $15$ arcsec diameter across the hexabundle.

Each field has galaxy and star positions drilled into a plug plate. Typically two different fields are drilled into the same physical plate. The 13 hexabundles and 26 sky fibres, along with 3 guide bundles are plugged by hand. SAMI makes use of the AAOmega spectrograph which is a flexible dual-arm workhorse spectrograph at the AAT \citep{2006SPIE.6269E..14S}. For the SAMI survey we used the 580V and 1000R gratings, delivering a wavelength range of $3750-5750$\,\AA\ and $6300-7400$\,\AA\ for the blue and red arms respectively. The spectral resolutions are $R=1808$ and $R=4304$ for the blue and red arms, equivalent to an effective velocity dispersion of $\sigma$ of 70.4 and 29.6\kms\ respectively  \citep{vandesande2017a,2018MNRAS.481.2299S}.

\begin{figure*}
    \centering
    \includegraphics[width=15cm]{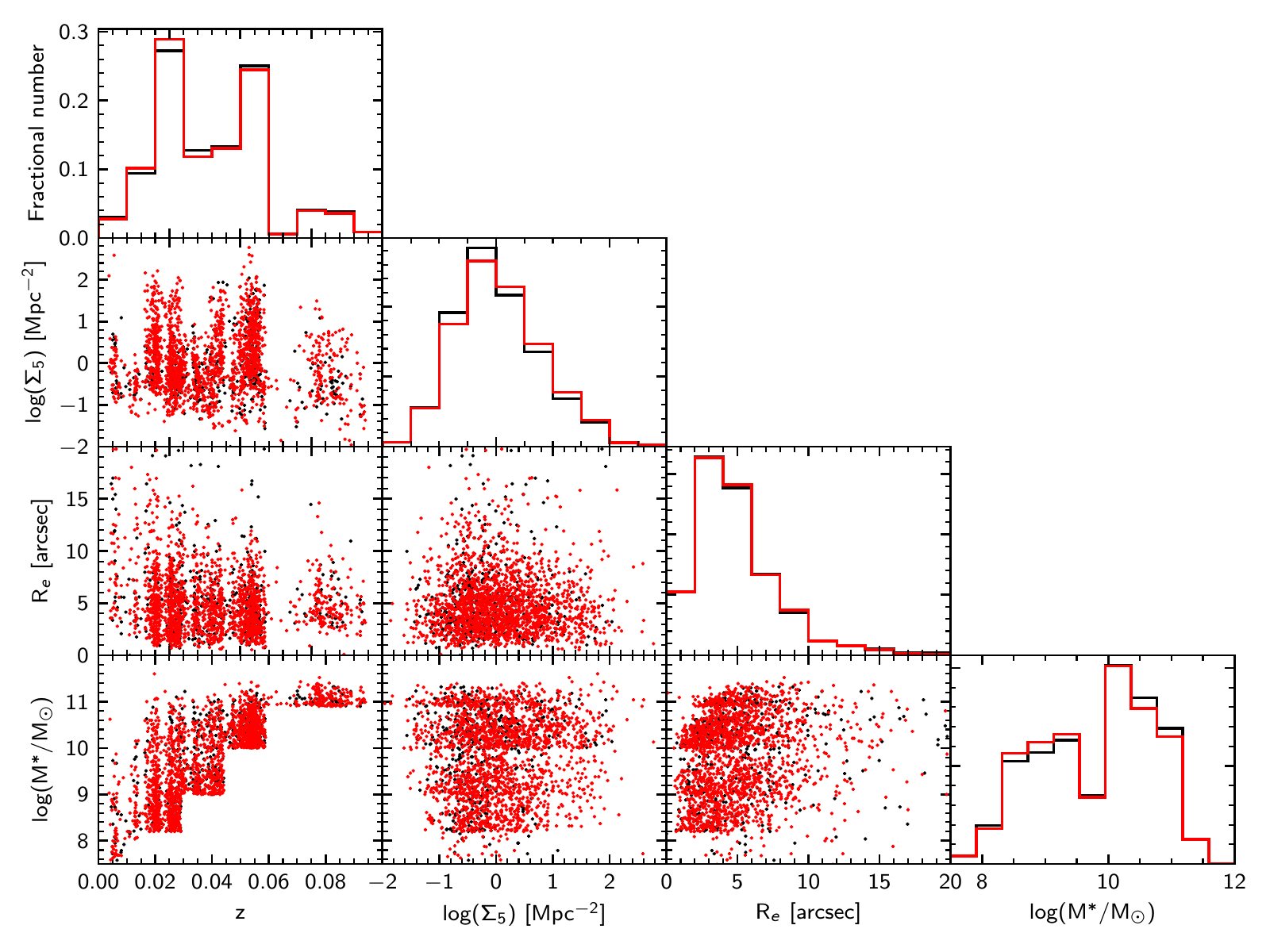}
    \caption{The distribution of good primary SAMI targets in the GAMA regions (black points) compared to the observed targets (red points) as a function of redshift, $\log(M^*/M_\odot)$, $R_{\rm e}$ (major axis in arcsec) and 5th nearest neighbour density $\log(\Sigma_5)$. Histograms are normalized separately for each sample.}
    \label{fig:gama_comp}
\end{figure*}

\subsection{Observations}\label{sec:obs}

The SAMI Galaxy Survey observations took place over 250 nights from March 2013 to May 2018. Each field, containing 12 galaxy targets and one calibration star, would typically be observed for 7 exposures, each of 1800s. Between exposures the field centre would be offset in a hexagonal pattern to provide uniform coverage of the target in the bundle, allowing for the small gaps between fibres \citep{2015MNRAS.446.1551S}. At least one arc frame and one dome flat field frame were taken for each field. Where possible twilight flats were also taken to aid various aspects of calibration. Primary flux standards were observed at the start and end of the night when conditions were photometric.

During observations data quality was checked both in terms of the spatial point spread function (PSF) and transmission. Any exposure where the PSF had FWHM $>3$ arcsec, or the transmission was less than 70 percent of the nominal system transmission was flagged for re-observation. Due to scheduling constraints not all flagged data could be re-observed. The threshold in data quality for generating cubes in data reduction was somewhat more relaxed than the above observational constraints (to allow all useful data to be included). Exposures were added to cubes if they had: i) exposure time $>600$ sec; ii) a relative transmission of $>33$ percent; iii) seeing FWHM of $<4$ arcsec.  Only targets that had at least 6 frames that passed these quality controls were made into reduced cubes.

Repeat observations of SAMI galaxies fall into two classes. The first is where galaxies are observed within the same plate configuration, but on different observing runs (typically separated by a month or more in time). In this case the individual exposures are combined into a single cube across all the observations. However, we also make a cube from the individual observing runs if there are sufficient exposures that meet our quality control limits. In this case there will be different cubes that share some of the same individual exposures. The main reason for such combinations was insufficient high quality exposures within a single observing run. 

The second class of repeat observations was where the same galaxy was observed in two different plates (and therefore usually different hexabundles). This could be during the same or different observing runs. For this class of repeat the data were not combined across the different plates, and hence the galaxy will have two completely independent cubes. There are 215 galaxies with multiple cubes from the same plate, that share some of their data. There are 70 galaxies that have repeated observations from different plates that are fully independent. The best cube, based on seeing FWHM and S/N ratio is identified with the ISBEST flag within the CubeObs catalogue that describes observations of all sources (see Section \ref{sec:incat} below).

\begin{figure*}
    \centering
    \includegraphics[width=15cm]{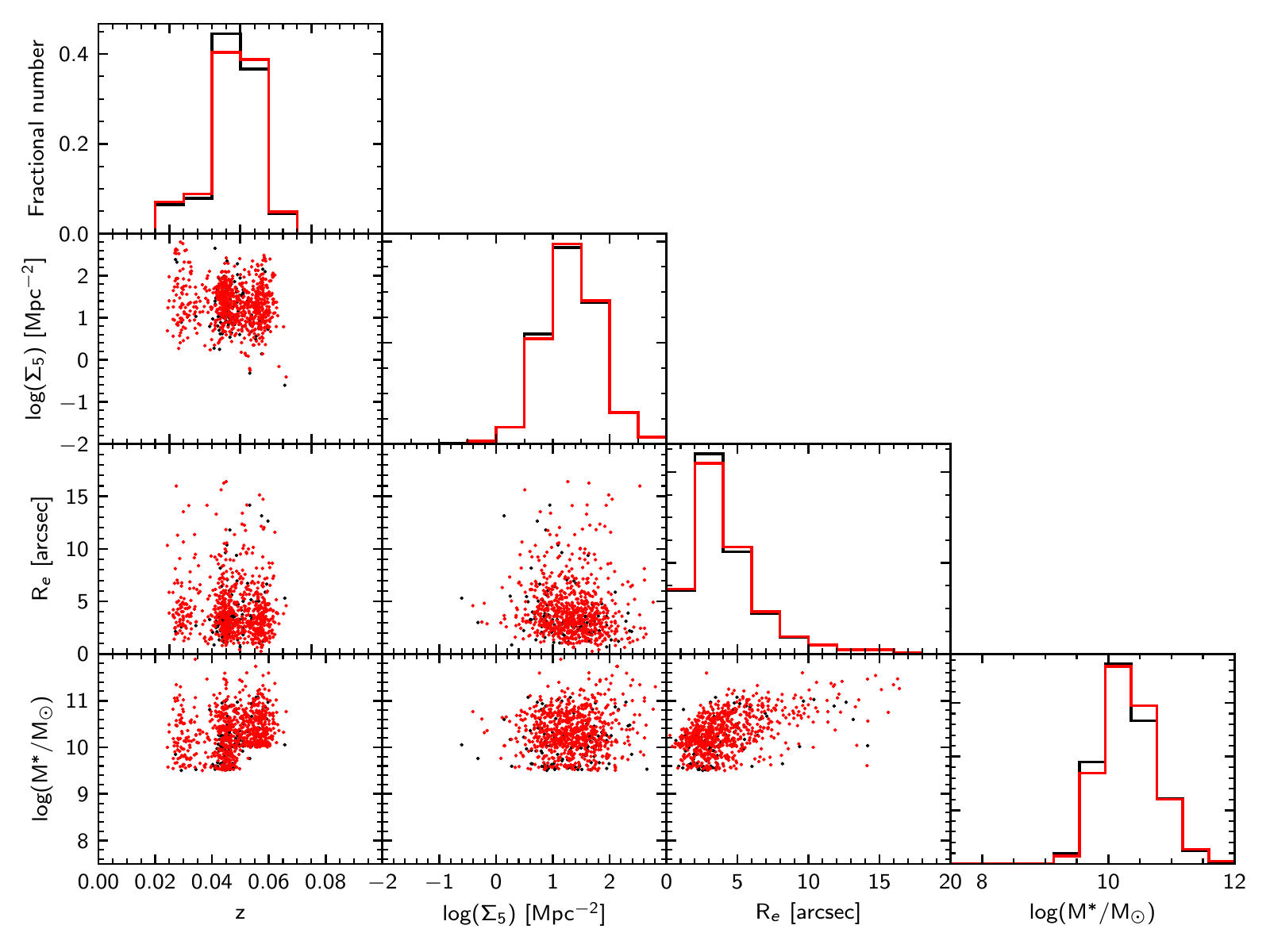}
    \caption{The distribution of good primary SAMI targets in the cluster regions (black points) compared to the observed targets (red points) as a function of redshift, $\log(M^*/M_\odot)$, $R_{\rm e}$ (major axis in arcsec) and 5th nearest neighbour density $\log(\Sigma_5)$. Histograms are normalized separately for each sample. Axes are on the same scale as Fig. \ref{fig:gama_comp}.}
    \label{fig:clusters_comp}
\end{figure*}

\subsection{Survey completeness}

Estimates of survey completeness are based on the number of galaxies for which we could successfully construct data cubes ($N_{\rm obs}$) compared to the number of potential targets that have good quality flags ($N_{\rm good}$, that is, BAD\_CLASS = 0, 5 or 8). The number of unique galaxies successfully observed in each of the GAMA regions along with the completeness is listed in Table \ref{table:gama}. In total 1940 unique primary galaxies from the GAMA regions were observed, with a completeness of 80.6 percent. There is some variation in completeness between GAMA regions with the 12h field being 87.5 percent complete, while the 15h field is 73.2 percent complete. This variation is largely driven by the larger number of targets in the 15h region, due to denser large-scale structure (see Fig.\ \ref{fig:gama_wedge}). As would be expected, the completeness of secondary targets in the GAMA field is much lower, only 160 of 2514 good targets are observed.

The distribution of good primary targets and observed primary targets in the GAMA regions are shown in Fig.\ \ref{fig:gama_comp} as a function of redshift, stellar mass, major-axis half-light radius (in arcsec) and 5th nearest neighbour density, $\Sigma_5$ (see Section~\ref{sec:env} for details). The histograms in Fig.\ \ref{fig:gama_comp} are normalized to the total numbers in each population, removing the overall difference in numbers and allowing relative differences to be more visible. In all the displayed parameters the observed distributions are representative of the underlying target distribution. The median stellar masses for the primary target and observed galaxies are $\log(M^*/M_\odot)=10.04\pm0.02$ and $10.02\pm0.03$ respectively. There is a marginally significant difference in the median $\log(\Sigma_5)$, with values of $0.01\pm0.02$ and $0.06\pm0.02$ for primary and observed galaxies. However, this difference is very small compared to the dynamic range of $\log(\Sigma_5)$, which is $\sim4$\,dex. The distributions of $\log(\Sigma_5)$ for secondary targets are also found to be slightly different, but not significantly so, with median values of $0.04\pm0.02$ and $0.09\pm0.07$ for all and observed objects respectively.  {\update We also note that, while not explicitly chosen to fall in regions of low galaxy density, the regularly-spaced redshift cuts used to select the sample were checked to make sure that they did not cut across specific large-scale structures.  As a result, the regions with a high density of galaxies in Fig.\ \ref{fig:gama_comp} tend to lie within a single $\log(M^*)$--redshift region \citep{2015MNRAS.447.2857B}.}

On average, the completeness of the primary targets in the cluster regions, defined as $N_{\rm obs}$/$N_{\rm good}$ is very good at 87\% (Table~\ref{table:clusters}). There are two clusters for which the completeness of observed primary targets is substantially lower than the average: A168 and A119 with 75 and 79 percent respectively. While A119 has one of the lower completeness values for primaries, it also contains the largest number of primary targets that have been observed. 
 
For the cluster A168, the completeness was affected by a change in the coordinates used to define the cluster centre as outlined in Section~\ref{subsec:clus_input}. This redefinition of the cluster centre led to the addition of 17 previously-defined secondary targets to the primary sample. Many of these redefined galaxies were ultimately not observed during the survey, leading to an excess of unobserved primaries in the southern part of the cluster (Figure~\ref{fig:clus_dist}).

The distribution of primary targets (black points) and those that have been observed (red points) in the clusters is shown in Fig.\ \ref{fig:clusters_comp}. As would be expected, these sit at higher 5th nearest neighbour density, $\log(\Sigma_5)$, than the GAMA region galaxies. We find no significant differences between the primary targets and the observed galaxies in any of the parameters shown in Fig.\ \ref{fig:clusters_comp}. The secondary targets that were observed are significantly different from their parent population in terms of redshift. This is driven by the varying number of secondary targets observed in different clusters. There are no other significant differences between the observed secondaries and their parent population.

The number of objects observed from filler samples can be seen in Table \ref{table:filler}. The completeness for filler samples is generally low, as is expected given they were used only when galaxies from the main sample were not available.

\section{Data Reduction and data quality}
\label{sec:qual}

The reduction of SAMI data for DR3 follows the procedures described in \citet{2015MNRAS.446.1567A} and \citet{2015MNRAS.446.1551S}, including the modifications and additional steps described in \citet{2018MNRAS.475.716G} and \citet{2018MNRAS.481.2299S}. Here we summarise the process of reducing SAMI data for DR3 and in the following subsections describe in detail significant changes from previous versions of the data. SAMI DR3 data are equivalent to the internal release v0.12. Note that the previous public release, SAMI DR2, used internal release v0.10.1 data.

SAMI data reduction can be neatly divided into two main phases: i) the extraction of row stacked spectra (RSS) from raw observations, and ii) the combining of the RSS frames into 3-dimensional data cubes. The first phase is largely carried out by the {\sc 2dfDR} data reduction package\footnote{https://www.aao.gov.au/science/software/2dfdr} \citep{2015ascl.soft05015A}, with the second phase and the overall process managed by the SAMI {\sc Python} package \citep{2014ascl.soft07006A} via the SAMI `Manager'.

\begin{figure*}
    \centering
    \includegraphics[width=16cm]{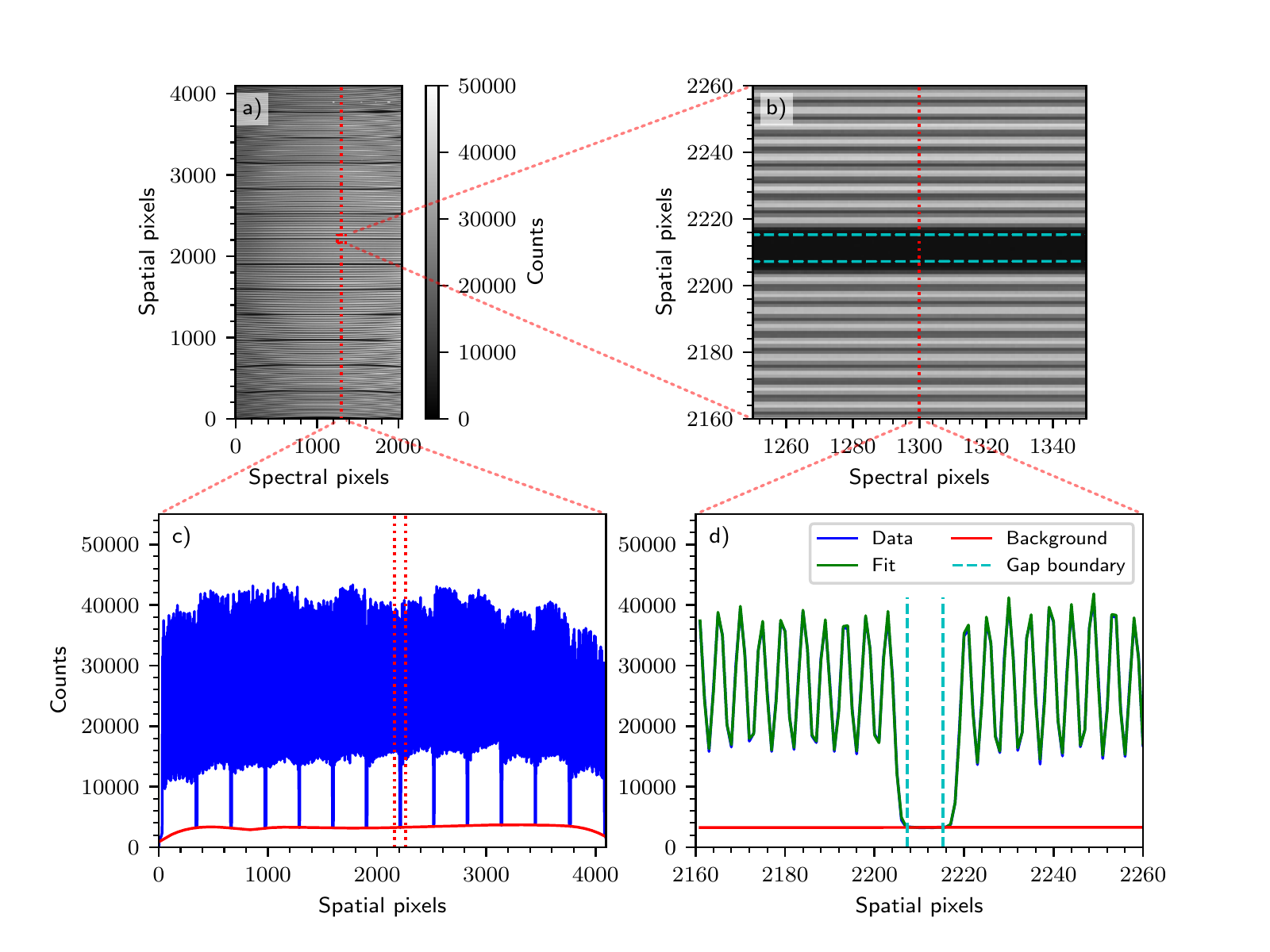}
    \caption{Fibre and scattered light background fits for a SAMI flat field frame. a) A full SAMI fibre flat field. Wavelength varies in the x-direction (spectral pixels) and fibres are arranged in the y-direction (spatial pixels). The red dotted line and box show sub-regions and slices that will be shown in subsequent plots (the aliasing seen in the image of the full detector is purely a display affect). b) A zoom in of the fibre flat field near a gap between slitlets (region indicated by red box in a). Dispersed light from individual fibres can be seen running horizontally. The cyan dashed lines indicate the region defined as the slitlet gap ($\pm4\sigma_{\rm spat}$ of the nearest fibre). c)  A vertical slice of the fibre flat field at column 1300 showing the observed counts as a function of position (blue) and the best fit scattered light  background model (solid red line).  The red dashed lines show the region plotted in d.  d) A vertical slice in a narrow range around a slitlet gap for column 1300 with the observed counts (blue, mostly hidden by the green line), scattered light background (red) and fitted fibre models (green). The cyan dotted lines show the region defined as the gap.}
        \label{fig:fib_profiles}
\end{figure*}

{\sc 2dfDR} applies the standard steps of overscan subtraction, spectral extraction, flat-fielding, fibre throughput correction, wavelength calibration and sky subtraction. The end result of these steps is a single RSS frame per observation, consisting of 819 1-dimensional spectra corresponding to the 819 fibres of the SAMI instrument. Each RSS frame contains spectra from twelve galaxies (61 spectra each), one secondary flux standard star (61 spectra) and twenty six sky spectra.

Telluric correction and relative and absolute flux calibration are applied to the individual RSS frames by the SAMI {\sc Python} package. The flux calibrated, telluric corrected RSS frames are then combined into 3-dimensional datacubes, one per galaxy or secondary standard star, with each cube using spectra from between 6 and 14 separate RSS frames. Fibre spectra from each frame are registered onto a regular grid, including a differential atmospheric refraction correction to align each wavelength slice. Aperture spectra and a set of pre-binned data cubes are also constructed at this stage. The PSF of each datacube is determined from the secondary standard star using a Moffat profile fit to the star datacube at 5000 \AA. A dust correction vector, calculated using the dust correction law of \citet{1989ApJ...345..245C} and the Planck Milky Way foreground thermal dust map of \citet{2014A&A...571A..11P} is also added to each cube (although not applied directly to the data).

For DR3 we have improved various aspects of the SAMI data reduction.  This includes modelling of scattered light during spectral extraction, wavelength calibration, sky subtraction, telluric absorption correction, flux calibration, world coordinate system (WCS) calculation and bad pixel rejection.  Details of each of these are given below.

\begin{figure*}
    \centering
    \includegraphics[width=6.25in]{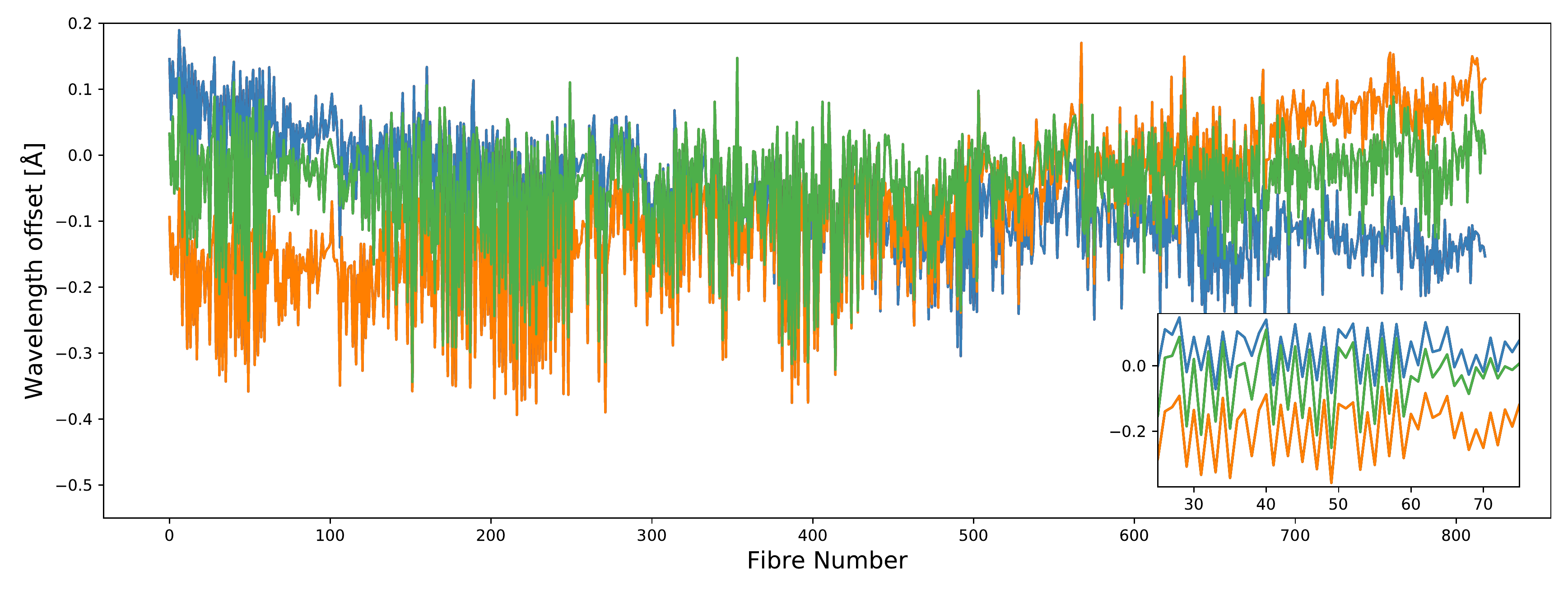}
    \caption{The measured wavelength offset with respect to a high-resolution solar spectrum in \AA\ as a function of fibre number for three twilight sky observations (indicated by the three different colours). The inset shows a zoom in over a small range of fibres to better illustrate the fibre-to-fibre variation.}
    \label{fig:twilight_wavelength_offsets}
\end{figure*}

 \begin{figure}
    \centering
    \includegraphics[width=3.25in]{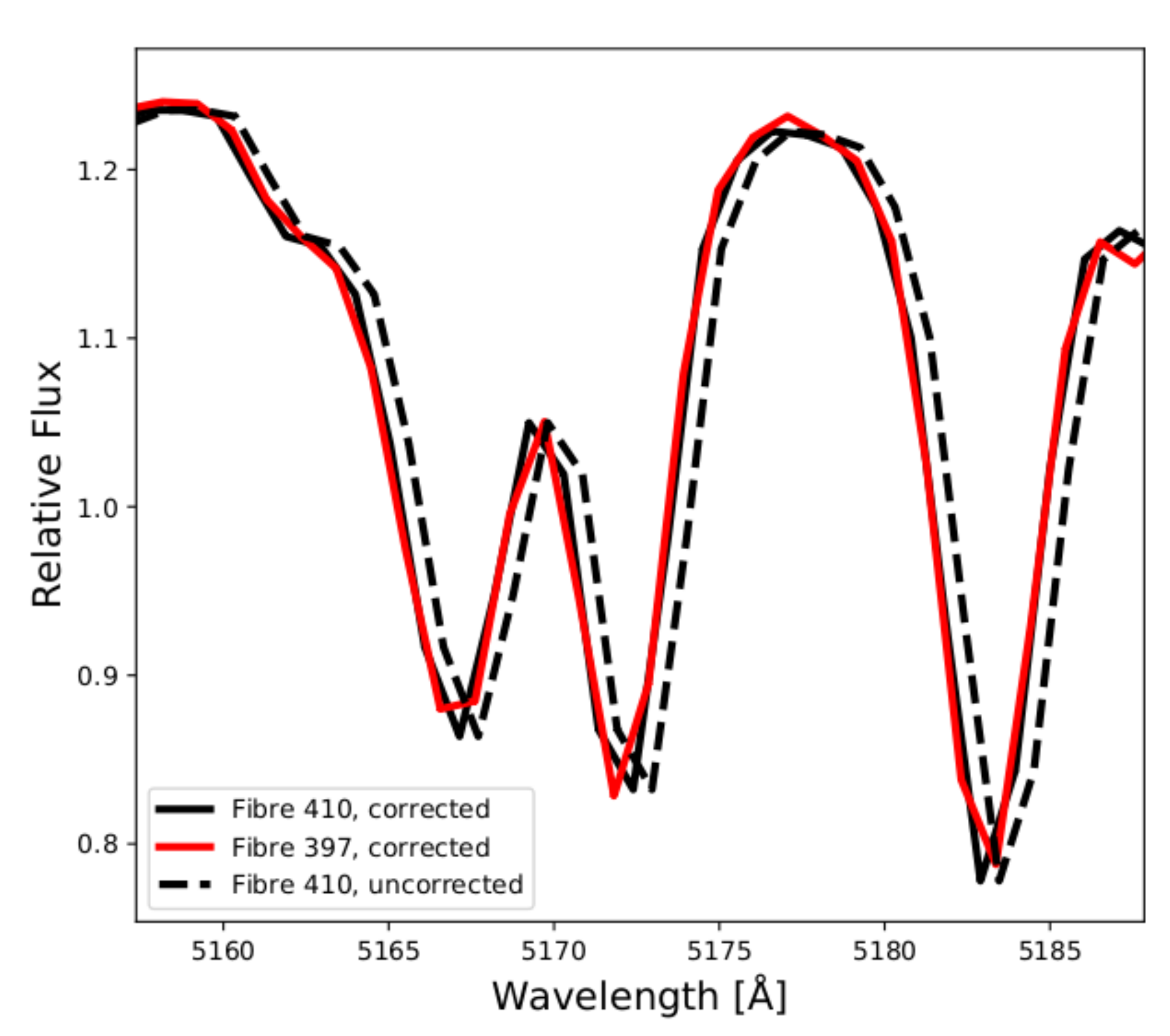}
    \caption{Illustration of the improvement in blue arm wavelength calibration around the Mgb 5177 \AA\ lines. The black solid and dashed lines show a corrected and uncorrected twilight sky spectrum from fibre 410. The solid red line shows a wavelength corrected spectrum from fibre 397. After wavelength correction the agreement between different fibres is significantly improved.}
    \label{fig:wavecorr_example}
\end{figure}

\subsection{Extraction and scattered light removal}\label{sec:extraction}

Extraction of spectra from the 2D CCD image is a fundamental step in data reduction of all fibre--fed spectrographs. The accuracy of this extraction impacts data quality in several ways. The standard approach used within \2dfdr\ is to use flat field or twilight exposures to model the width of the fibre profiles in 1D (as a Gaussian of width $\sigma_{\rm spat}$) and then simultaneously fit the amplitude of all fibres in a given column on the CCD (see Fig. \ref{fig:fib_profiles}). Each column is fit independently. Previous versions of SAMI data reduction \citep[e.g.][]{2015MNRAS.446.1551S} also fit a cubic spline simultaneously with the fibre amplitudes to model the scattered light.  The spline typically used 12 to 16 uniformly spaced knots across the entire CCD column of 4096 spatial pixels. 

The previously used approach suffered problems in two senses. First, the small number of pixels in gaps between blocks of fibres did not have sufficient weight or  S/N to place strong constraints on the scattered light model. Second, some data exhibited excess scattered light that was more localized and could not be adequately modelled. The excess in scattered light was particularly significant after the CCD in the blue arm of AAOmega was replaced in the first half of 2014. Contamination of the CCD dewar during the replacement led to low-level condensation on the field-flattening lens in front of the detector within the dewar. This was slowly reduced by repeated pumping over a number of months. The contamination led to extra scattering that was particularly visible around the bright 5577\AA\ night-sky emission line as a circular Lorentzian halo around each line.

To address the above issues we revised the scattered light modelling so that the new procedure was as follows:
\begin{enumerate}
\item Identify the location of gaps between slitlets on the CCD image (see Fig. \ref{fig:fib_profiles}) by selecting pixels that are $>4\sigma_{\rm spat}$ from the centre of fibres either side of the gap. The gaps are located between groups of 63 fibres (61 for the hexabundle, 2 for sky fibres).
\item For each gap calculate the average flux in bins 30 pixels wide (10 pixels for bright frames such as flat fields and twilights) along the spectral direction, clipping outliers. The averages are not variance weighted as the count rates in the gaps are low, which can lead to biases in the estimated variance.
\item Fit a cubic spline to the average flux along each gap, with 8 (48 for flat fields and twilights) uniformly spaced knots along the spectral direction. For the blue arm of AAOmega, we ignore the $\pm60$ pixels either side of the 5577\AA\ night-sky line when fitting the spline.
\item The smooth model for scattered light along the gaps is then fit using a cubic spline across the gaps (in the spatial direction) with 6 evenly spaced knots. This provides a full 2D model of the smooth component of the scattered light (red solid line in Fig.\ \ref{fig:fib_profiles}c, d).
\item For data from the blue arm of AAOmega the spline is then subtracted from the average flux in the fibre gaps and the residual compared to a 2D Lorentzian model with $\Gamma=17.6$ pixels. The value of $\Gamma=17.6$ pixels was chosen based on investigation of the typical width of the scattered light profile across different frames. The model consists of 819 2D Lorentzians, each one centred on the 5577\AA\ line for a fibre. The normalization of the model is allowed to vary smoothly in the spatial direction on the CCD, and this variation is parameterized by a 4th order polynomial fit to all the fibre gaps simultaneously. The model for scattering around the 5577\AA\ line is then added to the smooth 2D scattered light model generated in the previous step.
\end{enumerate}

The combined model is then subtracted from the 2D image prior to fitting the fibre profiles to extract the flux in each spectrum. The key outcome from this modified approach is a substantial improvement in sky subtraction that will be further discussed below in Section~\ref{sec:skysub}. In one particular case an individual field (Y15SAR3\_P006\_12T097) contained a bright extended source (galaxy 273336) that caused excess scattered light. In this case we fit a high-order (order 16 and 60 in the red and blue arms respectively) polynomial simultaneously with the fit to the fibre profiles using the old scattered light approach.

\begin{figure*}
    \centering
    \includegraphics[clip, trim=3.0cm 0.5cm 3.8cm 0cm,height=5.4cm]{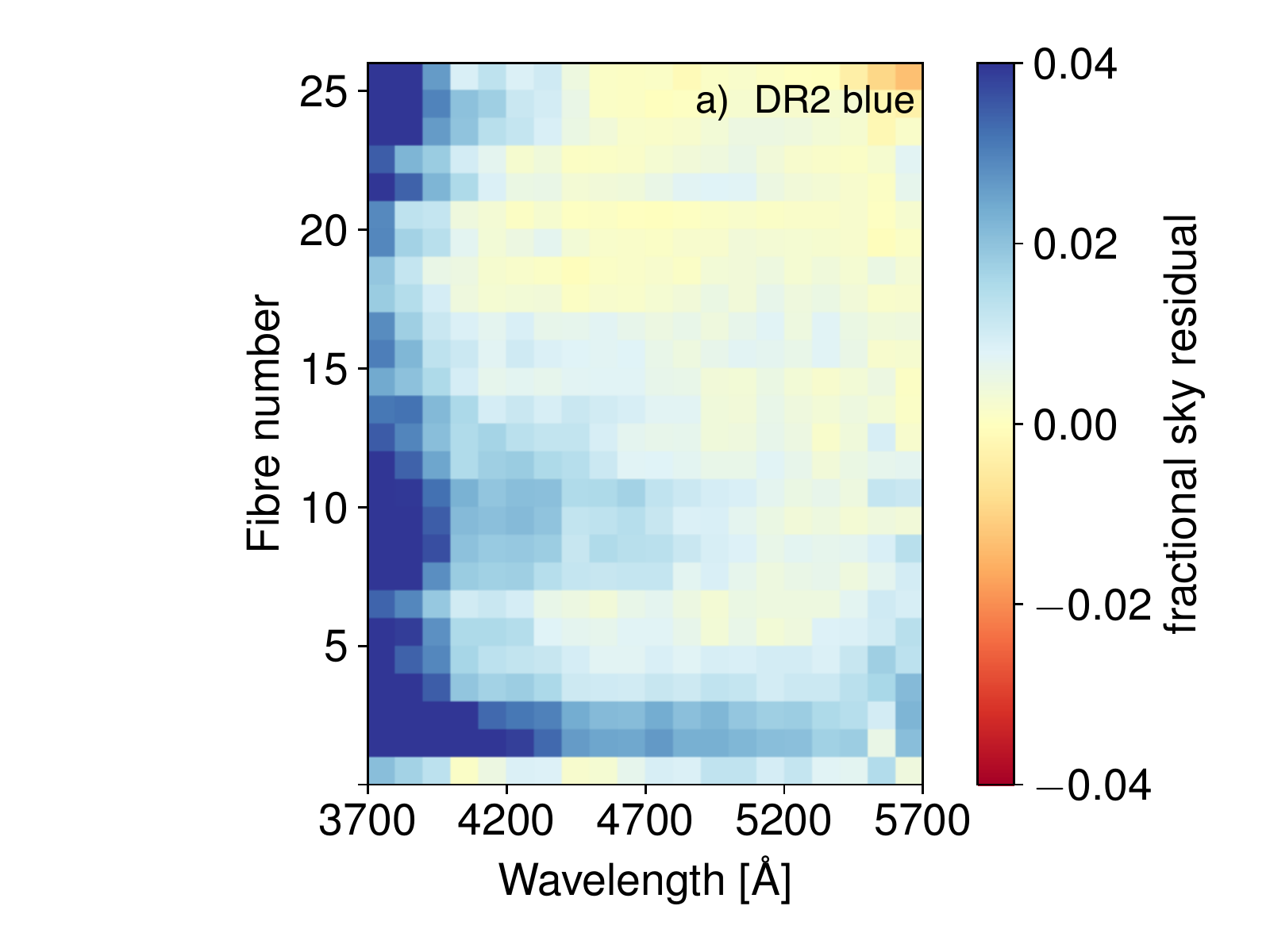}\includegraphics[clip, trim=3.8cm 0.5cm 3.8cm 0cm, height=5.4cm]{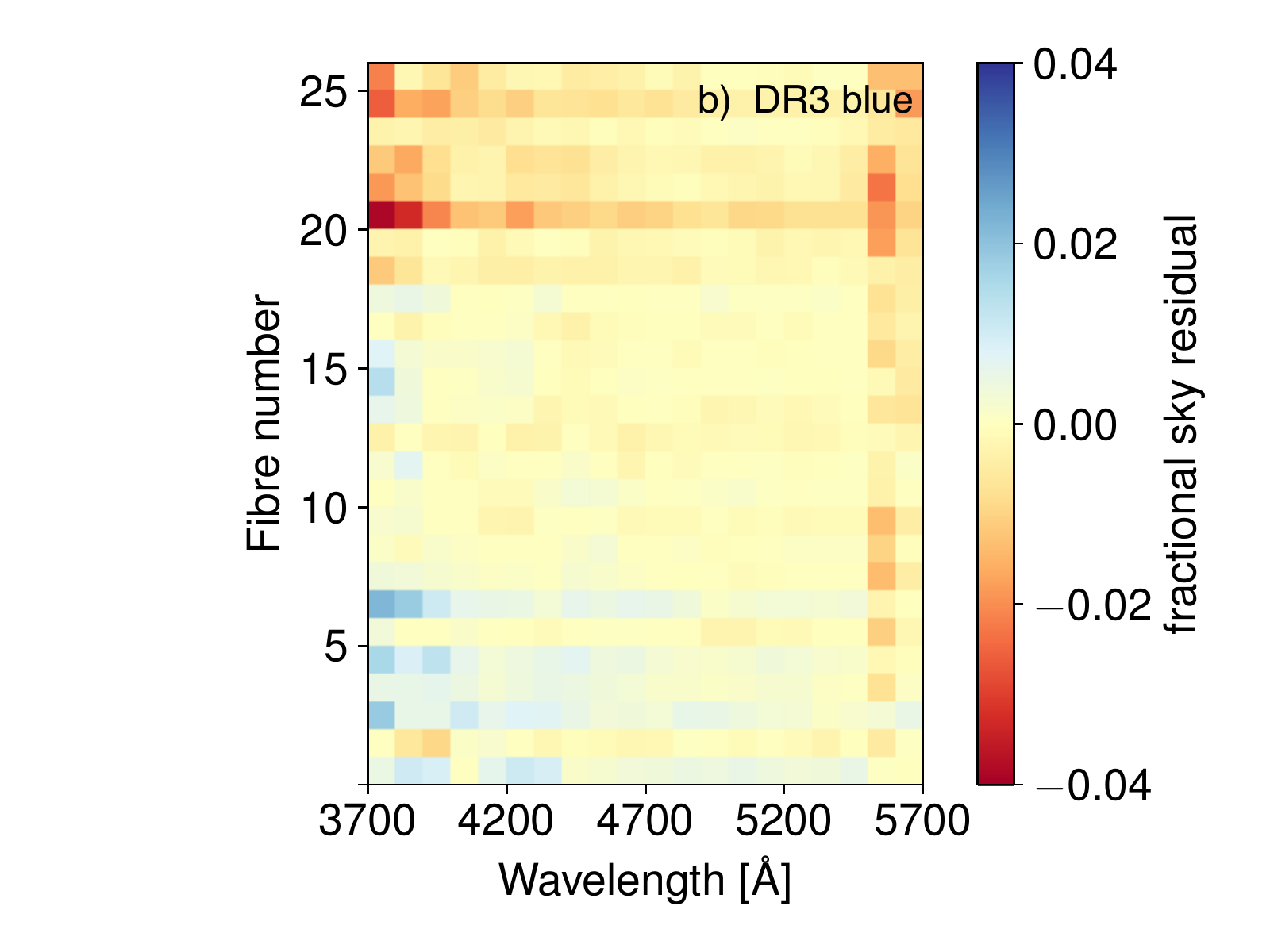}\includegraphics[clip, trim=3.8cm 0.5cm 3.8cm 0cm, height=5.4cm]{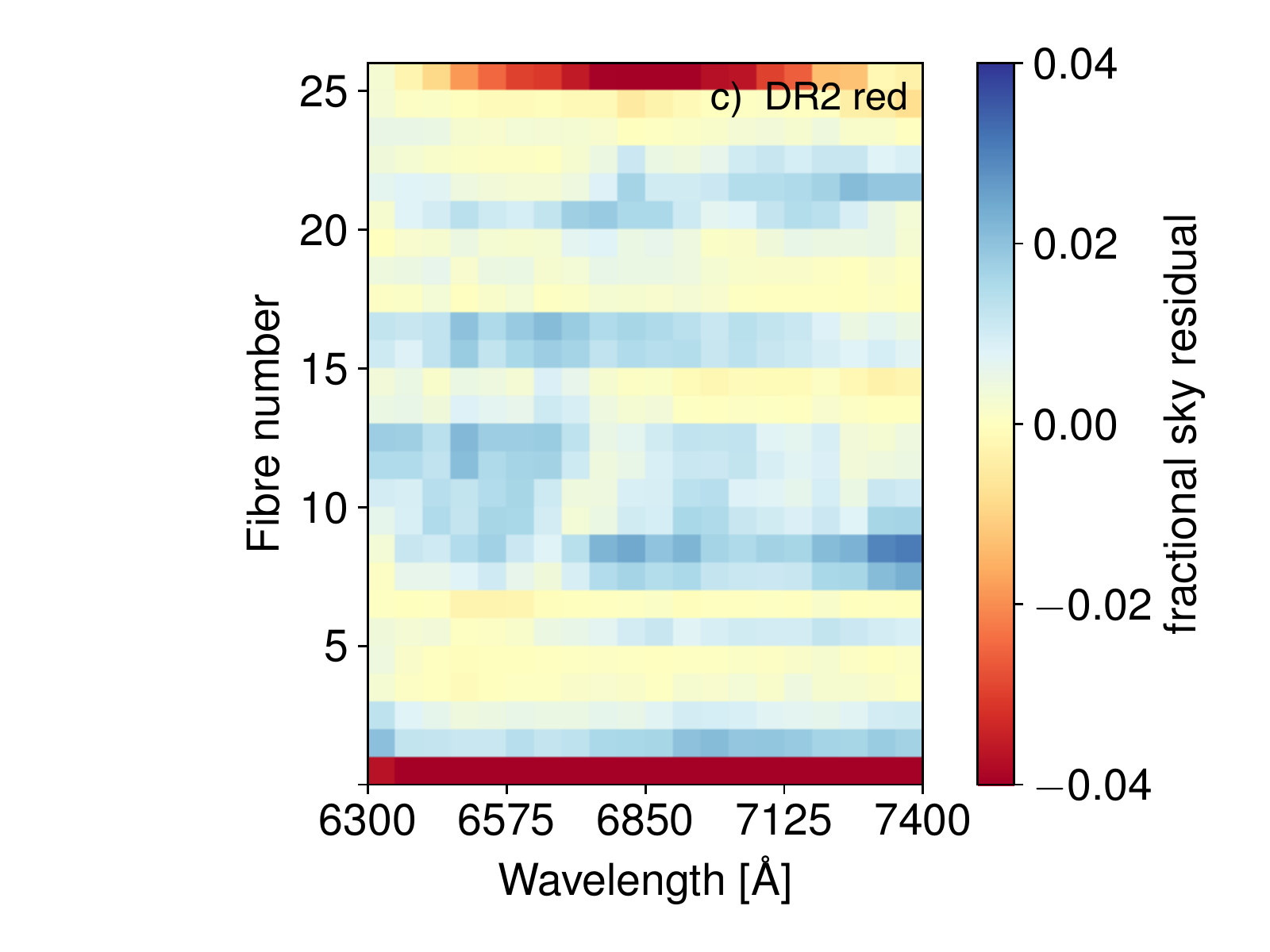}\includegraphics[clip, trim=3.8cm 0.5cm 0.3cm 0cm, height=5.4cm]{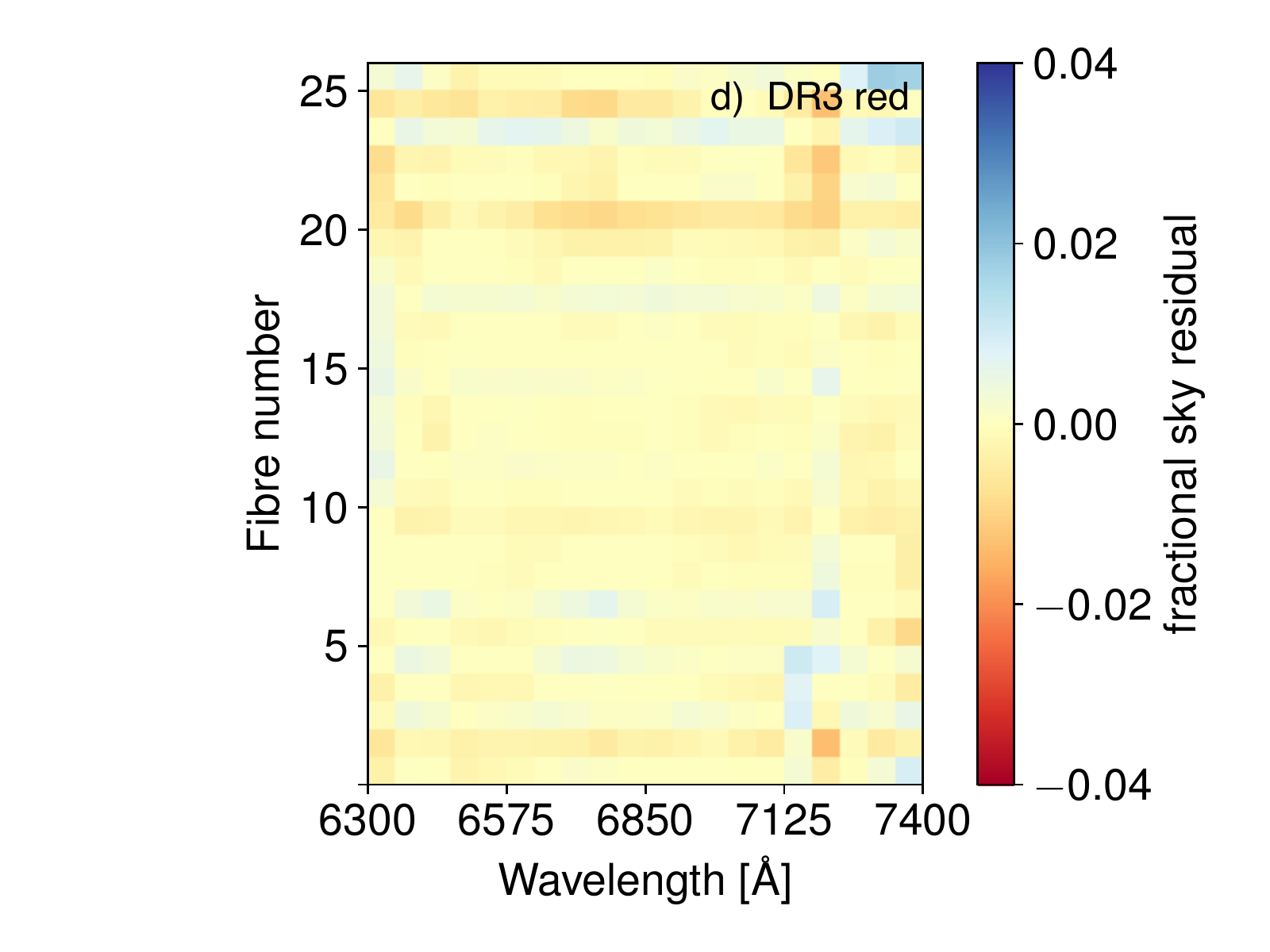}
    \caption{Median fractional sky subtraction residuals as a function of wavelength and fibre number (equivalent to slit location) for sky fibres.  {\update  On the left we show the blue arm data for DR2 (a) and DR3 (b).  On the right we show the red arm data for DR2 (c) and DR3 (d).}  The median is calculated over all data frames with exposures $>900s$ from the entire set of survey observations.}
    \label{fig:skysub}
\end{figure*}

\begin{figure}
    \centering
    \includegraphics[width=8cm]{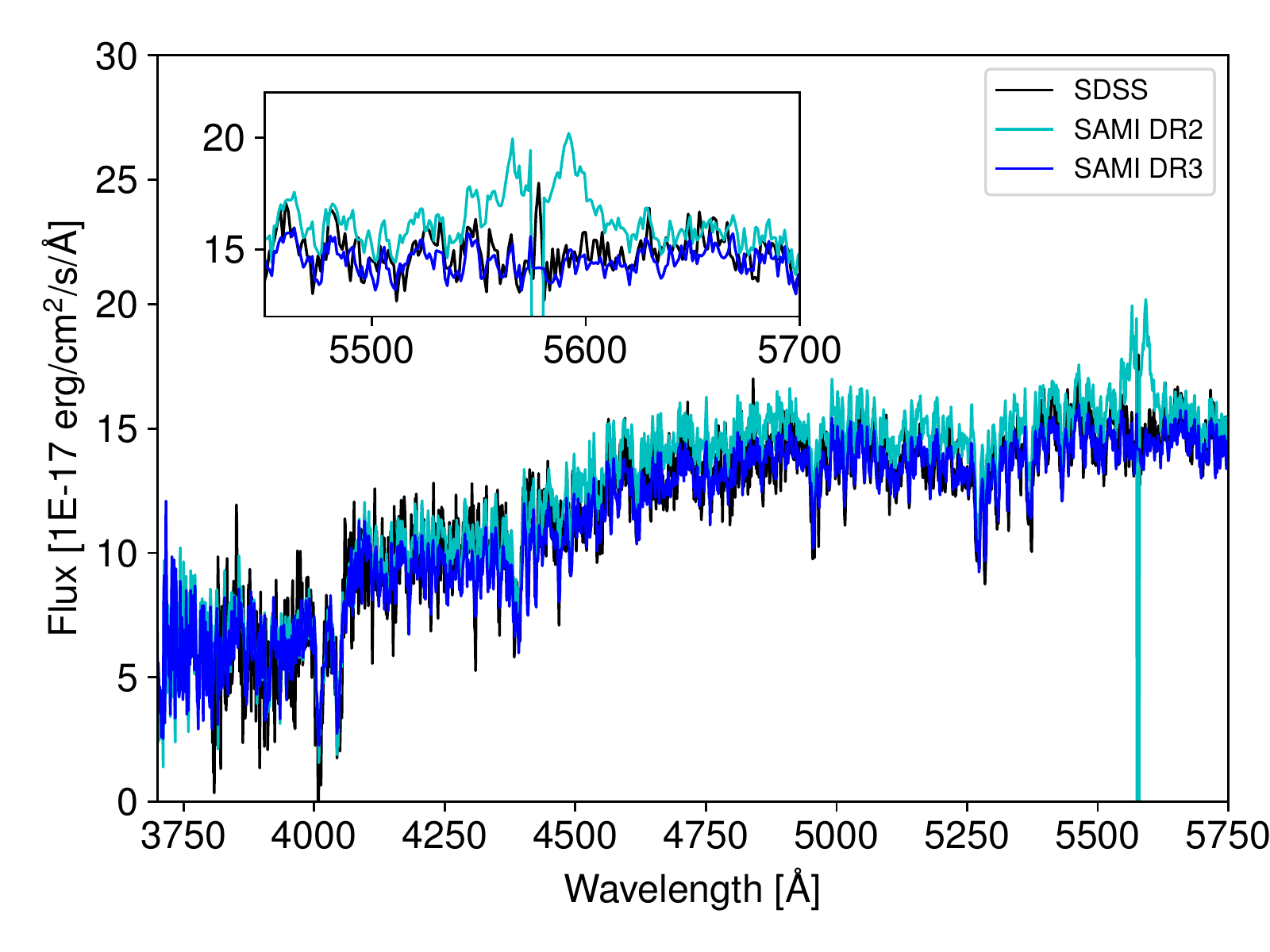}
    \caption{A comparison of SAMI blue arm spectra (3 arcsec diameter aperture) from DR2 (cyan) and DR3 (blue) for object 289198 that showed particularly strong residual scattered light around the 5577\,\AA\ night-sky line in DR2. We also compare the spectra to an SDSS spectrum of the same object (black). Inset is a close--up of the region around the 5577\,\AA\ line.}
    \label{fig:sky5577}
\end{figure}

\subsection{Wavelength calibration}

When performing a combined analysis of data from the red and blue arms we identified a small ($\lesssim 0.5$ \AA) offset in the wavelength solution between the two arms. Further investigation identified residual wavelength calibration errors in the blue arm spectra of $\sim 0.3$ \AA. While the blue arm wavelength calibration relies solely upon the arc frame solution, a further refinement is applied to the red arm from a fit to night sky emission lines. Therefore an offset between the two is not unexpected. This offset varies from fibre to fibre but is constant with wavelength and shows no variation in the instrumental dispersion, consistent with the detailed analysis presented in \citet{2018MNRAS.481.2299S}.

In the DR3 data release we implemented an additional wavelength calibration refinement step based on twilight sky observations for the blue arm only. We convolve the extremely high resolution solar spectrum of \citet[R $> $ 300,000]{1999SoPh..184..421N} to the resolution of the SAMI blue arm. We then normalise both the solar spectrum and the SAMI twilight sky spectrum using a 20$^\mathrm{th}$ order polynomial to remove the continuum shape. We interpolate the SAMI twilight spectrum onto the high spectral sampling wavelength scale of the solar spectrum (0.01 \AA\ pixel$^{-1}$), then cross-correlate each individual fibre twilight spectra with the solar spectrum to determine the wavelength offset between the two, repeating this process for all 819 fibres.

Inspection of the fibre wavelength offsets showed two distinct variations: a pattern of fibre-to-fibre wavelength offsets that was consistent between different observations, and an overall shape variation between observations taken on different nights (or between evening and morning twilight on the same night). This overall shape variation is well approximated by a simple linear fit. In Fig.\ \ref{fig:twilight_wavelength_offsets} we illustrate these two trends of the wavelength offset as a function of fibre number for a small subset of all the twilight sky observations taken as part of the SAMI Galaxy Survey. {\update  The constant, small-scale, fibre-to-fibre variations are caused by small misalignments between fibres in the slit, combined with the arc lamps feeding the fibres at an f-ratio that is different to the sky (due to the location of the arc lamps in front of the corrector in the AAT top end). The larger-scale effect that changes with time is caused by small shifts in the relative position of camera and slit through the night \citep{2015MNRAS.446.1551S}.}

Given these two observed patterns we implemented a three-step correction for the blue arm wavelength solution. For each observing run, we first perform and subtract a linear fit to the wavelength offset vs.\ fibre number relation for each twilight sky observation. We then average these offsets across all observations for each fibre, resulting in an accurate measurement of the fibre-to-fibre wavelength offsets for each observing run. For each object frame we identify the closest-in-time twilight sky observation and combine the linear fit to the wavelength offset vs.\ fibre measurements for that twilight sky observation with the averaged fibre-to-fibre wavelength offset described above.  As a final correction to account for any linear shift between the nearest twilight and the data frame in question we use a robust linear fit to the shift of the 5577\AA\ sky line as a function of fibre number.  This combined wavelength shift is then applied to the wavelength solution for each blue object frame as an additional refinement to the original arc-derived wavelength calibration. 

The wavelength correction is illustrated in Fig.\ \ref{fig:wavecorr_example}, which shows a wavelength corrected spectrum from fibre 397 (red line), and a wavelength corrected and uncorrected spectrum from fibre 410 (black solid and dashed lines respectively) for a typical twilight sky observation. As well as improving the wavelength solution in the blue arm, we note that the improved fibre-to-fibre wavelength calibration results in a reduction of the residuals around the 5577 \AA\ sky line. See the following section for further details. 

\begin{figure*}
    \centering
    \includegraphics[width=16cm]{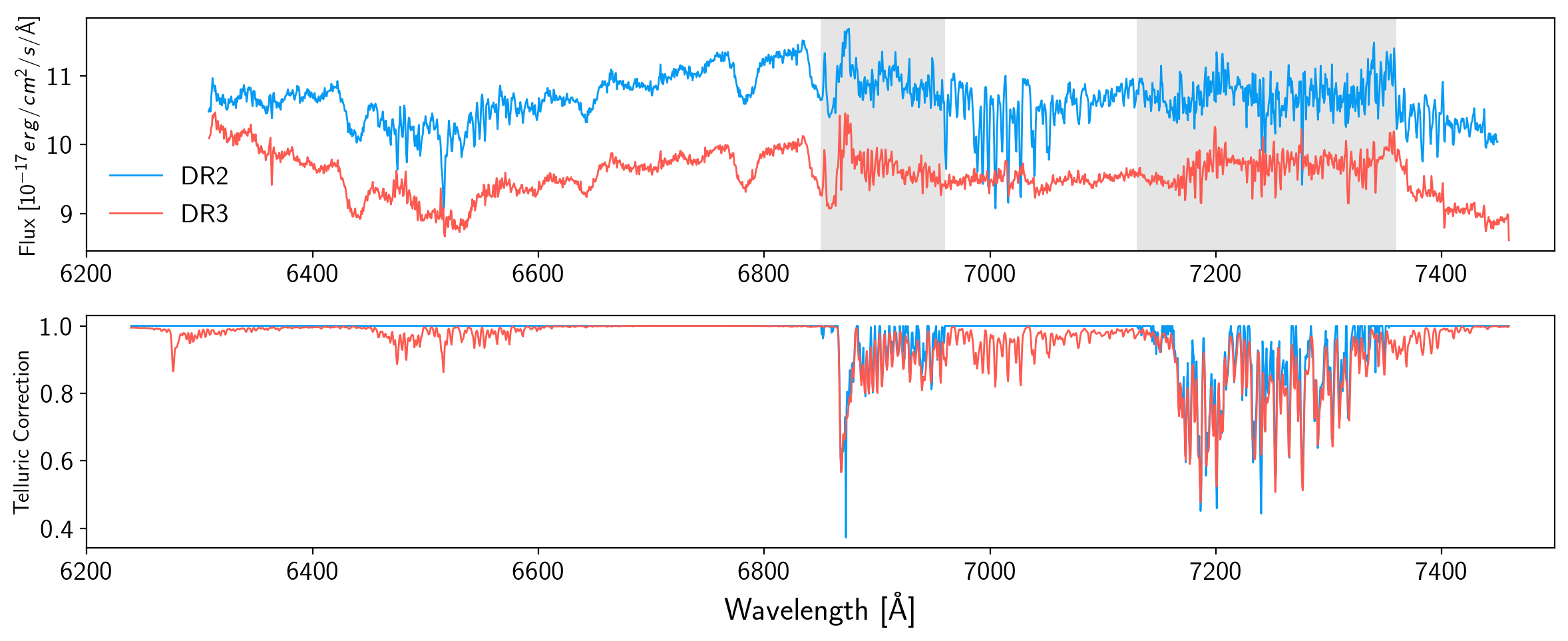}
    \caption{A comparison of the red arm 3 arcsec aperture spectrum for galaxy 136602 from DR2 (blue) and DR3 (orange). The top panel shows the flux, while the bottom panel shows the telluric correction function applied to each spectrum. The grey shaded regions indicate the two bands that telluric correction was applied to in DR2, between 6850 and 6960\AA, and between 7130 and  7360\AA. Comparatively, in DR3 we allow for telluric correction to the entire red spectrum. The offset between the DR2 and DR3 fluxes is due to differences in the flux calibration.}
    \label{fig:telluric}
\end{figure*}

\subsection{Sky subtraction}\label{sec:skysub}

A number of improvements have been made to the sky subtraction for SAMI DR3. The first of these are changes discussed above concerning scattered light lead to improved continuum sky subtraction accuracy. 

A second modification is related to the way the sky fibre spectra are combined before subtraction. Previous versions of \2dfdr\ used inverse variance weighting to combine sky spectra. However, at very low count rates in the far blue ($\sim4000$\,\AA\ and below) this leads to a small but significant bias in the combined sky (typically $\sim0.5$ counts). The cause of the bias is that pixels that scatter low will have estimates of their variance that are smaller, and so have greater weight. The DR3 version of \2dfdr\ uses an unweighted combination of the sky spectra that leads to lower systematic sky--subtraction residuals below 4000\,\AA.

To quantify these improvements we carry out a similar analysis to previous SAMI data releases, using residuals in the sky fibres to estimate sky subtraction residuals. In Fig.\ \ref{fig:skysub} we show the median sky--subtraction residuals as a function of sky fibre number (equivalent to distance along the slit) and wavelength. {\update We compare the new sky subtraction residuals (Figs.\ \ref{fig:skysub}b and d) to those from DR2 [Figs.\ \ref{fig:skysub}a and c; see also Fig.\ 3 of \citet{2018MNRAS.481.2299S}]}. In both the blue and red arms of AAOmega the systematic sky residuals are lower in the DR3 reductions. For example, the increased residuals near the edge of the CCD seen in DR2 data are substantially reduced. The systematic residuals below 4000\.\AA\ are also reduced, largely because of the revised sky combination.

A weak residual gradient from top to bottom of the CCD remains in the sky subtraction. This is at the level of $\simeq\pm0.5$ percent. We suspect that this is caused by a small ($\ll 0.1$ pixels from top to bottom) relative stretch of the fibre locations as the slit moves through the night. We previously noted \citep{2015MNRAS.446.1551S} that due to the boiling off of liquid nitrogen the AAOmega cameras systematically shift through the night by $\sim0.03$ pixels per hour. To account for this we apply a shift between the fibre `tramlines' measured on a flat field and those observed in the on--sky data. However, for each 0.1 pixel shift, the optics of the spectrograph also cause a relative $\simeq0.03$ pixel stretch in the fibre locations from top to bottom of the CCD. This small misalignment leads to slightly different fluxes extracted. Future updates to \2dfdr\ should incorporate a correction for this effect. 

To further quantify our sky subtraction accuracy we estimate the median fractional continuum residual in each sky fibre (this minimizes the impact of shot-noise on the measurement).  We then calculate the median of the absolute value of all these residuals across all sky fibres in the entire SAMI DR3 data set. In the blue arm we find a median continuum sky residual of 0.76 percent (and a 90th percentile of 2.9 percent), compared to the DR2 value of 1.2 percent (90th percentile of 4.6 percent). In the red arm the median continuum sky residual is 0.72 percent (90th percentile 2.8 percent) compared to 0.90 percent (90th percentile of 3.1 percent) in DR2. 

A third improvement focused on the 5577\,\AA\ night sky line. As discussed in Section~\ref{sec:extraction}, contamination in the dewar of the blue arm of AAOmega causes enhanced scattered light. The scattered light modelling approach described in Section \ref{sec:extraction} reduced the residual scattered light, typically by a factor of 2 or more. However, for the more severely affected frames a significant residual was still visible. To remove this we applied a two-stage principal component analysis (PCA) in the reduced wavelength calibrated and sky--subtracted data that makes use of the algorithm built into \2dfdr\  \citep{2010MNRAS.408.2495S}. 

In the first stage of the PCA correction we generate principal components of the 5577\,\AA\ line residual in a narrow 6\AA\ window around the line. A smooth (median filtered) continuum is subtracted from the spectra and then the faintest 10 per cent of fibres (which will be dominated by sky emission) are used to generate principal components. Only the first three principal components are used at this point. Weights for each component are calculated to minimize the sky line residual in the core of the line. The aim of this first stage is to minimize the residual in the core of the line so that it does not dominate the PCA in the next stage.

The second stage of the PCA correction generates components over a 100\,\AA\ window centred on the 5577\AA\ line. Again only the 10 percent faintest fibres are used to generate the principal components.  In this second stage the first ten principal components are used to minimize the sky residuals. Once the model is generated for the sky residuals it is smoothed using a median filter in the fibre direction (of width 101 fibres). This is because in a small number of cases very bright galaxies can have real small-scale structure in their spectra removed by the PCA. However, the scattered light residual around the 5577\,\AA\ line varies slowly across the slit. The median filtering maintains excellent correction of scattered light, and removes any negative impact on bright galaxies. An example is shown in Fig.\ \ref{fig:sky5577} for object 289198 that compares SAMI 3 arcsec diameter aperture spectra from DR2 and DR3, as well as an SDSS spectrum of the same object. The DR2 data shows a significant residual around the 5577\,\AA\ line (cyan line) that is completely removed in DR3 (blue line). It is also worth noting that the overall shape of the spectrum from DR3 more closely matches the SDSS spectrum due to improvements in flux calibration that will be detailed below. 

\subsection{Telluric Correction}

The handling of telluric correction has changed substantially since previous data releases, going from a comparatively straightforward polynomial interpolation approach, to using the dedicated telluric feature-fitting software \texttt{molecfit} \citep{2015A&A...576A..77S,2015A&A...576A..78K}. While the two methods both rely on the secondary standard star observations, \texttt{molecfit} uses a physically motivated model to fit and then correct for atmospheric absorption, therefore requiring additional atmospheric and meteorological data. We fit for two molecules, $\rm H_2O$ and $\rm O_2$, using the \texttt{equ.atm} reference atmospheric profile supplied in the software installation. The improvement in the telluric correction is illustrated in Fig.\ \ref{fig:telluric} for galaxy ID 136602. Previously only two regions were corrected for telluric features, between 6850 and 6960\AA, and between 7130 and 7360\AA\ (illustrated by the shaded grey regions in Fig.\ \ref{fig:telluric}). In contrast, for DR3 we allow for correction to the whole red arm spectrum. Therefore, in addition to improving the telluric correction applied to these regions by using a physically motivated model, we now correct for additional features outside these bands that were not accounted for in previous data releases. Most notably, residual telluric absorption around 7000\AA\ present in DR2 (blue line in Fig.\ \ref{fig:telluric}), are now modelled and removed in DR3 (orange line). 

\subsection{Flux calibration}

\begin{figure*}
    \centering
    \includegraphics[width=8cm]{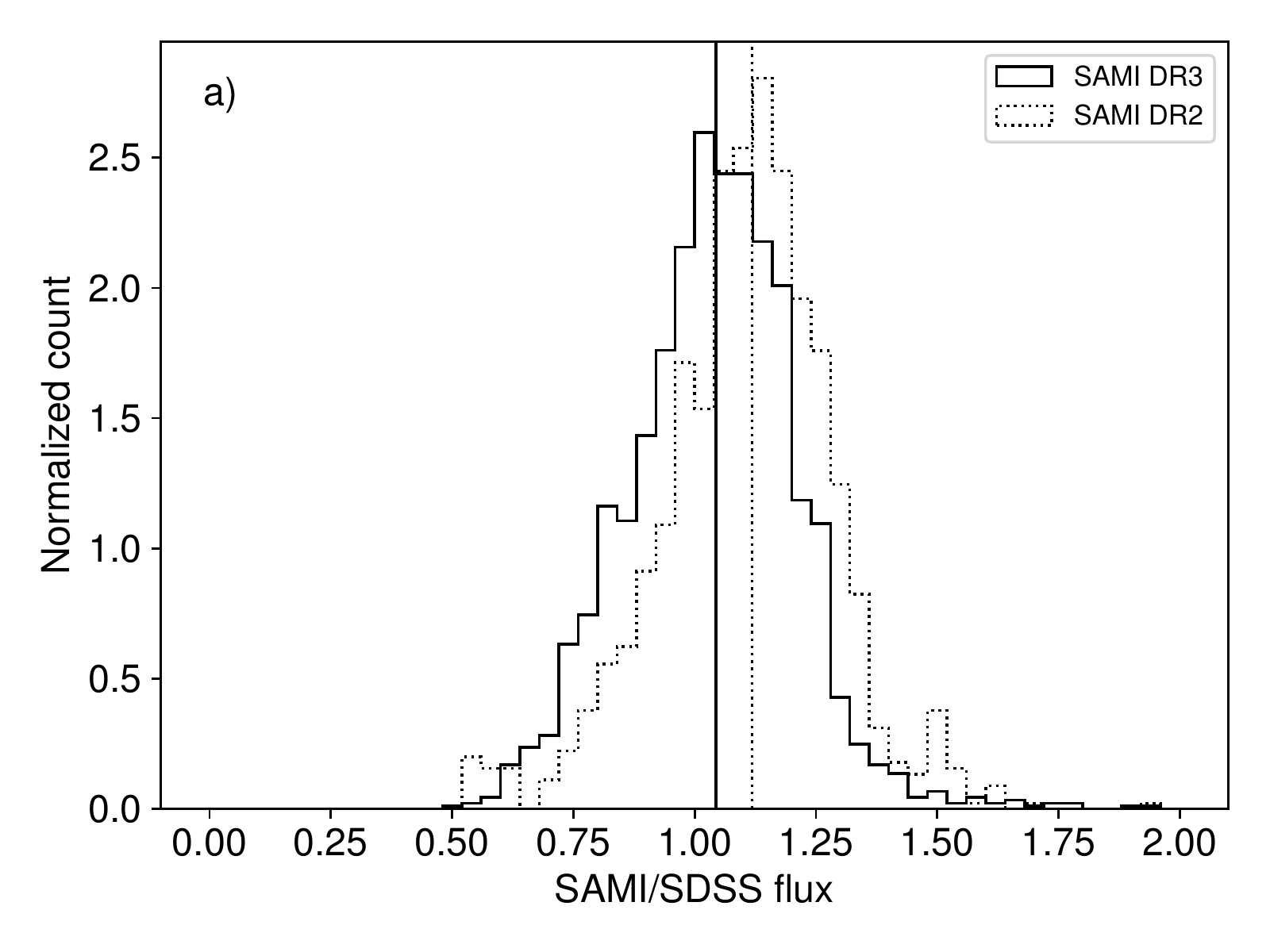}\includegraphics[width=8cm]{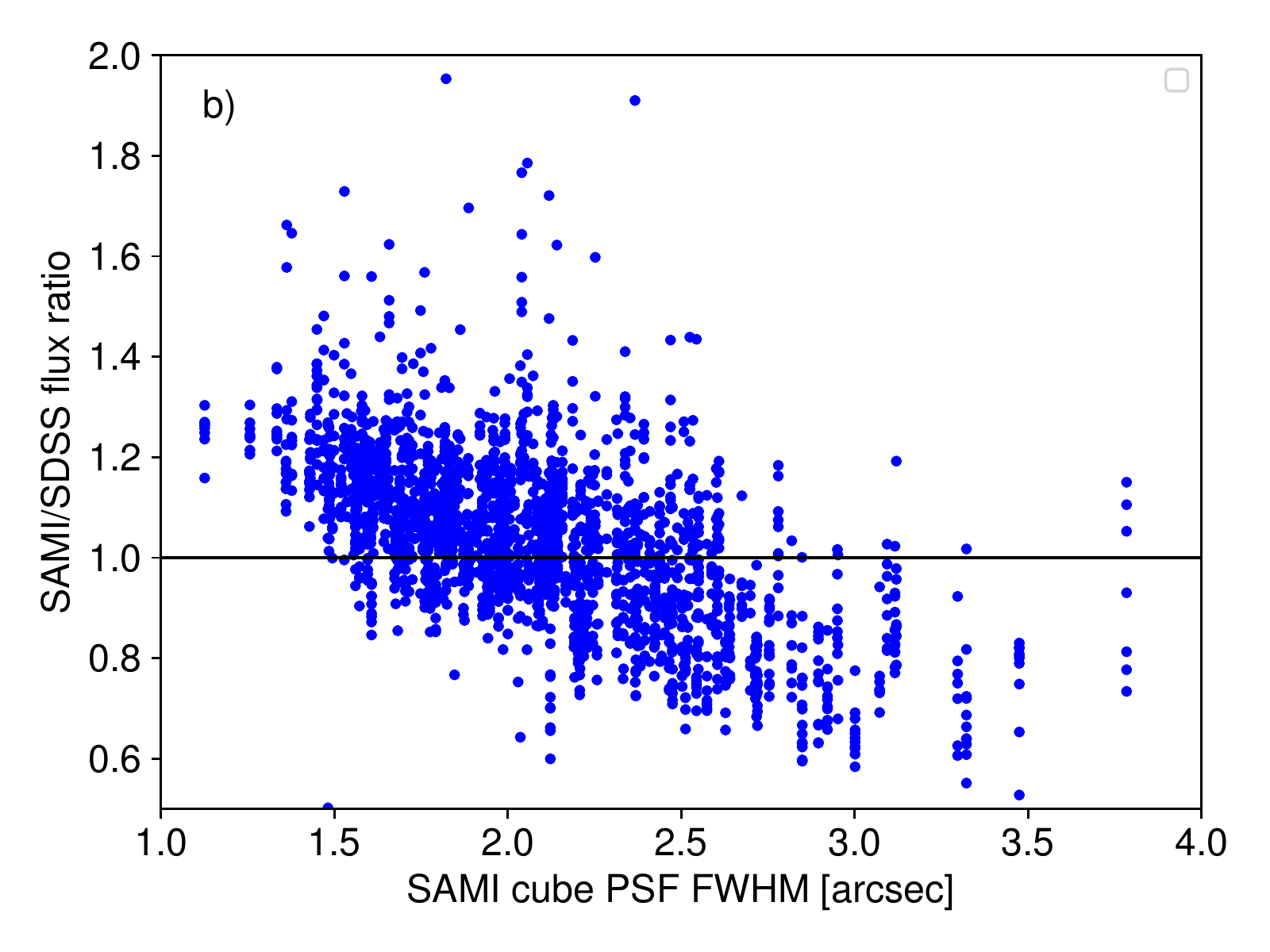}
    \caption{a) the distribution of SAMI to SDSS flux ratio from 3 arcsec diameter aperture spectra for SAMI DR3 (solid line) and DR2 (dotted line). The vertical lines show the median for each histogram. b) The SAMI to SDSS flux ratio for DR3 as a function of the FWHM of the SAMI PSF in each cube.}
    \label{fig:fluxcal_scale}
\end{figure*}

Flux calibration for SAMI makes use of `primary' standards\footnote{These were selected as A or F stars from the ESO spectrophotometric standards list: https://www.eso.org/sci/observing/tools/standards/spectra.html} observed separately to the galaxies and `secondary' standards observed at the same time as galaxies using a hexabundle. The secondary standards are colour--selected to be early F-stars. 

The flux calibration of DR2 and earlier was based on the primary flux standards that were typically observed at the start and end of a night. As a result there could be several hours' difference between observations of galaxy data and standard stars. In some cases (where nights were not photometric) no standard observations were made and the nearest in time and date was chosen to use as calibration. This approach provided relative flux calibration, but to obtain the correct normalization the secondary standard stars observed at the same time as the galaxies were compared to their photometric magnitudes to obtain a scaling for each observed frame. Once cubes were made this scaling was re-applied so that all SAMI cubes should in principle be normalized to the input imaging data (either SDSS or VST).

\begin{figure*}
    \centering
    \includegraphics[width=8cm]{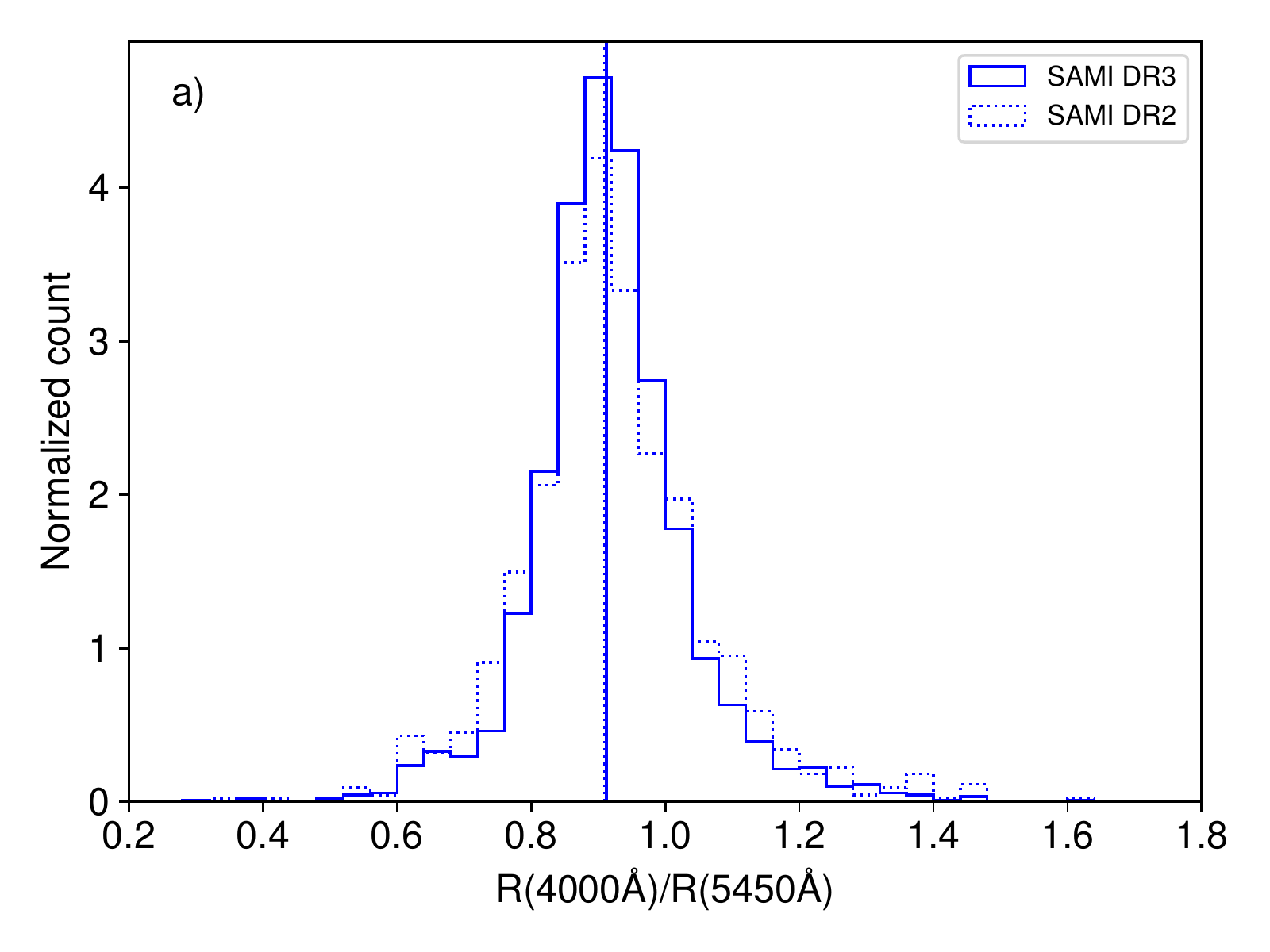}\includegraphics[width=8cm]{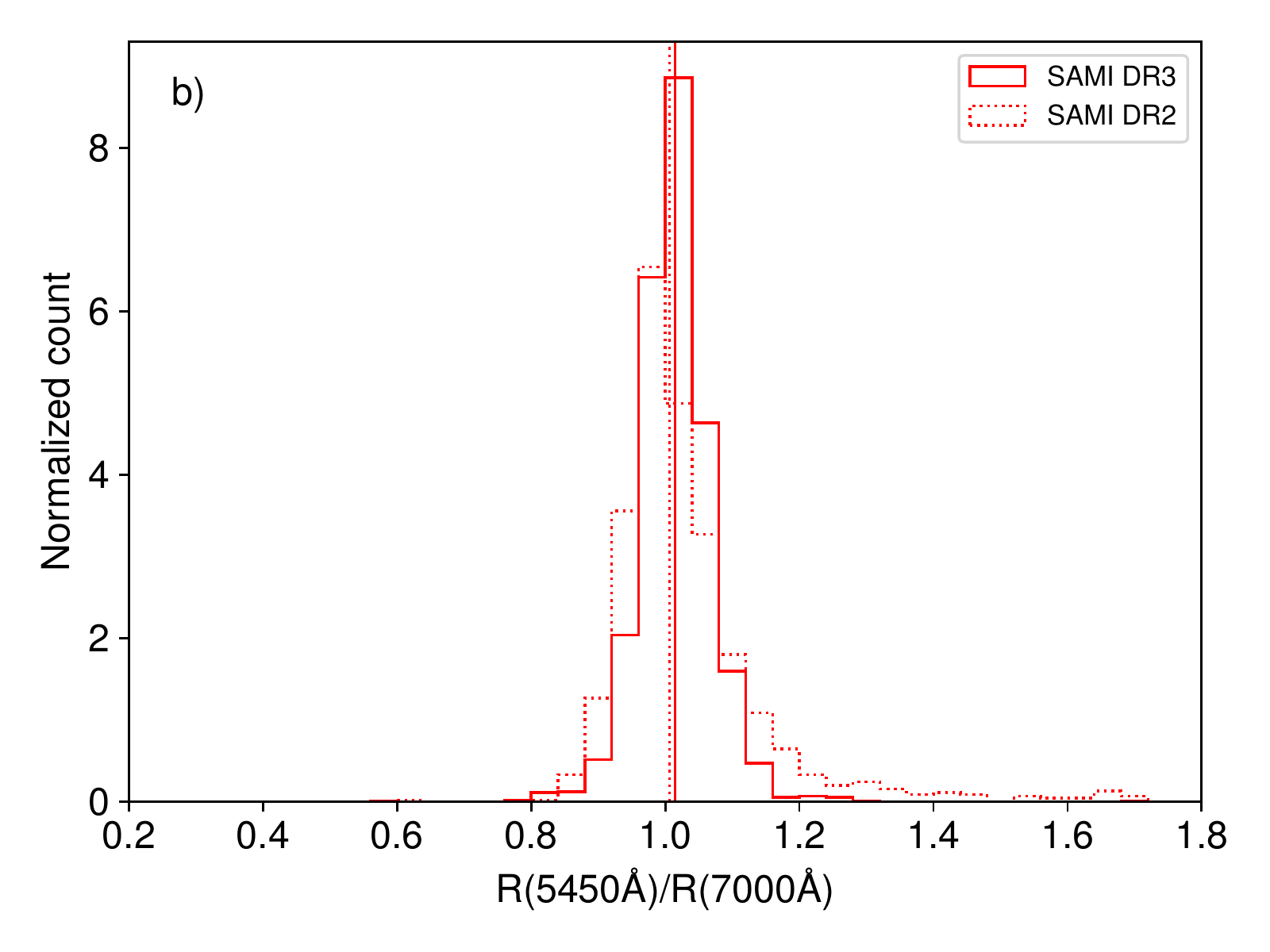}
    \caption{The ratio of SAMI to SDSS spectral flux ratios at different wavelengths, showing difference in spectral shape between 4000\,\AA\ and 5450\,\AA\ (a) and 5450\,\AA\ and 7000\,\AA\ (b). The distributions are shown for both DR2 (dotted lines) and DR3 (solid lines). Vertical lines denote the medians of each distribution.}
    \label{fig:fluxcal_colour}
\end{figure*}

For DR3 a number of modifications were made to flux calibration procedures. In the processing of the primary calibrators, three changes were made. First, the  original reference calibration spectra, that are relatively low resolution (often 50\,\AA\ bins) and sometimes had residual telluric absorption still present, were replaced with higher resolution telluric--corrected versions from the Supernovae Factory project \citep{2002SPIE.4836...61A}\footnote{https://snfactory.lbl.gov/snf/snf-specstars.html}. Second, \texttt{molecfit} was applied to the primary standards before determining the spectrophotometric transfer function, allowing us to better sample the transfer function in the red arm. Third, the fitting of the standard star flux within the individual RSS frames was modified to use differential atmospheric refraction (DAR) parameters based on atmospheric conditions and zenith distance. Previous versions had directly fit the positional shift of the star with wavelength, but the under--sampling of the PSF by the fibres meant that in some cases the direct fit was not robust.   

The processing of the secondary flux calibration stars was substantially revised, so that a full transfer function could be estimated from them. This followed a similar approach to that used by SDSS \citep{2004AJ....128..502A}.  In detail the procedure was as follows:
\begin{enumerate}
\item Individual RSS frames were corrected for atmospheric extinction, approximately flux calibrated using the primary standards and then telluric absorption corrected.
\item For each reduced RSS frame the flux of the secondary standard was extracted fitting a Moffat profile allowing for DAR.
\item The extracted secondary spectra were fit to model template spectra based on \citet{1992IAUS..149..225K} model atmospheres. We used Penalized Pixel-Fitting \citep[pPXF;][]{2004PASP..116..138C} and this was done in two stages. The first stage fitted individual templates in a grid of effective temperature ($T_{\rm eff}$), metallicity ([Fe/H]) and surface gravity. For the best fit surface gravity the nearest 4 templates in $T_{\rm eff}$ and [Fe/H] were then refit allowing a linear combination of templates. The fitting was only done on the blue arm data (3700--5700\,\AA) and included a multiplicative polynomial of 8th order to take out residual transfer function errors. The weights of the templates were saved to the RSS frame.
\item The weights of the templates are averaged across all the observations of a field, as these typically contain 7 (or more) observations of the same star. 
\item From the average weights a best fit template spectrum was derived, including Galactic extinction from {\it Planck} \citep{2014A&A...571A..11P}, as the secondary flux calibration stars are sufficiently distant that they are all in the Galactic halo. The best fit template was normalized by comparing to the observed $g$- and $r$-band photometry for the star in question and applying the average normalization from the two bands.
\item Transfer functions for each RSS frame were derived by comparing the observed spectrum to the best fit template. While in principle this could be applied to the individual frames, we only have a single secondary calibration star observed per field, and this can lead to some scatter in the transfer function. Instead we average the transfer functions across all the observations of a field in a given night.
\item The average transfer function was applied to the data, but allowing for an individual normalization for each frame to account for variations in transmission.
\end{enumerate}
Once the transfer function was applied, no further scaling of the data for transmission was carried out.

\begin{figure}
    \centering
    \includegraphics[width=8cm]{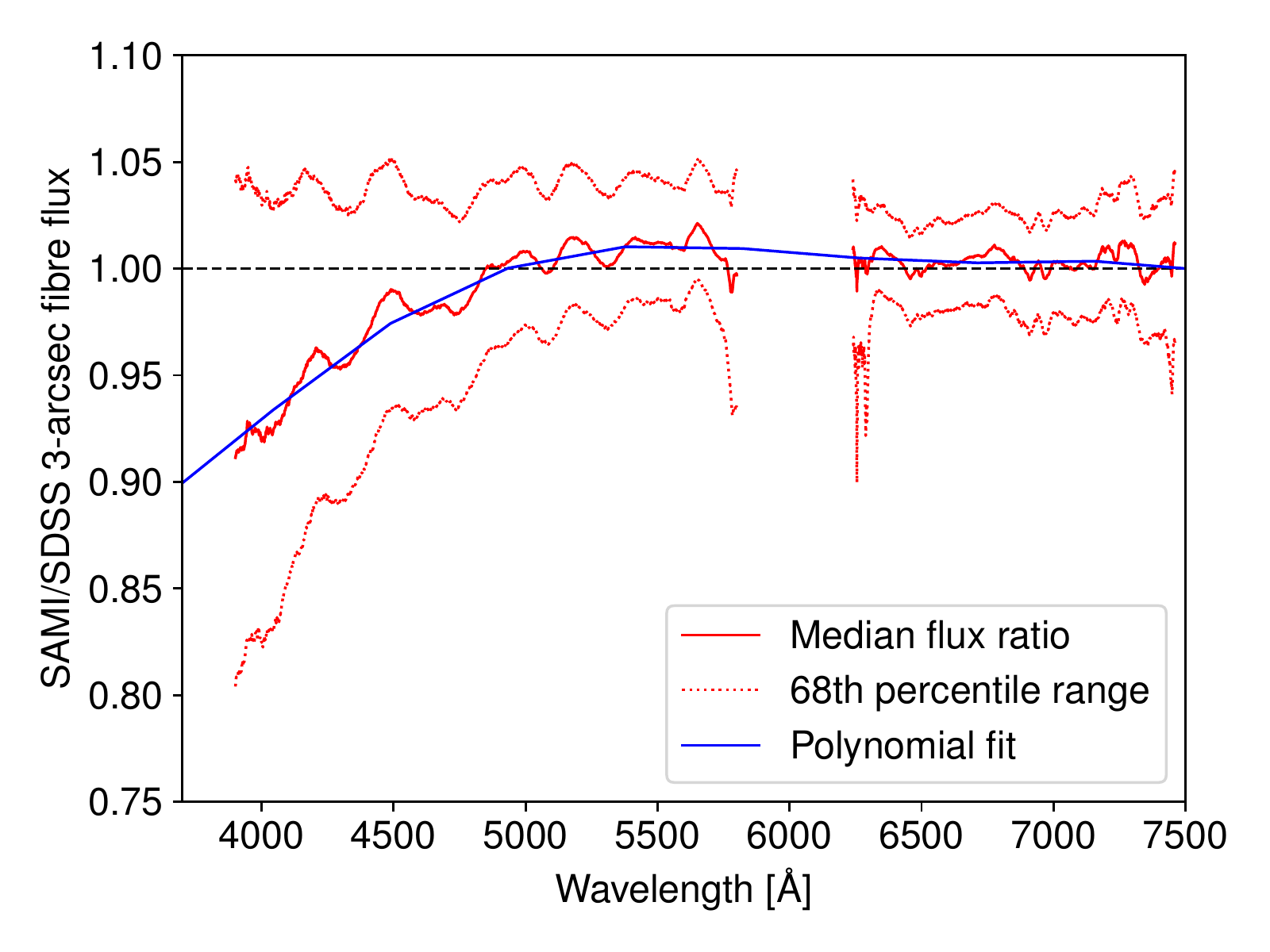}
    \caption{The median ratio of SAMI to SDSS 3-arcsec aperture spectra in the AAOmega blue and red arms. The dotted lines show the 68th percentile range. Spectra are smoothed using a 51 pixel median filter prior to combining, to remove the impact of individual bad pixels.  The blue line is a 5th order polynomial fit to the median ratio.}
    \label{fig:fluxcal_spec}
\end{figure}

\begin{figure}
    \centering
    \includegraphics[width=8cm]{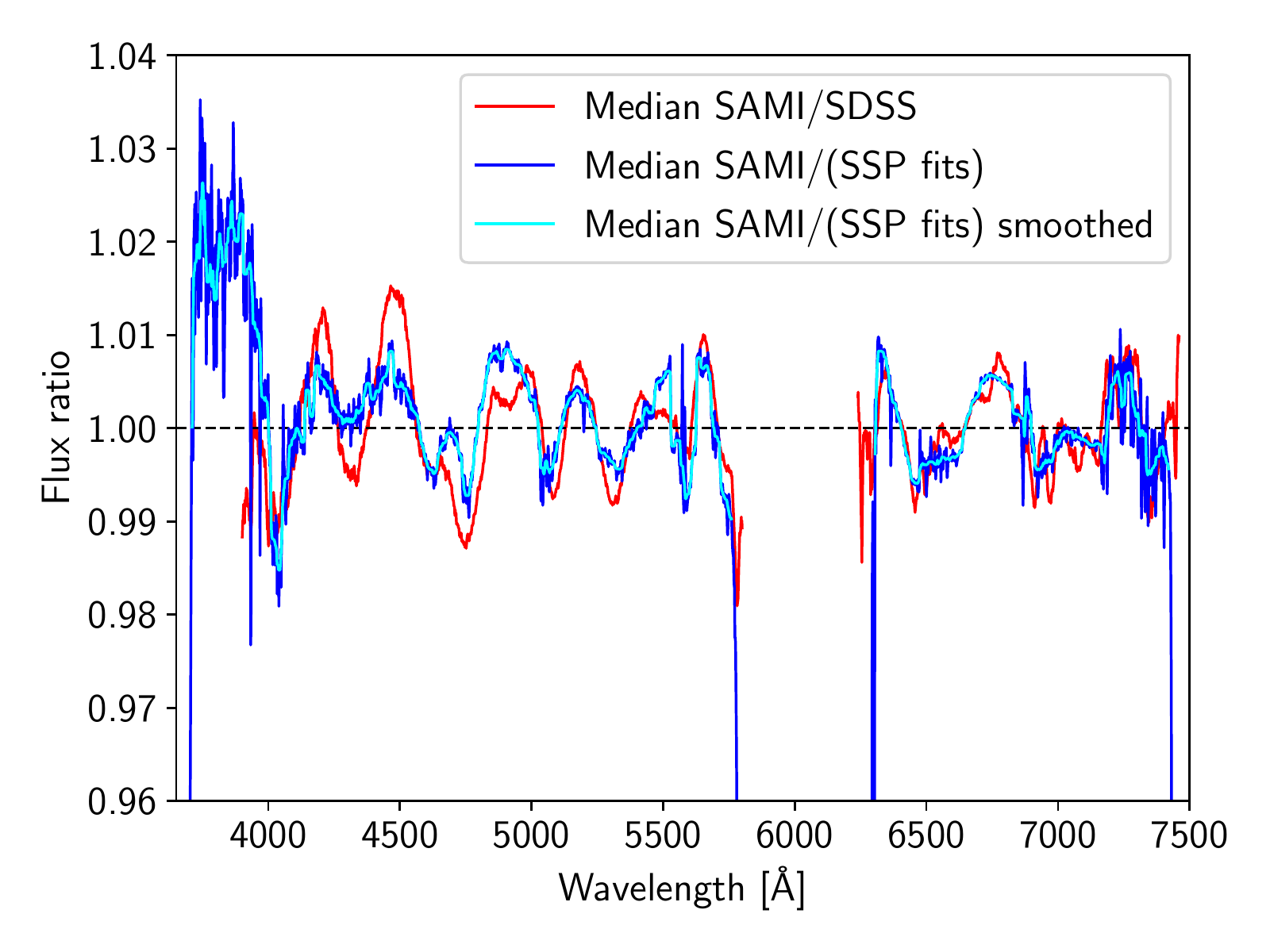}
    \caption{The median ratio of SAMI to SDSS 3-arcsec aperture spectra (red line), compared to the ratio of SAMI to stellar population fits (blue line).  Also shown is a median filtered version of the SAMI to stellar population ratio (cyan line).}
    \label{fig:fluxcal_spec_sp}
\end{figure}

Various tests were carried out to examine the robustness of this approach. We investigated whether variations in the assumed atmospheric extinction curve caused residuals, but there was no correlation found between spectrophotometric residuals and airmass. Residual flux calibration differences (see below) also did not appear to have the same shape as known occasional changes due to excess dust and smoke in the atmosphere (Mike Bessell, private communication). We also found that the scatter in spectrophotometric calibration increased when we applied transfer functions derived from individual frames.

To test the improvement of the spectrophotometric calibration we compare 3 arcsec aperture spectra derived from SAMI cubes (see Section~\ref{sec:primedata}) in DR2 and DR3 to single fibre SDSS spectra of the same object. This is not an absolute test, as varying seeing and aperture effects (including differences in pointing) can influence the aperture spectra, particularly for single fibre spectra. However, differences between SDSS and SAMI spectra provide a lower limit on the accuracy of our spectrophotometric calibration. 

As a first test we compare the median SAMI flux (across both SAMI arms) to the median SDSS flux in the same spectral windows. As  SDSS 3-arcsec diameter spectra are renormalized to the PSF magnitudes we first correct them (on average) back to fibre fluxes with a scaling of 0.35 mag. The distribution of SAMI to SDSS flux ratios is shown in Fig.\ \ref{fig:fluxcal_scale}a. The median offset (solid vertical line) is 1.04 with a 68th percentile width of 0.16. For DR2 the median offset was larger at 1.12, but with the same width of 0.16. The width of the distribution is dominated by variations in seeing, as can be seen in Fig.\ \ref{fig:fluxcal_scale}b.  The SAMI/SDSS flux ratio declines with increased SAMI seeing as less flux falls within the 3-arcsec diameter aperture defined within the SAMI cubes.  There is not the same trend with SDSS seeing, as the SDSS spectra are normalized to PSF magnitudes.  Given that the median seeing for the SDSS and SAMI data in this comparison are very similar, 1.96 and 2.04 arcsec respectively, it is not surprising that we have good agreement between the flux scales of the two data sets.

In Fig. \ref{fig:fluxcal_colour} we make a comparison of the spectral shapes of SDSS and SAMI aperture spectra. We calculate the median flux in 100\,\AA\ bands centred at 4000\,\AA, 5450\,\AA\ and 7000\,\AA. We then define the ratio
\begin{equation}
R(\lambda) = F(\lambda)_{\rm SAMI}/F(\lambda)_{\rm SDSS}
\end{equation}
for each wavelength, where $F(\lambda)$ is the median flux in a 100\,\AA\ window centred at a wavelength of $\lambda$. The distribution of the ratio of $R(\lambda)$ at different wavelengths is then shown in Fig.\ \ref{fig:fluxcal_colour}. The ratio $R(4000)/R(5450)$ has a median value of 0.91, and the values for both DR2 and DR3 are similar. This implies that on average SAMI spectra have less blue flux than SDSS spectra. The 68th percentile width of the $R(4000)/R(5450)$ distribution is 0.09 for DR3, slightly narrower than the value for DR2 (0.12). The ratio $R(5450)/R(7000)$ (Fig.\ \ref{fig:fluxcal_colour}b) is very close to 1, with a median of 1.01 for DR3. The scatter is significantly reduced from DR2 to DR3, with the 68th percentile width reducing from 0.073 to 0.045. The new flux calibration in DR3 has in particular removed the tail to high $R(5450)/R(7000)$ seen in DR2. This shows that the blue and red arms are now better matched spectrophotometrically. 

Fig.\ \ref{fig:fluxcal_spec} shows the median ratio of SAMI to SDSS flux, once each spectrum is scaled by the median flux ratio (to take out normalization differences due to seeing and other aperture effects). The trend seen in Fig.\ \ref{fig:fluxcal_spec} is consistent with that in Fig.\ \ref{fig:fluxcal_colour}a. At $\sim4000$\,\AA\ SAMI spectra have $\simeq7$ percent less blue flux. The 68th percentile range of flux ratios (dotted lines in Fig.\ \ref{fig:fluxcal_spec}) is seen to increase towards the blue.  To enable a direct mapping to the SDSS spectral flux scale we fit a 5th order polynomial to the flux ratio (blue line in Fig.\ \ref{fig:fluxcal_spec}).  The best fit function is
\begin{equation}
\begin{aligned}
R(x) = &\, 8.25975 -41.8455 x + 90.374 x^2 -93.2674 x^3 \\
    & + 46.5418 x^ 4 -9.0518 x^5,
  \end{aligned}
  \label{eq:sami_to_sdss}
\end{equation}
where $x=\lambda/5500$ and $\lambda$ is the wavelength in \AA.  We have investigated various explanations of the difference between SDSS fibre spectra and SAMI aperture spectra that could be caused by the flux calibration of either or both data sets.  Small known differences in the pipeline approaches, such as SAMI using Planck dust extinction estimates compared to SDSS using estimates from \citet{1998ApJ...500..525S}, can only account for $\sim1$ percent differences.  A small number of individual SDSS plates show differences in the relative flux calibration that are factors of $2-3$ larger than the median relation in Fig.\ \ref{fig:fluxcal_spec}, but removing these does not noticeably modify the median trend.  We do not find correlations of the difference with airmass or seeing.  We do not apply Eq. \ref{eq:sami_to_sdss} directly to the SAMI DR3 data products, as the source of the difference between SAMI and SDSS flux calibration is unclear. However, we provide it here to allow users to correct (on average) between the two data sets.

Once the SAMI/SDSS flux ratio is corrected using Eq.\ \ref{eq:sami_to_sdss} there are weak (less than 1 percent) residual small-scale features remaining.  To examine whether these are real we create a flux ratio between SAMI and model stellar population fits made to 69428 SAMI spectra (each binned to a minimum S/N ratio of 20 \AA$^{-1}$). These fits are performed using the penalised pixel fitting code \textsc{pPXF} \citep{2004PASP..116..138C, cappellari2017}. We use templates from the simple stellar population (SSP) models of \citet{vazdekis2015}, as well as including templates representing a number of common optical emission lines in the (rest-frame) wavelength range of 3500 \AA{} to 7000 \AA. Full details of these stellar population fits will be presented in an upcoming paper (Vaughan et al. in prep).  

A comparison between the SAMI/SDSS flux ratio (red line) and the SAMI/(SSP fits) flux ratio (blue line) is shown in Fig.\ \ref{fig:fluxcal_spec_sp}.  The small-scale variations in the flux ratios are in relatively good agreement, suggesting that most of the ripples seen are low-level systematic features in the SAMI flux calibration.  In most cases these flux calibration residuals will only impact spectral fits of very high S/N spectra.  However, as part of DR3 we provide a correction vector to remove these residuals based on the SAMI/(SSP fits) flux ratio.  This is median filtered using a 21 pixel window and only specified within the wavelength ranges $3710-5760$\AA\ and $6305-7420$\AA\ where the correction is well defined.  The correction vector is shown as the cyan line in Fig.\ \ref{fig:fluxcal_spec_sp} and SAMI data should be divided by this vector to apply the correction.

We perform an independent fit of the $\sim3000$ SAMI aperture spectra (see Section \ref{sec:ap_spec}) to test the above correction using the {\sc starlight} code \citep{2005MNRAS.358..363C} and templates from \citet{2003MNRAS.344.1000B}.  The correction leads to very slightly reduced residuals, but these are not significant for individual spectra.  For example, using the correction the mean fractional residuals are reduced from $-0.029\pm0.187$ to $-0.028\pm0.177$ in the window 4000-4500\,\AA. At 4500-5000\,\AA\ the change is slightly larger from $0.017\pm0.084$ to $0.014\pm0.082$.  The errors quoted are the RMS scatter about the mean.  Tests using other code and templates lead to similar results.  There is also no significant difference in the estimated ages and metallicities.  In summary, correcting the low-level ripples in flux calibration does not significantly impact our spectral fitting.

\begin{figure}
    \centering
    \includegraphics[width=8cm]{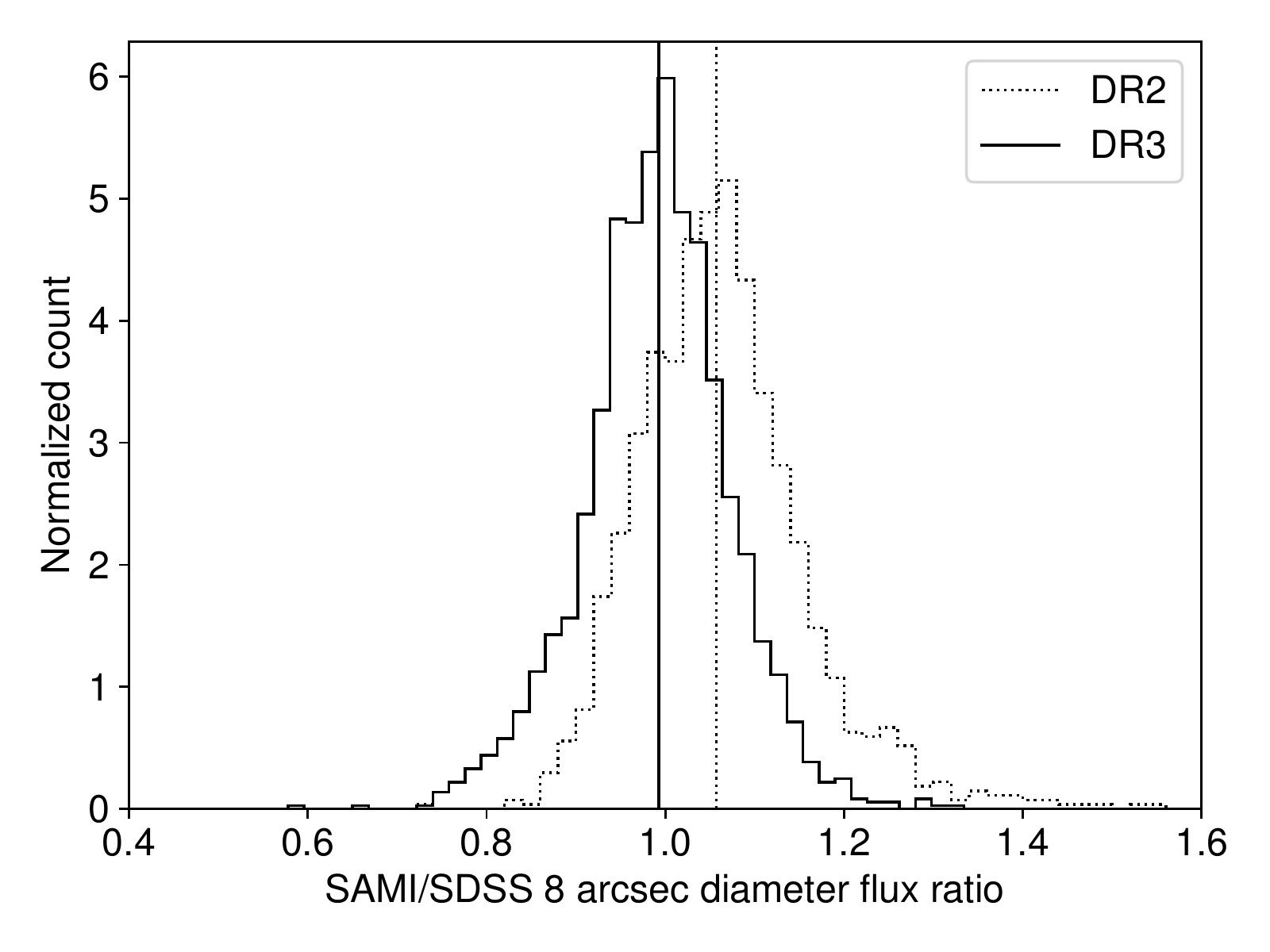}
    \caption{The distribution of flux ratios of SAMI blue cubes to SDSS $g$-band imaging for DR3 (solid lines) and DR2 (dotted lines). The vertical lines show the median values in each case.}
    \label{fig:fluxcal_ratio}
\end{figure}

Because of the impact of aperture effects when comparing to SDSS 3-arcsec fibre spectra, we also make a comparison to the SDSS $g$-band imaging. We follow the same procedure described by \citet{2018MNRAS.481.2299S}, taking an 8 arcsec diameter aperture to measure the flux in the SAMI cubes and SDSS $g$-band imaging data. The flux from the SAMI cube is convolved with the $g$-band filter response prior to making the comparison. The ratio of SAMI to SDSS flux is shown in Fig.\ \ref{fig:fluxcal_ratio}. Consistent with the comparison to SDSS spectra, SAMI DR3 data is a closer match to SDSS photometry. The median ratio for DR3 is $0.991\pm0.002$, compared to $1.051\pm0.003$ for DR2. The 68th and 95th percentile ranges of the flux ratio are $\pm0.074$ and $\pm0.135$ for DR3 compared to $\pm0.081$ and $\pm0.144$ for DR2. Hence the scatter is also slightly reduced in DR3 relative to DR2. 

\subsection{Centring and WCS}

The observed flux of each galaxy in each individual exposure was fit with a two-dimensional Gaussian to identify the centre of the source. This approach sometimes generates an inaccurate centroid when there is a bright star or a secondary object within the field-of-view of the hexabundle. In such a case, we masked out the flux from the secondary object before fitting a Gaussian, which reduces the mis-identification of the galaxy of interest. After this, only 0.5\% of the entire survey (12 galaxies) is assessed to have inaccurate centroids due to a bright secondary object/star (catalogue IDs 218717, 228278, 273256, 380682, 509892, 549182, 610816, 760733, 91579, 93384, 9008500120, and 9016800113). The dithers are combined after aligning the centroids. The galaxy centre is located at cube spaxel coordinates (25.5, 25.5) where we assign the catalogue coordinate of the galaxy. Then, we defined a world-coordinate system (WCS) for each cube using the relative position to the spatial coordinate of the galaxy centre.

\begin{figure}
\centering
\includegraphics[width=\columnwidth]{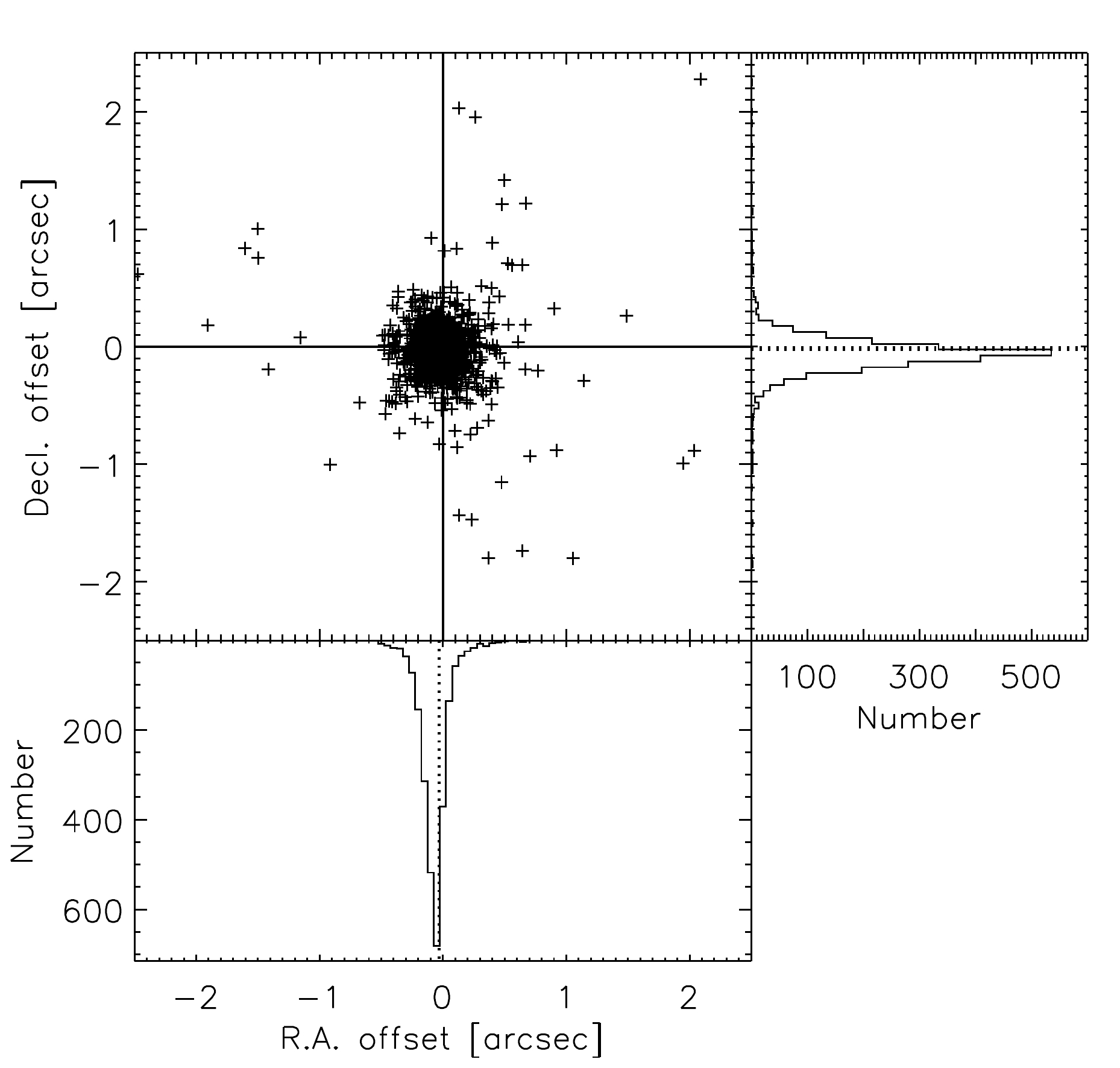}
\caption{The distribution of R.A. and Decl. offset between SAMI and SDSS with histograms of the offset in R.A. and Decl. along the axes. Dotted lines in the histograms show the median offset in R.A. (-0.030 arcsec) and Decl. (-0.017 arcsec). Note that 12 mis-identified galaxies are not presented.}
\label{fig:wcs}
\end{figure}

We characterise the accuracy of the SAMI WCS by cross-correlating the reconstructed SAMI images with $g$-band images from the SDSS. We applied the SDSS $g$-band filter response to each cube to reconstruct a two-dimensional image. The SDSS images have been re-sampled to have the same pixel scale as the SAMI cubes. Fig.~\ref{fig:wcs} presents the offset in RA and Dec.\ between the SAMI and SDSS WCS. The median offset is $-0.030 \pm 0.004$ arcsec in RA and $-0.017 \pm 0.005$ arcsec in Dec. We visually inspected 51 outliers whose offset is greater than 0.5 arcsec (one SAMI pixel) either in RA or Dec. We found 8 galaxies whose catalogue coordinates do not correspond to the object centre in the SDSS images though they are well centred on the SAMI cubes. The circular Gaussian distribution used for centroiding may not represent the centre of highly disturbed or edge-on galaxies very well. We found an offset greater than 0.5 arcsec from 31 disturbed and 12 edge-on galaxies. We corrected the cube WCS of 12 mis-identifications (listed above) and the 51 outliers to match that estimated from the SDSS imaging.

\subsection{Cubing and bad pixel rejection}\label{sec:cubing}

The procedure for cubing is based on a Drizzle-like algorithm, described by \citet{2015MNRAS.446.1551S}. For DR3 the only modification to the cubing was to implement a more robust bad pixel rejection algorithm. In DR2 and prior, the bad-pixel rejection was a simple clipping of pixels across multiple frames that are more than $5\sigma$ from the median value in a given output pixel. However, in some cases this was not stable, particularly for very bright objects, where small seeing variations between frames might lead to spurious rejection. The result of this was that a small number of individual spaxels in the output cubes had sudden steps in their data as a function of wavelength.  

The bad pixel rejection as part of cubing is particularly important for the red arm of AAOmega. This is because the thick detector in the red camera causes some cosmic ray events to be spread out, to be as broad as the instrumental PSF. As a result, regular edge-detection algorithms \citep{2001PASP..113.1420V,2012A&A...545A.137H} are not effective. 

To improve the bad-pixel rejection we implemented a more robust algorithm that compares fibre spectra prior to projecting them into a cube. For each fibre spectrum we identify the six fibres spatially nearest to the fibre in question (using all RSS frames that will contribute to the cube). Each fibre has the median filtered continuum subtracted and they are then combined to generate a median spectrum (from seven fibres including the fibre being tested for bad pixels). The median absolute deviation is estimated from the seven continuum subtracted spectra. The median spectrum is then compared to the individual spectra, allowing for a scaling between the two, and pixels that were more than 5 times the median absolute deviation away from the median spectrum were flagged to be rejected. {\update Application of this new method resolved the problems that the previous algorithm had caused.}

\section{Primary data products}\label{sec:primedata}

\subsection{Cubes}

The main data products from the SAMI Galaxy Survey are spectral cubes for each observed galaxy. The structure of these is unchanged from DR2 \citep{2018MNRAS.481.2299S}, but we give a brief description here for completeness. There is a separate cube for the blue and red arms of the spectrograph. The cubes have $50\times50$ spatial pixels (spaxels), each one $0.5\times0.5$ arcsec in area, and 2048 wavelength slices. Each cube contains the measured flux, a variance cube, a weight map and the compressed covariance between spaxels \citep{2015MNRAS.446.1551S}.

In DR3 we release all cubes that pass our quality control limits (see Section~\ref{sec:obs}). The release includes repeat observations of galaxies as separate cubes. There are a total of 3426 pairs of galaxy cubes of 3068 unique galaxies. We also include the calibration star cubes used for flux calibration and PSF estimation. There are 286 pairs of calibration star cubes of 177 unique stars.

As well as the default cubes with $0.5\times0.5$ arcsec spaxels we release cubes with different binning schemes. These are described in detail by \citet{2018MNRAS.481.2299S}. In summary these are:
\begin{itemize}
\item `adaptive' binning, using the Voronoi method of  \citet{2003MNRAS.342..345C} to a median blue arm S/N of 10.
\item `annular' binning into five linearly spaced elliptical annuli.  The ellipses are defined using the {\sc find\_galaxy} routine \citep{cappellari2002} applied to images generated by collapsing the cubes in the wavelength direction.
\item `sector' binning that azimuthally subdivides each of the annular bins into eight equal area regions.
\end{itemize}

All the binned cubes are generated using the binning module within the SAMI data reduction pipeline \citep{2014ascl.soft07006A}.

\subsection{Aperture spectra}\label{sec:ap_spec}

To allow easy comparison to single aperture data (e.g. single fibre surveys), and to measure quantities in consistent physical or relative apertures, we generate aperture spectra. The data structure for aperture spectra is unchanged compared to DR2 \citep{2018MNRAS.481.2299S}. The aperture spectra are calculated using the binning module in the SAMI data reduction pipeline. The apertures used are the following:
\begin{itemize}
\item circular 1.4, 2.0, 3.0 and 4.0 arcsec diameter apertures.

\item a circular 3\,kpc diameter aperture based on the flow corrected redshift, z\_tonry (for the GAMA regions), or the cluster redshift (for cluster galaxies).

\item an elliptical 1$R_{\rm e}$ aperture, based on Sersic fits to SDSS or VST imaging data, as described by \citet{2012MNRAS.421.1007K} and  \citet{2019ApJ...873...52O}.

\item an elliptical 1$R_{\rm e}$ aperture based on multi-Gaussian expansion (MGE) fitting of SDSS or VST photometry. The MGE fitting is described in Section~\ref{sec:cat_phot} and is new to DR3.
\end{itemize}
For a small number of galaxies aperture spectra are not available in all apertures.  For example MGE measurements have not been made for galaxies in the filler samples.

\subsection{Input and observational catalogues}\label{sec:incat}

\begin{table*}
\centering
\caption{A summary of the catalogues provided as part of DR3.  We list the catalogue name, the number of rows ($N_{\rm rows}$), a brief description of the catalogue and the primary references for the catalogue.\label{table:cats}}
    \begin{tabular}{ccccccc}
    \hline
Name & $N_{\rm rows}$ & Description & Reference\\
\hline
InputCatGAMADR3 & 5536 & Targets and their properties in GAMA regions & \citet{2015MNRAS.447.2857B}\\
InputCatClustersDR3 & 1433 & Targets and their properties in cluster regions & \citet{2017MNRAS.468.1824O}\\
InputCatFiller & 2980 & Filler targets & This paper\\
FstarCatGAMA & 2578 & Colour selected calibration stars & \citet{2015MNRAS.447.2857B}\\
FstarCatClusters & 183 & Colour selected calibration stars & \citet{2017MNRAS.468.1824O}\\
VisualMorphologyDR3 & 3068 & Visual morphological classification & \citet{2016MNRAS.463..170C}\\
CubeObs & 3712 & Observed cubes and quality flags & This paper\\
MGEPhotomUnregDR3 & 3150 & MGE photometric measurements & D'Eugenio et al.\ in prep\\
samiDR3Stelkin & 3426 & Stellar kinematic measurements & \citet{vandesande2017a}\\
samiDR3gaskinPA & 3426 & Gas kinematics position angle & \citet{2018MNRAS.481.2299S}\\
IndexAperturesDR3 & 3375 & Stellar continuum index measurements & \cite{2017MNRAS.472.2833S}\\
SSPAperturesDR3 & 3375 & SSP age, [Z/H] and [$\alpha$/Fe] & \cite{2017MNRAS.472.2833S}\\
DensityCatDR3 & 6969 & Local density estimates & This paper\\
EmissionLine1compDR3 & 3245 & Aperture em.\ line measurements, 1-component fits & \citet{2018MNRAS.481.2299S}\\
EmissionLineRecomcompDR3 & 3245 & Aperture em.\ line measurements, recommended component fits &  \citet{2018MNRAS.481.2299S}\\
\hline
\end{tabular}
\end{table*}

Input catalogues for all sources are provided as part of DR3. This is a total of 5 catalogues listed in Table \ref{table:cats}. The input catalogue for the GAMA regions (named InputCatGAMADR3) is identical to that presented by \cite{2015MNRAS.447.2857B} with two exceptions.  The first is that two sets of coordinates are now listed. One set, named RA\_OBJ and DEC\_OBJ are the object coordinates taken from the input photometric catalogues. A second set of coordinates is also provided, named RA\_IFU and DEC\_IFU. These give the nominal IFU pointing for each target. In most cases these two sets of coordinates are identical, but in a small number of cases the IFU pointing was modified (see Section~\ref{section:Input_cats}).  The second change for the GAMA input catalogue is that the source position angle (PA) has been corrected to the standard on-sky North through East definition.  Previous versions of the input catalogue had PA defined in image coordinates, taken directly from the GAMA Sersic fit catalogues \citep{2012MNRAS.421.1007K}.

The input catalogue for the SAMI cluster regions (named InputCatClustersDR3) is newly presented in this data release. It is based on the SAMI Cluster Redshift Survey \citep[][]{2017MNRAS.468.1824O} carried out using the 2-degree Field (2dF) instrument on the AAT. Selection is described in Section~\ref{subsec:clus_input} and \citet{2017MNRAS.468.1824O}. The cluster input catalogue includes similar measurements to the GAMA input catalogue, including redshift, stellar mass, $g-i$ colour. It also includes $R_{\rm e}$, ellipticity and PA derived from Sersic photometric fits from \citet{2019ApJ...873...52O}.

With DR3 we release the complete catalogues of all potential filler targets for the SAMI Survey (catalogue name InputCatFiller, see also Section \ref{sec:input_filler}).  These catalogues do not have the same extensive set of parameters as the main survey targets.  The input catalogues for colour selected F-stars is also provided as part of DR3 in both the GAMA and cluster regions (these catalogues are named FstarCatGAMA and FstarCatClusters).

Details of all the observations that passed our quality criteria and that led to cubes being produced are listed in the catalogue CubeObs.  The primary role of this catalogue is to provide quality assessments such as seeing and relative transmission for each observed cube.  The CubeObs catalogue includes observations of calibration stars.  Where repeat observations have led to multiple cubes being made, the CubeObs catalogue separately lists each set of cubes.  The flag ISBEST is used to indicate in the case of repeats which of the observations is considered the highest quality.  The CubeObs catalogue also contains a number of flags indicating a variety of issues, including missing value-added products, multiple objects in the IFU field-of-view and various calibration and measurements problems.   Description of these flags is contained as part of the schema in the online database.

\section{Photometry based catalogues}\label{sec:cat_phot}

\subsection{Visual Morphology}

All galaxies in the survey have been visually classified by several members of our team taking advantage of RGB combined colour images from either the SDSS Data Release 9 or VST ATLAS surveys. The classification scheme is simple, with just 4 types (ellipicals, S0s, early- and late-spirals), and it is based on the prominence/presence of a bulge component and spiral arms. For a small fraction of the sample (148 galaxies, $\sim$5\% of the sample), no classification is provided as it was considered too uncertain/subjective. We refer the reader to \cite{2016MNRAS.463..170C} for a more extensive description of our classification procedure.  The morphology measurements are available in catalogue VisualMorphologyDR3.       

\subsection{MGE fits}

For each galaxy, we measure the photometric parameters of magnitude, position angle,
projected ellipticity and projected half-light radius using the Multi Gaussian 
Expansion algorithm \citep[MGE,][]{emsellem1994}. The measurements are based on
$r-$band SDSS and VST images. To characterise the PSF, we use the source 
detection software {\sc SExtractor} \citep{bertin1996} to retrieve all the stars
in a 400$\times$400 arcsec cutout image around the galaxy in question. We then use the Python package {\sc mgefit} to obtain the
MGE of each star \citep{cappellari2002}. To model the galaxy, the adopted PSF is 
the best-fit MGE of
the star that is the best compromise between goodness of the MGE, magnitude and
distance to the target. To mask regions contaminated by neighbours and 
interlopers, we use segmentation maps from {\sc SExtractor}. The position angle
is measured using either the method of moments, or {\sc SExtractor}, whichever
gives the lowest $\chi^2$. The total magnitude $m$ and the circularised half-light
radius $R_e$ are calculated from the best non-regularised MGE model. The projected
ellipticity
is the ellipticity of the isophote of area $\pi R_e^2$. Our measurements are
validated against the corresponding GAMA measurements, that are based on 
2-dimensional S{\'e}rsic fits \citep{2012MNRAS.421.1007K}. In log-space, the best-fit
linear relation between S{\'e}rsic- and MGE-based half-light radii has $rms =
0.067 \, \mathrm{dex}$. Based on this result, we estimate the measurement
uncertainty of the MGE half-light radii as $0.067 / \sqrt{2} = 0.047 \,
\mathrm{dex}$, similar to the estimate of ATLAS3D \citep{cappellari2013a}.   In the Abell 85 cluster both SDSS and VST photometry is available and so we provide measurements for both sets of data within the MGE catalogue.  More
information on these measurements will be provided in a separate paper (D'Eugenio~et~al., in prep.)

\section{Stellar kinematics value added products}\label{sec:kin}

\subsection{Method}\label{subsubsec:kinematic_method}

We extract the line-of-sight stellar velocity distribution (LOSVD) from SAMI cubed data using the penalised Pixel Fitting code ({\sc pPXF}) to fit all spectra \citep{2004PASP..116..138C,cappellari2017}. The method is described in detail by \citet{vandesande2017a} with the various value added products outlined by \citet{2018MNRAS.481.2299S}. We give a brief overview below.

In DR3 we provide two stellar kinematic data products: one where we assume a Gaussian LOSVD (two moments: \vmt, \smt), and an alternative where the LOSVD is parametrized with a truncated Gauss-Hermite series \citep{vandermarel1993,gerhard1993} with four kinematic moments: \vmf, \smf, $h_3$ and $h_4$. Here, $h_3$ and $h_4$ are related to the skewness and kurtosis of the LOSVD. Local optimal templates are derived from annular binned spectra (Section~\ref{sec:primedata}) using the MILES stellar library \citep{sanchezblazquez2006,falconbarroso2011} that consists of 985 stars covering a large range in stellar atmospheric parameters. 
 
To extract the LOSVD, we fit each spaxel three times with  \textsc{pPXF}. Once for determining a precise measure of the noise scaling from the residual of the fit, a second time to clip outliers using the CLEAN parameter in \textsc{pPXF}, and a third time where \textsc{pPXF} is provided with the optimal templates from the annular bin in which the spaxel is located (derived from the previous fits to the annular bins), as well as the optimal templates from neighbouring annular bins. A 12th order additive Legendre polynomial is used to remove residuals from small errors in the flux calibration. Uncertainties on the LOSVD parameters are estimated using a Monte-Carlo approach. The same method is applied to the binned data (Section~\ref{sec:primedata}). For the aperture spectra (Section~\ref{sec:primedata}), optimal templates are constructed for each individual aperture and we then use the same procedure as described above to extract the LOSVD.

For SAMI DR3 data we recommend applying the following quality criteria to the stellar kinematic maps. For the Gaussian LOSVD maps: S/N$>3$\ \AA$^{-1}$, \sobs $>$ FWHM$_{\rm{instr}}/2 \sim 35$\kms, $V_{\rm{error}}<30$\kms\ \citep[Q$_1$ from][]{vandesande2017a}, and $\sigma_{\rm{error}} < \sobs \times 0.1 + 25$\kms\ \citep[Q$_2$ from][]{vandesande2017a}. For the four-moment LOSVD fits, a reliable estimate of $h_3$ and $h_4$ requires a higher S/N quality cut of  $>20$\,\AA$^{-1}$ and $\sigma>70$\kms \citep[Q$_3$ from][]{vandesande2017a}. 

\subsection{Products}\label{subsubsec:kinematic_products}

Maps of the two- and four-moment kinematic fitted parameters are provided as part of DR3 for all the binning schemes we release.  Additionally we provide four-moment fits for another binning scheme, adaptively binned to S/N = 20\,\AA$^{-1}$.  From the two-moment kinematic maps we derive the following value added products. 
\begin{enumerate}
    \item The kinematic asymmetry of the galaxy velocity fields. We determine the amplitudes of the Fourier harmonics $k_5/k_1$ on all velocity data that pass the quality cut Q$_1$, measured using the \kinemetry\ routine \citep{krajnovic2006,krajnovic2008} following the method outlined in \citet{krajnovic2011} and \citet{vandesande2017a}. 
    \item The position angle (PA) of the stellar rotation. We measure the PA from  the stellar velocity maps on all spaxels that pass the quality cut Q1 using the \textsc{fit\_kinematic\_pa} code that is based on the method described in Appendix C of \citet{krajnovic2006}. 
    \item The ratio of ordered versus random motions {\update given by \vse\ and \lre.   We use the definitions described by \citet{cappellari2007} and \citet{emsellem2007}. We calculate the sum of the unbinned flux, velocity, and velocity dispersion maps over all spaxels that pass the kinematic quality cuts Q$_1$ and Q$_2$ \citep{vandesande2017a} within an ellipse with semi-major axis \re\ and axis ratio $b/a$. In the calculation of \lr\, we adopt a radius,  $R$,  definition as described by \citet[see][]{2016MNRAS.463..170C}; i.e.\ $R$ is the semi-major axis of the ellipse on which each spaxel sits, not the polar radius. A fill factor of 95 per cent of good spaxels within the aperture is required for producing \lre\ and \vse\ measurements}. When the largest kinematic aperture radius is smaller than the effective radius, we apply an aperture correction to \vs\ and \lr\ as described in \citet{vandesande2017b}.
\end{enumerate}
For these kinematic data products, we provide a flag derived from a visual inspection where we have identified maps with irregular kinematics due to nearby objects or mergers that influence the stellar kinematics of the main object.  The main stellar kinematic catalogue is samiDR3Stelkin, however the flags for problem objects are contained within the main observational catalogue CubeObs, as they can influence measurements other than the stellar kinematics.

\section{Stellar population value added products}\label{sec:pop}

In DR3 we provide a table of single stellar population (SSP) equivalent measurements of the light-weighted age, metallicity ([Z/H]) and alpha-enhancement ([$\alpha$/Fe]) derived from the SAMI DR3 aperture spectra included in this release. We provide SSP measurements for all apertures with sufficient S/N, averaged over the entire blue arm spectrum. We impose a minimum S/N limit of 10 per \AA\ for age and [Z/H] and a minimum S/N of 20 per \AA\ for [$\alpha$/Fe] measurements -- we consider measurements that satisfy this criterion `successful'. The value -99 is used to indicate measurements that do not satisfy these S/N criteria, or are otherwise missing. We also provide Lick absorption line index measurements for 20 different indices for all apertures.

We attempt to measure SSP equivalent parameters for all apertures of all galaxies that have 3-kpc aperture spectra in DR3. The Lick absorption line index and SSP measurement tables provided as part of DR3 include 3375 objects, of which 358 are duplicate observations. We successfully measure age and [Z/H] ([$\alpha$/Fe]) for 2989 (2754) unique galaxies in the 4 arcsec circular aperture (the highest average S/N aperture), decreasing to 2772 (2241) in the 1.4 arcsec circular aperture (the lowest average S/N aperture). Median formal uncertainties in the 4 arcsec aperture are 1.4 Gyr in age, 0.20 dex in [Z/H] and 0.18 dex in [$\alpha$/Fe]. Comparison of 70 galaxies with independent duplicate observations within DR3 gives rms scatters in reasonable agreement with the formal uncertainties (0.15, 0.27 and 0.12 dex in log age, [Z/H] and [$\alpha$/Fe]); however, we note that these estimates are less reliable given the relatively small number of duplicate observations of any given galaxy. When comparing DR2 and DR3 population measurements for galaxies in common between the two we find rms scatters of 0.12 dex in log age, 0.21 dex in [Z/H] and 0.14 dex in [$\alpha$/Fe].

The method used to measure absorption line indices and SSP-equivalent parameters is identical to that outlined in \citet{2018MNRAS.481.2299S}, with details provided in \citet{2017MNRAS.472.2833S}. Briefly, absorption line indices are measured on emission and Milky Way-extinction corrected blue-arm aperture spectra using a {\sc Python} implementation of the {\sc EZ\_Ages IDL} software package \citep{2014ascl.soft07019G}, which includes a correction for the intrinsic velocity dispersion of each galaxy. We adopt index definitions from \citet{1998ApJS..116....1T}. From the Lick index measurements we determine stellar population parameters by comparing to the stellar population synthesis models of \citet[for metallicity and alpha-enhancement]{2011MNRAS.412.2183T} and \citet[for age]{2007ApJS..171..146S}.  For a discussion of why these two different sets of models are used, see \citet{2017MNRAS.472.2833S}. We select the best-fitting SSP-equivalent age, [Z/H] and [$\alpha$/Fe] using a $\chi^2$-minimisation approach following \citet{2004MNRAS.355.1327P}.

\section{Emission line value added products}\label{sec:lzifu}

The emission line value-added products have been one of the key parts of the previous data releases, and thus a systematic process has been built to create these. The final SAMI galaxy survey data release follows a similar fitting process, and full details of the process can be found in the earlier data releases \citep{2018MNRAS.475.716G,2018MNRAS.481.2299S}. We briefly summarise that process here and emphasize key differences with the previous data releases.

\begin{figure*}
    \centering
    \includegraphics[width=16cm]{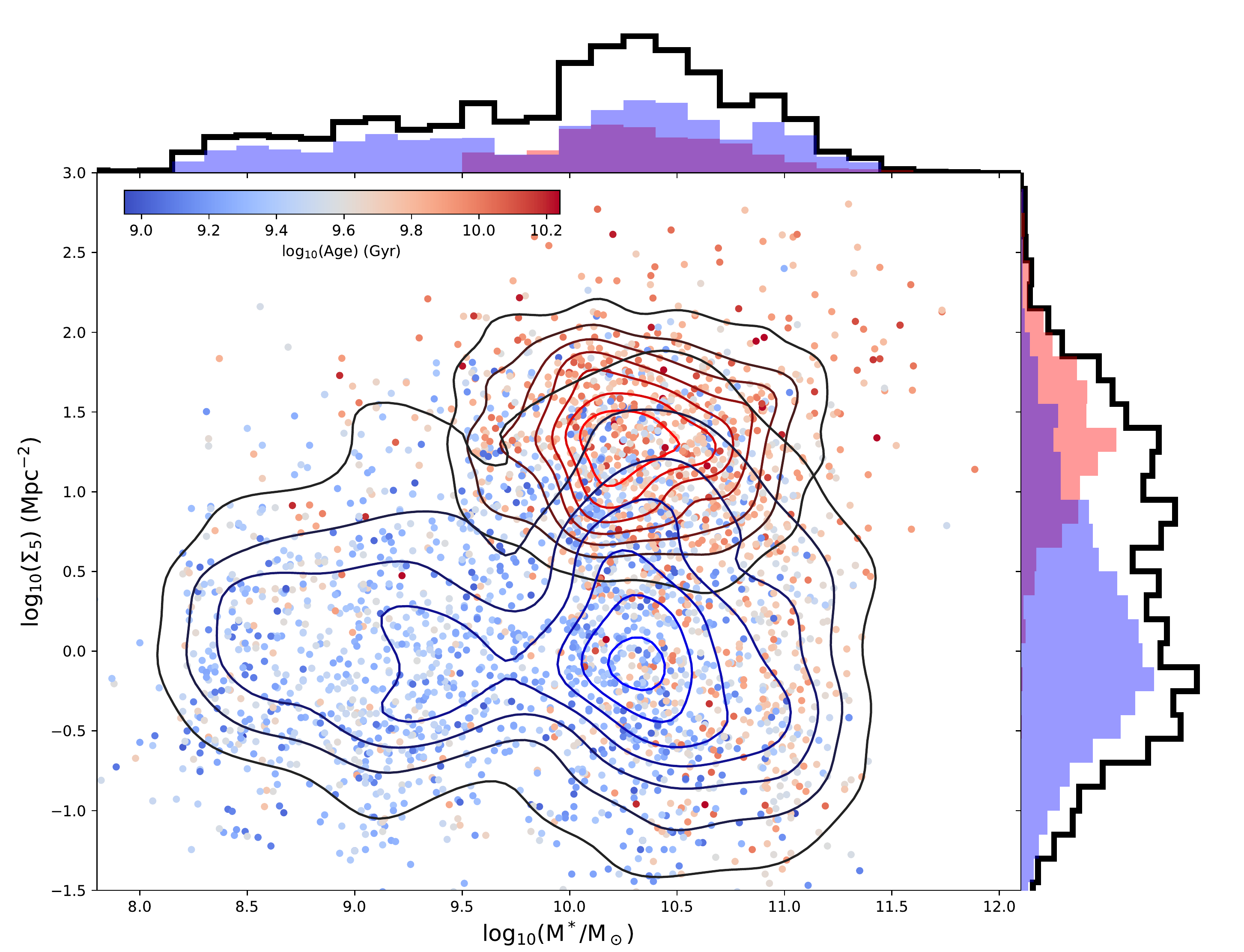}
    \caption{The filled circles show the distribution of SAMI DR3 galaxies in fifth-nearest-neighbour surface density ($\Sigma_5$, Mpc$^{-2}$) versus stellar mass (M$^*$/M$_\odot$). For each galaxy, the filled circles are colour-coded based on the light-weighted age determined from an aperture spectrum containing light from within $R_e$ in Section~\ref{sec:pop}. The  colourbar inset at the top left-hand corner shows the mapping from age to colour. The blue and red contours show the kernel-density estimates for the separate distributions of the GAMA and cluster samples, respectively, highlighting the contribution of the cluster regions, which allows the SAMI survey to probe the full range in environmental density. Histograms show the marginal distributions for the full DR3 sample (black line), the GAMA galaxies (blue shaded), and the cluster galaxies (red shaded). Note that only cluster galaxies with $\log_{10} (M^*/M_{\odot}) > 9.5$ were observed (see Section~\ref{subsec:clus_input}).}
    \label{fig:mstar_density}
\end{figure*}

For each of the data products we described in Section~\ref{sec:primedata} (the cubes, the various binned cubes and the aperture spectra) we fit the strong emission lines in each unique spatial element with one to three Gaussian profiles using the fitting code LZIFU \citep{2016ApSS.361..280H}. As in DR2, we first continuum subtract each spatial element using \textsc{pPXF} with the MILES simple stellar population (SSP) spectral library \citep{2010MNRAS.404.1639V} supplemented with younger SSP templates drawn from \citet{2005MNRAS.357..945G}. However, for the full resolution cubes there is not sufficient S/N in the continuum in each spatial element for a good fit, so we first fit the continuum and lines simultaneously in the Voronoi-binned cubes and then use this fit to refit the full-resolution cube using a limited set of templates with priors on the weights, as described in \citet{2019ApJ...873...52O}. However, even with this approach, there are spaxels (and even Voronoi-binned elements) where the spectra has a zero or even negative (due to sky-subtraction residuals) signal-to noise. In these cases we set the continuum to 0, but still attempt to fit any faint emission line that may exist. Due to differential atmospheric refraction some spaxels will not have complete spectral coverage at the edge of the SAMI field-of-view.  For emission line fitting we flag and mask (set to NaN) any spaxel that has more than 500 pixels without spectral data in either the blue or red cubes.  This is because the missing data can prevent a good continuum fit, and at times prevents lines from being fit as well.  

Once the continuum is subtracted, LZIFU then fits the emission lines in each spatial element in the data products with 1, 2, and 3 Gaussian components. The velocity centroids of the components are measured relative to the redshift of the galaxy (as defined by the input catalogue), and components are ordered by increasing velocity dispersion. As with previous data releases, we fit all emission lines simultaneously, tying the velocities and velocity dispersions together, but allowing the relative strengths of these components to vary. The emission lines we fit are \oii\,$\lambda\lambda 3727,3729$, \hb, \oiii\,$\lambda 5007$, \ha, \nii\,$\lambda 6583$, \sii\,$\lambda 6716$, and \sii\,$\lambda 6731$.  For the aperture spectra we also fit for the higher order Balmer lines, \hc, \hd\ and H\,$\epsilon$, as well as [Ne\,{\small III}]\,$\lambda3869$, due to the generally higher S/N in these spectra.

The resulting spectral fits are then run through a trained Neural Network LZCOMP \citep{Hampton2017} that determines whether 1, 2, or 3 Gaussian components are necessary to describe the observed emission line structure. This is returned as our "recommended" fit where each spatial element is described by the necessary number of components. As with previous data releases, we only provide the different fluxes in each component for the \ha\ line as it is generally the strongest.  The \ha\ line also lies in the higher spectral resolution red spectrograph, providing the most information on the spectral decomposition of the emission lines. For all other lines we only provide the total flux. The central velocity and velocity dispersion for each necessary component are given for each spatial element.

An exception to this is the annular-binned data and the aperture spectra. In this instance for many objects the annular binning leads to double-peaked (or horned) profiles due to the sampling of the rotational profile which LZCOMP is unable to process. Thus for all annular binned data we only present our 2-component Gaussian fits with LZIFU and return the flux in all emission lines for these data. Similarly for the aperture spectra we only report the 1 component fits within the tables. Given the generally higher S/N in these larger bins, most emission lines have sufficient flux for significant decomposition.

The final DR3 released emission line products for full-resolution and binned SAMI data cubes are: total line flux maps for \oii\,$\lambda\lambda 3727,3729$, \hb, \oiii\,$\lambda 5007$, \nii\,$\lambda 6583$, \ha, \sii\,$\lambda 6716$, and \sii\,$\lambda 6731$ using 1-component and multi-component Gaussian fits, velocity and velocity dispersion based on the 1-component fits and for each recommended component of the multi-Gaussian fits, and for the multi-Gaussian fits the decomposed flux of each component for \ha. 

\subsection{Emission line derived value-added products}

In addition to the emission line properties extracted from the SAMI data, we also derive standard value added products from this emission line data, namely: Balmer decrement derived extinction maps, classification maps based on standard emission line diagnostic diagrams, and star-formation maps. Full details of the derivations of these can be found in the DR2 paper and in \citet{2018MNRAS.475.5194M}.

The extinction maps are given as an \ha\ correction factor and assume an intrinsic Balmer decrement of ${\rm H\alpha/H\beta}=2.86$ and a \citet{1989ApJ...345..245C} extinction curve with $R_{V}=4.1$.  We note, as described in \citet{2018MNRAS.475.716G} and in \citet{2018MNRAS.475.5194M}, that the Balmer decrement is sensitive to the aliasing introduced by DAR and the sampling of the PSF of the SAMI hexabundles. For this reason, our extinction maps have been smoothed by a Gaussian kernel with a FWHM of 1.6 spaxels (0.8 arcsec) which corrects this issue \citep[with full details of the Kernel and the results shown in][]{2018MNRAS.475.5194M}. 

The classification maps distinguish star-formation dominated spaxels from spaxels where another ionising mechanism, such as AGN, supernova or diffuse ionised gas contribute significantly to the overall emission using emission-line ratio based diagnostics. As emphasized in previous papers, these classification maps are a conservative mask, capturing all spaxels where the SFR can be confidently determined. While star-formation can, and is likely to, contribute to the emission in masked spaxels, the other contributing sources such as AGN need to be accounted for in a non-straightforward process \citep[see, e.g.,][for examples of this]{2016MNRAS.462.1616D,2019MNRAS.485L..38D}. A full description of the classification algorithm is in \citet{2018MNRAS.475.5194M}.

The star-formation maps are based on the extinction-corrected H$\alpha$ flux for all star-forming dominated spaxels and use the \citet{1994ApJ...435...22K} relation, corrected to a \citet{2003PASP..115..763C} stellar initial mass function, to convert to a SFR;
\begin{equation}
    {\rm SFR\ [M_{\odot}\,yr^{-1}]}=5.16\times10^{-42}\ F_{\rm H\alpha}\ [{erg\, s^{-1}}].
\end{equation}
For full SFR maps where the impact of AGN etc.\ are neglected, the same equation can be applied directly to the extinction-corrected H$\alpha$ maps.

A final emission line related value-added product is the measurement of the gas kinematic PA.  This is measured in an identical way to the stellar kinematic PA discussed in Section \ref{subsubsec:kinematic_products}, but applied to the 1-component gas kinematic maps.

\section{Environmental metrics}\label{sec:env}

As outlined in Section~\ref{section:Input_cats}, DR3 includes SAMI data for the cluster regions for the first time. Therefore, when combined with the SAMI data drawn from the GAMA regions, the DR3 data span the full range of environmental densities, from the low-density field to the high-density regions found in the cores of rich clusters. In this Section, we outline the environment metrics included as value-added data products in DR3.

\subsection{Fifth-nearest neighbour surface density estimates}\label{subsection:nn_dense}

Fifth-nearest neighbour surface density estimates are obtained for targets in both the GAMA and cluster regions. The surface density estimates are determined in a similar manner to that described in \citet{2013MNRAS.435.2903B} and \citet{2017ApJ...844...59B}. Briefly, we use the redshift information provided by the GAMA survey \citep{2011MNRAS.413..971D,2015MNRAS.452.2087L} and the SAMI Cluster Redshift Survey \citep{2017MNRAS.468.1824O} to specify a density-defining population consisting of galaxies that have absolute $r$-band magnitudes $M_r < -18.6$ or $M_r < -19.0$\,mag, with the latter limit included to allow densities to be estimated for the secondary targets with $z>0.1$ in the GAMA regions. For the GAMA sample, the magnitudes of the density-defining population are $k$-corrected as outlined in \citet{2012MNRAS.420.1239L}, while for the cluster sample we use the {\sc CALC\_KCORR} software \citep{2010MNRAS.405.1409C}. 

\begin{figure}
    \centering
    \includegraphics[width=9cm]{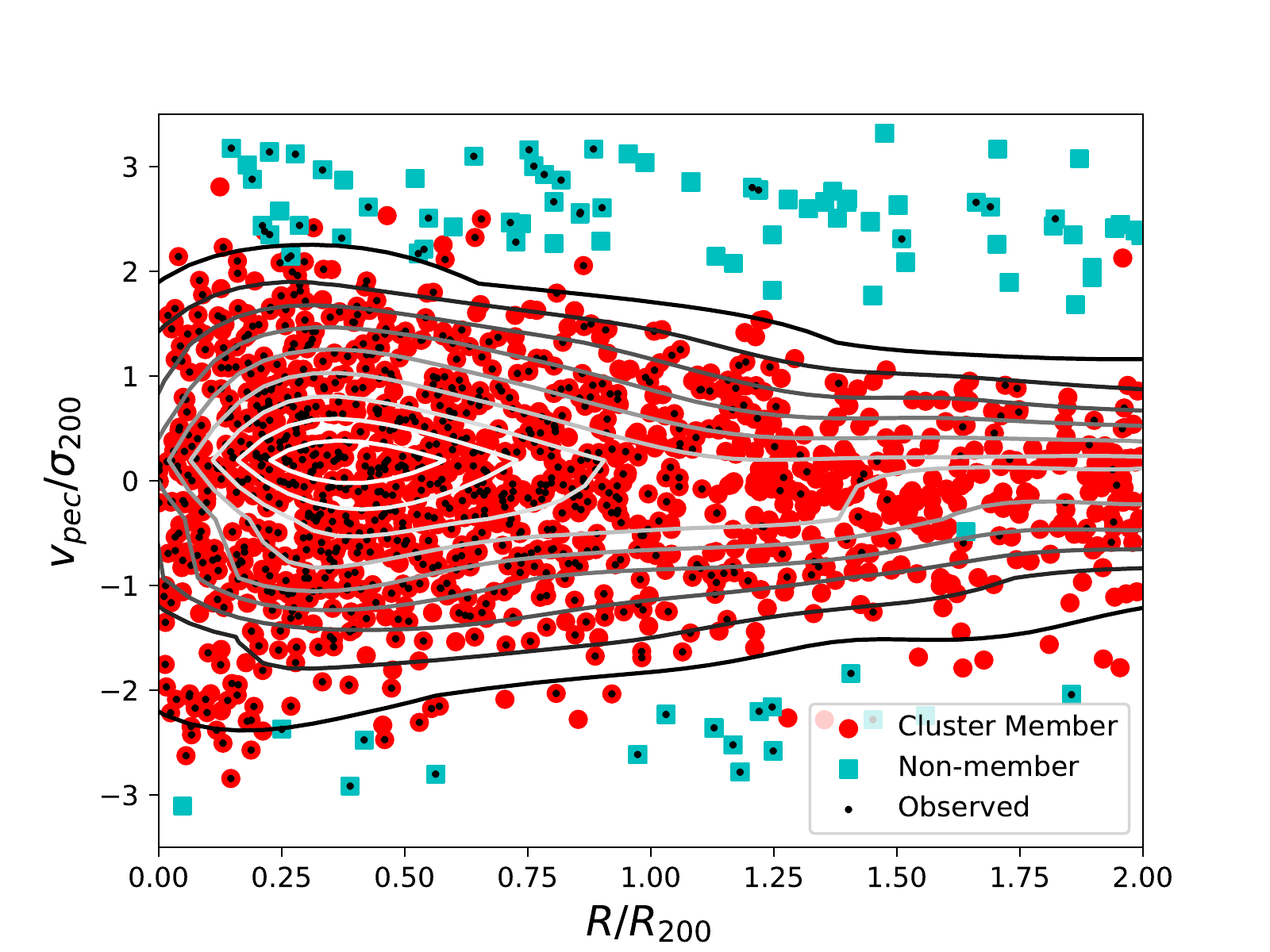}
    \caption{The combined projected-phase-space diagram for all galaxies in the input catalogue described in Section~\ref{subsec:clus_input}. Galaxies assigned cluster membership {\update in \citet{2017MNRAS.468.1824O}} are shown as red circles, while those {\update non-members} with $|v_{\rm pec}/\sigma_{200}| < 3.5$ {\update are shown as} cyan squares. Targets that were observed during the survey are highlighted by black points, while the contours show the kernel density estimate derived from all spectroscopically confirmed cluster members in the SAMI Cluster Redshift Survey. }
    \label{fig:PPS}
\end{figure}

\begin{figure*}
    \centering
    \includegraphics[width=17.5cm]{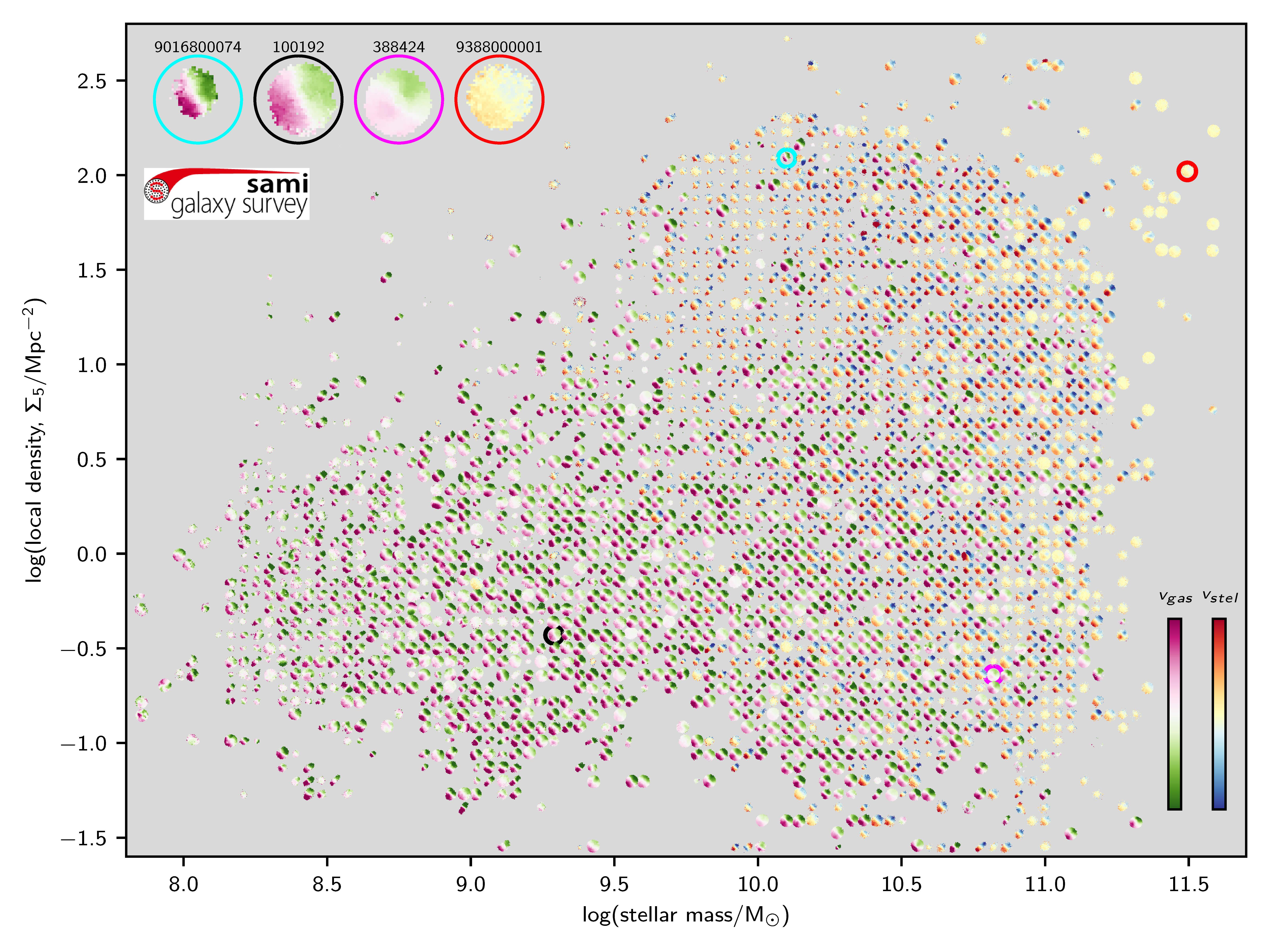}
    \caption{Velocity fields for SAMI galaxies in the plane of fifth-nearest-neighbour surface density ($\Sigma_5$) vs.\ stellar mass.  The velocity scale for each galaxy is normalized to a maximum of $0.7\times$ the velocity inferred from the stellar mass Tully-Fisher relation (hence there is no scale on the velocity colour bars).  Stellar velocities are shown for early-type galaxies, while gas kinematics are shown for late-type galaxies, as defined by the visual morphology catalogue.  Each velocity map has the same image scale in arcseconds.  Top {\update left} we show zoom-ins of four example galaxies, where their locations in $\Sigma_5$ and stellar mass are indicated by the coloured circles.  Zoom-ins of the same galaxies are shown in Figs.\ \ref{fig:mass_den_ha} and \ref{fig:mass_den_n2ha}.}
    \label{fig:mass_den_vel}
\end{figure*}

\begin{figure*}
    \centering
    \includegraphics[width=17.5cm]{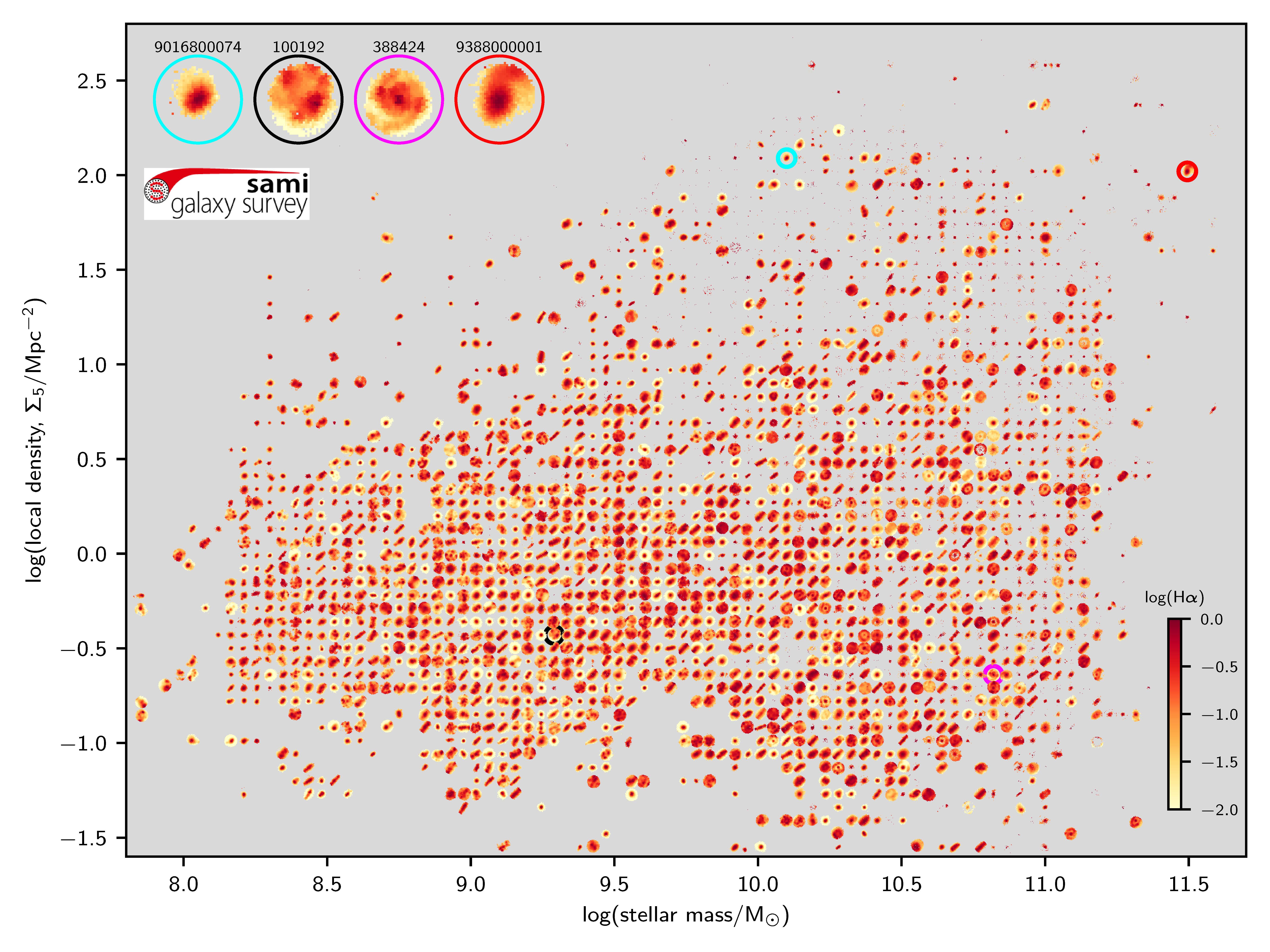}
    \caption{\ha\ maps for SAMI galaxies in the plane of fifth-nearest-neighbour surface density ($\Sigma_5$) vs.\ stellar mass.  The \ha\ flux is on a log scale and normalized such that $\log({\rm flux})=0$ for the maximum \ha\ in each galaxy.  The zoom-ins in the top {\update left} are the same galaxies as in Figs.\ \ref{fig:mass_den_vel} and \ref{fig:mass_den_n2ha}.}
    \label{fig:mass_den_ha}
\end{figure*}

\begin{figure*}
    \centering
    \includegraphics[width=17.5cm]{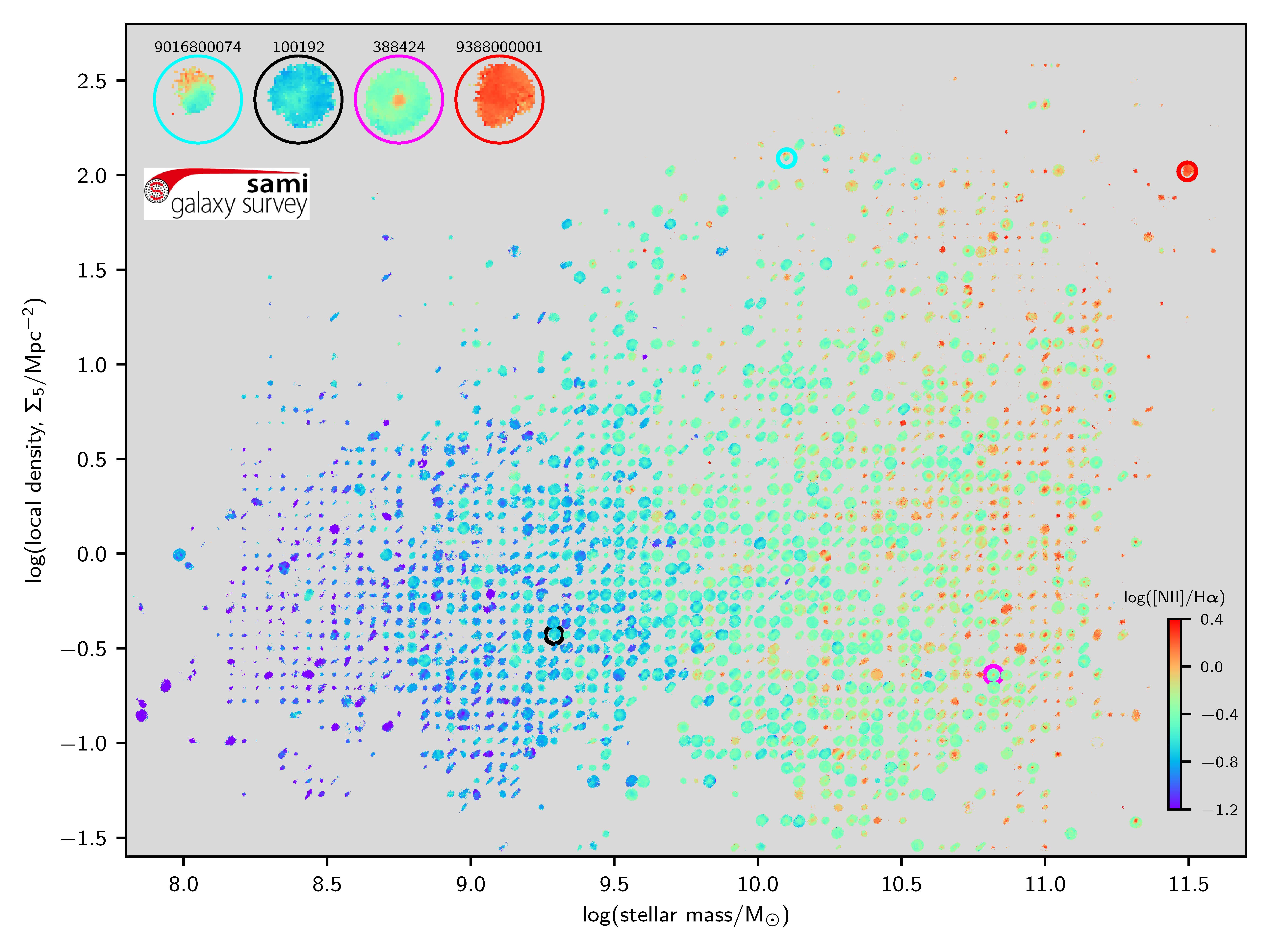}
    \caption{Maps of log([N\,{\small II}]/H$\alpha$) flux ratio for SAMI galaxies in the plane of fifth-nearest-neighbour surface density ($\Sigma_5$) vs.\ stellar mass.  The zoom-ins in the top {\update left} are the same galaxies as in Figs.\ \ref{fig:mass_den_vel} and \ref{fig:mass_den_ha}.}
    \label{fig:mass_den_n2ha}
\end{figure*}

The surface density for each SAMI target is determined as $\Sigma_5 = 5/\pi d_n^2$, where $d_n$ is the projected comoving distance to the 5th nearest galaxy in the density-defining population that is within $\pm 1000\,{\rm km/s}$ of the SAMI target redshift. The $\Sigma_5$ values are corrected to account for the effect of  spectroscopic incompleteness.  In 95\% of the cases that require a correction, the $\Sigma_5$ values increase by a factor that is smaller than 1.15. Uncertainties on the $\Sigma_5$ values are determined as $\Sigma_{5, err} = \max (|\Sigma_5-\Sigma_4|,|\Sigma_5-\Sigma_6|)$, where $\Sigma_4$ and $\Sigma_6$ are the surface densities determined to the 4th and 6th nearest neighbours, respectively. For the GAMA $\Sigma_5$ measurements, quality flags are assigned based on proximity to survey edges and bright star exclusion regions. For the cluster $\Sigma_5$ measurements, flags are assigned based on proximity to survey edges, although in practice this only occurs for a small number of cluster targets ($<1\%$).

Fig.\ \ref{fig:mstar_density} shows the distribution of observed SAMI targets on the $\log_{10} (M^*/M_{\odot})$ versus $\log_{10} (\Sigma_5)$ plane, where the points have been colour-coded based on the light-weighted SSP ages measured within $1\,R_e$ apertures as described in Section~\ref{sec:pop}. Contours show the kernel-density estimates separately for the GAMA and cluster targets, while the histograms show the marginal distributions of the full sample (black) and the GAMA (blue shaded) and cluster (red shaded) samples. Fig.\ \ref{fig:mstar_density} demonstrates the breadth of environmental density that the combined DR3 sample probes.

\subsection{Cluster-specific measurements}\label{subsection:cluster_env}

To complement the newly released cluster targets in DR3, within the cluster input catalogues we include three cluster-specific environment metrics: a member/non-member flag, normalised clustercentric distances ($R/R_{200}$), and normalised peculiar velocities ($v_{\rm pec}/\sigma_{200}$). The description of the measurements of these metrics can be found in \citet{2017MNRAS.468.1824O}, and we provide a brief overview here.

The $R/R_{200}$ measurements are determined as the projected angular distance of the target to the cluster centre listed in Table~\ref{table:clusters}, where the conversion from angular to a physical separation is determined using the angular diameter distance to the cluster. The normalisation by the virial radius estimator, $R_{200}$, allows for more physically meaningful comparisons between the clustercentric distance of galaxies in clusters of disparate halo mass. Here, $R_{200}=0.17\sigma_{200}/H(z)\,{\rm Mpc}$ is determined iteratively using the cluster velocity dispersion, $\sigma_{200}$, which is derived from the cluster members within $R_{200}$ using the biweight estimator \citep{1990AJ....100...32B}. The peculiar velocities for the SAMI targets are determined as $v_{\rm pec}=c[(1+z_{\rm pec})^2-1]/[(1+z_{\rm pec})^2+1]$ where $z_{\rm pec}=(z-z_{\rm clus})/(1+z_{\rm clus})$, $z$ is the redshift of the SAMI target determined from the SAMI Cluster Redshift Survey, and $z_{\rm clus}$ is the cluster redshift determined using the biweight mean of the cluster members ($z_{\rm clus}$). The values of $R_{200}$, $z_{\rm clus}$, and $\sigma_{200}$ for each cluster are listed in Table~\ref{table:clusters}. 

As noted in Section~\ref{subsec:clus_input}, the selection of SAMI targets used a more relaxed cut in $v_{\rm pec}$ when compared with that used to define cluster members in \citet{2017MNRAS.468.1824O}. This less-conservative cut was designed to allow for the inherent uncertainty in assigning cluster membership, which may exclude important high-velocity galaxies infalling onto the cluster for the first time. We therefore include a flag in our catalogue that indicates whether a SAMI target was allocated as a cluster member in \citet{2017MNRAS.468.1824O}. In Fig.\ \ref{fig:PPS} we show the projected-phase-space (PPS) diagram for the SAMI cluster targets.  The cluster environmental metrics are contained within the input catalogue for the cluster sample, InputCatClustersDR3.

Also listed in Table~\ref{table:clusters} are estimates for the virial mass of the clusters, which are determined using the spectroscopically confirmed members as outlined in \citet{2017MNRAS.468.1824O}. Halo masses for groups in the GAMA regions will be made available in a forthcoming release (Robotham et al. {\it in prep.}), and the SAMI DR3 galaxies in the GAMA regions can be linked to these groups via their CATIDs. It is worth reiterating that the virial mass determinations for the clusters in \citet{2017MNRAS.468.1824O} differ from those determined by \citet{2011MNRAS.416.2640R} for the groups in the GAMA survey both in terms of the assumed cosmology and the methodology. As outlined in \citet{2017MNRAS.468.1824O}, care must therefore be taken in combining the catalogues. After rescaling masses to the same assumed cosmology, \citet{2017MNRAS.468.1824O} recommend that the cluster virial mass estimates should be scaled by a factor of 1.25 in order to match the virial-like mass estimates determined for the GAMA groups in \citet{2011MNRAS.416.2640R}.

\subsection{SAMI galaxies across the mass-environment plane}

A fundamental aim of the SAMI Galaxy Survey was to measure the internal structure of galaxies across mass and environment.  To visualise this Figs.\ \ref{fig:mass_den_vel}, \ref{fig:mass_den_ha} and \ref{fig:mass_den_n2ha} present velocity maps, \ha\ flux maps and $\log$(\nii/\ha) maps in the plane of stellar mass vs. $\Sigma_5$.   Below we provide a qualitative description of some of the trends seen when visualizing the SAMI data in this way.

In Fig.\ \ref{fig:mass_den_vel} we show gas velocity maps for late-type galaxies (green to purple colour scale) and stellar velocity maps for early-type galaxies (blue to red colour scale).  For each plotted velocity map the maximum and minimum velocity is set to be $0.7\times$ the velocity inferred from the stellar mass of the galaxy, using the Tully-Fisher \citep{1977A&A....54..661T} relation as measured by the SAMI Galaxy Survey \citep{2017MNRAS.472.1809B}.  The position angle of the velocity maps is set by the stellar kinematic PA, or the gas kinematic PA if the error in the stellar kinematic PA is larger than 20 degrees.  The location of the maps approximately coincides with the value of stellar mass and $\Sigma_5$ for each galaxy, but to reduce overlapping of the maps each galaxy was positioned in the nearest available location in a grid.

Above $\Sigma_5\simeq10$\,Mpc$^{-2}$ the majority of SAMI galaxies are from the cluster fields.  The increased fraction of early-type galaxies above this value can be clearly seen, as can the increased number of early types towards high stellar mass.  At all masses and environments there are some galaxies with low velocity gradients relative to their expected Tully-Fisher rotational velocity.  These low gradients are largely due to inclination effects.  However, at the highest masses [above $\log(M_*/M_\odot)=11$] there is an increased number of galaxies with low velocity gradients, consistent with the increased fraction of slow rotators in this mass range \citep[e.g.][]{2017ApJ...844...59B}.  It is less clear from Fig.\ \ref{fig:mass_den_vel} whether the galaxies with the highest $\Sigma_5$ have lower rotation.  {\update In fact, we see suggestions of an increased fraction of slow rotators with mass at all values of $\Sigma_5$.  This will be investigated fully by van de Sande et al.\ (in preparation).}  At the lowest masses ($\log(M_*/M_\odot)<8.5$), more galaxies also tend to have lower rotation compared to their expected values from the Tully-Fisher relation, as pointed out by \citet{2017MNRAS.472.1809B}. 

Fig.\ \ref{fig:mass_den_ha} shows \ha\ maps across the stellar mass vs.\ $\Sigma_5$ plane.  The \ha\ maps are log-scaled and normalized relative to the brightest \ha\ flux in each galaxy.  As $\Sigma_5$ increases, fewer galaxies show evidence of extended \ha\ emission, with either no emission at all, or the emission appearing more compact.  The \ha\ morphology also varies with stellar mass.  As mass increases, galaxies often have a bright central peak, that is sometimes surrounded by a ring, providing a `target-like' morphology.

The ionization state of the gas can be visualized using $\log$(\nii/\ha) flux ratio maps, as in Fig.\ \ref{fig:mass_den_n2ha}.  The most significant trend visible here is the change from low $\log$(\nii/\ha) (blue colours) at low mass, to high $\log$(\nii/\ha) (green colours) at high mass.  This is driven by increasing gas-phase metallicity with increasing mass. $\log$(\nii/\ha) values above $\sim-0.2$ cannot be caused by star formation, but must be driven by other sources of ionization, such as AGN or shocks.  This non-star-formation emission becomes more dominant at high stellar mass.
The \nii/\ha\ ratio maps show a diversity of structure, including examples of galaxies with enhanced \nii/\ha\ at large radius, indicative of shocks in galactic-scale outflows or diffuse ionized gas.  At high stellar mass many of the galaxies with central \ha\ flux peaks (sometimes surrounded by rings) have high central \nii/\ha\ ratios, indicating LINER or AGN-like activity.  This is likely to be driven by increased bulge fraction in high mass disks that have central regions that are partially or completely quenched. 

In Figs.\ \ref{fig:mass_den_vel}, \ref{fig:mass_den_ha} and \ref{fig:mass_den_n2ha} we show zoom-ins of four example galaxies (top {\update left}), using coloured circles to indicate their location in $\Sigma_5$ and stellar mass.  The first example (galaxy 9016800074, cyan circle) is a late-type galaxy in cluster Abell 168 with regular rotation, but assymetric \ha\ emission on one side of the galaxy.  The assymetric gas has a high \nii/\ha\ ratio, suggestive of shocks, possibly caused by ram-pressure stripping \citep{2019ApJ...873...52O}.  Our second example (galaxy 100192, black circle) is a lower-mass late-type galaxy in a low density environment.  100192 has a regular rotating disk, including clumpy star formation and a radial \nii/\ha\ gradient, indicating a metallicity gradient.  The third example (galaxy 288424, magenta circle) is a higher mass disk galaxy that is close to face-on.  The central \nii/\ha\ ratio for 288424 is high, suggestive of AGN or LINER-like emission. Our final example (galaxy 9388000001, red circle) is the brightest cluster galaxy in Abell 3880.  Its stellar kinematics show it to be a slow rotator, but the galaxy has strong \ha\ emission.  The \ha\ emission is centrally concentrated, but with an extended plume that was also noted by \citet{2016MNRAS.460.1758H}.  All the ionized gas emission in 9388000001 has a high \nii/\ha\ ratio, consistent with shocks or AGN emission.  This galaxy is a known bright radio source \citep[PKS 2225-308;][]{1974AuJPA..32....1S}, and the ionization is likely related to this.

The maps presented here give just a few different examples of the richness of large-scale IFS data sets such as SAMI.

\section{SAMI DR3 products and access}\label{sec:access}

All SAMI DR3 data and products are available through Australian Astronomical Optics' Data Central service at https://datacentral.org.au/.  Data Central provides query tools based on a SQL framework to search and connect data from different SAMI catalogues and also federate it with data from other surveys such as GAMA \citep{2011MNRAS.413..971D}.  A simple query to select observed galaxies in a particular stellar mass range is given below:
\begin{verbatim}
SELECT t1.CATID, t1.RA_OBJ, t1.DEC_OBJ, t1.Mstar,
t2.CUBEID, t2.CUBEIDPUB
FROM sami_dr3.InputCatGAMADR3 as t1
INNER JOIN sami_dr3.CubeObs as t2 on 
t2.CATID = t1.CATID
WHERE  (t1.Mstar BETWEEN 10.5 and 11.0) and
(t2.ISBEST = True)
\end{verbatim}

Data products related to the output of a given SQL search can then be downloaded.

Cubes, images and spectra can be downloaded as FITS files.  Tables can be downloaded in a range of formats including FITS, CSV and VOTABLE files.   Data Central contains a full schema browser that describes the data products as well as detailed documentation.

\section{Summary}\label{sec:summary}

With this paper we have released data for all the observations made as part of the SAMI Galaxy Survey, as well as a range of value-added data products, including gas and stellar measurements.  This includes data for 3068 unique galaxies.  For the first time, the SAMI Galaxy Survey is releasing data within the eight clusters targeted as part of the survey.   These data provide a unique view of the role of dense environments in influencing galaxy evolution, as seen in the local Universe.  The SAMI Galaxy Survey has already enabled a diverse array of investigations into galaxy evolution, and we hope that the public release of this data set will enable many more.

There are a number of important next steps beyond the SAMI Galaxy Survey that can further elucidate the complexities of galaxy formation.  In the local Universe, larger samples and higher resolution (both spatial and spectral) are key directions.  Higher spectral resolution is particularly important to understand the kinematic signatures in low-mass galaxies, as well as high-mass disks and in cases with complex kinematics such as outflows.  The Hector instrument, to be commissioned on the Anglo-Australian Telescope in 2021 will provide this higher spectral resolution for over 10000 galaxies \citep{2016SPIE.9908E..1FB}.  Larger telescopes with greater sensitivity are also starting to make IFS measurements outside of the local Universe, although the number that have included stellar measurements as well as gas is limited.  New projects such as the  Middle-Ages Galaxy Properties with Integral Field Spectroscopy Survey \citep[MAGPI;][]{2020arXiv201113567F} are using the Multi-Unit Spectroscopic Explorer (MUSE) instrument \citep{2010SPIE.7735E..08B} to push stellar kinematics measurements to higher redshift.  In the future larger telescopes still will be required to overcome the surface brightness limitations of current facilities.  The prospect of multiplexing extremely large telescopes, for example using the MANIFEST facility on the Giant Magellan Telescope \citep{2010SPIE.7735E..68S}, opens up the possibility of carrying out SAMI-like surveys in the distant Universe.

\section*{Data availability}

The data presented in this paper are available from Astronomical Optics' Data Central service at https://datacentral.org.au/. 

\section*{Acknowledgements}

The SAMI Galaxy Survey is based on observations made at the
Anglo-Australian Telescope. The Sydney-AAO Multi-object Integral field
spectrograph (SAMI) was developed jointly by the University of Sydney
and the Australian Astronomical Observatory. The SAMI input catalogue
is based on data taken from the Sloan Digital Sky Survey, the GAMA
Survey and the VST ATLAS Survey. The SAMI Galaxy Survey website is
{http://sami-survey.org/}. The SAMI Galaxy Survey is supported by the
Australian Research Council Centre of Excellence for All Sky
Astrophysics in 3 Dimensions (ASTRO 3D), through project number
CE170100013, the Australian Research Council Centre of Excellence for
All-sky Astrophysics (CAASTRO), through project number CE110001020,
and other participating institutions.

Based on data acquired at the Anglo-Australian Telescope under programs A/2013B/012 and A/2016B/16. We acknowledge the traditional owners of the land on which the AAT stands, the Gamilaraay people, and pay our respects to elders past and present.

GAMA is a joint European-Australasian project based around a
spectroscopic campaign using the Anglo-Australian Telescope. The GAMA
input catalogue is based on data taken from the Sloan Digital Sky
Survey and the UKIRT Infrared Deep Sky Survey. Complementary imaging
of the GAMA regions is being obtained by a number of independent
survey programmes including GALEX MIS, VST KiDS, VISTA VIKING, WISE,
Herschel-ATLAS, GMRT and ASKAP providing UV to radio coverage. GAMA is
funded by the STFC (UK), the ARC (Australia), the AAO, and the
participating institutions. The GAMA website is
http://www.gama-survey.org/ .

SMC acknowledges the hospitality of the Astronomy, Astrophysics and Astrophotonics Research Centre at Macquarie University while preparing parts of this work.
MSO acknowledges the funding support from the Australian Research
Council through a Future Fellowship (FT140100255).  JJB acknowledges
support of an Australian Research Council Future Fellowship
(FT180100231).  FDE acknowledges funding through the H2020 ERC Consolidator Grant
683184.  JvdS acknowledges support of an Australian Research Council
Discovery Early Career Research Award (project number DE200100461)
funded by the Australian Government. NS acknowledges support of an Australian Research Council Discovery Early Career Research Award (project number DE190100375) funded by the Australian Government. TMB is supported by an Australian Government Research Training Program Scholarship. DO is a recipient of an
Australian Research Council Future Fellowship (FT190100083) funded by
the Australian Government. LC is the recipient of an Australian Research Council Future Fellowship (FT180100066) funded by the Australian Government. SB acknowledges funding support from the
Australian Research Council through a Future Fellowship
(FT140101166).  This work was supported by the UK Science and Technology Facilities Council through the `Astrophysics at Oxford' grant ST/K00106X/1. RLD acknowledges travel and computer grants from Christ Church, Oxford and support from the Oxford Hintze Centre for Astrophysical Surveys which is funded by the Hintze Family Charitable Foundation.  The work was undertaken in collaboration with the
Melbourne Data Analytics Platform (MDAP) at the University of
Melbourne. This material is based in part upon work supported by the National Science Foundation under Grant No. 2009416. 
We thank Mike Bessell for discussions regarding extinction
at Siding Spring Observatory.

This paper made use of the \texttt{astropy} \textsc{python} package \citep{astropy}, as well as the \texttt{matplotlib} plotting software \citep{matplotlib} and the scientific libraries \texttt{numpy} \citep{numpy}, \texttt{IPython} \citep{ipython} and \texttt{scipy} \citep{scipy}. 




\bibliographystyle{mnras}
\bibliography{bibliographies}







\bsp	
\label{lastpage}
\end{document}